\documentclass[%
 reprint,
nofootinbib,
 amsmath,amssymb,
 aps,
 longbibliography,
]{revtex4-1}

\usepackage{graphicx}
\usepackage{dcolumn}
\usepackage{bm}
\usepackage{xspace}
\usepackage{xcolor}

\usepackage{units}

\newcommand{\sibyll}{{\textsc{Sibyll}}\xspace}
\newcommand{\qgsjet}{{\textsc{QGSJet}}}
\newcommand{\eposlhc}{{\textsc{EPOS-LHC}}\xspace}

\def\xmax{\ensuremath{\langle X_{\rm max} \rangle}\xspace}

\def\nmu{\ensuremath{N_\mu}\xspace}

\def\Pom{{I\!\!P}}

\def\Reg{{I\!\!R}}

\begin{document}

\preprint{APS/123-QED}

\title{Hadronic interaction model \sibyll~2.3d
       and extensive air showers}

\author{Felix Riehn}
\email{friehn@lip.pt}
\affiliation{Instituto Galego de F\'isica de Altas Enerx\'ias (IGFAE),
Universidade de Santiago de Compostela, 15782 Santiago de Compostela, Spain}
\affiliation{Laborat\'{o}rio de Instrumenta\c{c}\~{a}o e F\'{i}sica Experimental de Part\'{i}culas (LIP) - Lisbon, Avenida Professor Gama Pinto 2, 1649-003 Lisbon, Portugal}
\affiliation{Department of Physics and Astronomy, University of Delaware, Newark, DE 19716, USA}
\affiliation{Karlsruher Institut f\"{u}r Technologie, Institut f\"{u}r Kernphysik, Postfach 3640, 76021 Karlsruhe, Germany}

\author{Ralph Engel}
\email{ralph.engel@kit.edu}
\affiliation{Karlsruher Institut f\"{u}r Technologie, Institut f\"{u}r Kernphysik, Postfach 3640, 76021 Karlsruhe, Germany}

\author{Anatoli Fedynitch}
\affiliation{Institute for Cosmic Ray Research, the University of Tokyo,
5-1-5 Kashiwa-no-ha, Kashiwa, Chiba 277-8582, Japan}
\affiliation{DESY, Platanenallee 6, 15738 Zeuthen, Germany}
\affiliation{Karlsruher Institut f\"ur Technologie, Institut f\"ur Kernphysik,
Postfach 3640, 76021 Karlsruhe, Germany}

\author{Thomas K. Gaisser}
\affiliation{Bartol Research Institute, Department of Physics and Astronomy, University of Delaware, Newark, Delaware 19716, USA}

\author{Todor Stanev}
\affiliation{Bartol Research Institute, Department of Physics and Astronomy, University of Delaware, Newark, Delaware 19716, USA}

\date{\today}

\begin{abstract}
We present a new version of the hadron interaction event generator \sibyll. While the core ideas of the model have been preserved, the new version handles the production of baryon pairs and leading particles in a new way. In addition, production of charmed hadrons is included. Updates to the model are informed by high-precision measurements of the total and inelastic cross sections with the forward detectors at the LHC that constrain the extrapolation to ultrahigh energy. Minimum-bias measurements of particle spectra and multiplicities support the tuning of fragmentation parameters. This paper demonstrates the impact of these changes on air-shower observables such as $X_{\rm max}$ and $N_\mu$, drawing comparisons with other contemporary cosmic-ray interaction models.

\end{abstract}

\pacs{Valid PACS appear here}
\maketitle


\section{\label{sec:intro}Introduction}

Studying cosmic rays at energies above $100\,$TeV imposes a challenge since the intensity is too low for direct measurements with high-altitude balloons or spacecraft. Instead the properties of the primary cosmic-ray nucleus must be inferred indirectly from the properties of extensive air showers (EAS) that can be observed with large, ground-based detectors. At energies in excess of several tens or hundreds of PeV (so-called ultrahigh energy cosmic-rays (UHECR)) the event rate per unit area and solid angle quickly drops, requiring ever larger and more sparsely instrumented detectors. Therefore, the interpretation of these air-shower data has necessarily to rely on detailed Monte Carlo simulations of the shower development and the experimental observables. The main challenge in these simulations is the modeling of nuclear and hadronic interactions that can occur at all possible energies ranging from the MeV up to ultrahigh energies $E \sim 10^{21}\,$eV. While interactions of hadrons with protons and nuclei are well studied up to several hundreds of GeV (in target rest frame) at fixed target detectors, at the highest energies it is necessary to rely on model extrapolations from collider experiments that measure primarily the central region.
This leads to the subclass of event generators in high-energy physics called cosmic-ray interaction models.

\sibyll is one of the first microscopic event generators for EAS~\cite{Fletcher:1994bd} and it is based in its core on the dual parton model (DPM)~\cite{Capella:1992yb} and the minijet model~\cite{Gaisser:1984pg,Pancheri85a,Pancheri:1986qg,Durand:1987yv}. Particle formation (or hadronization) is adopted from the Lund algorithms~\cite{Bengtsson87,Sjostrand:1987xj} and shares in this sense many ideas about the interactions of color strings with the popular {\sc Pythia} event generators \cite{Sjostrand:2006za}. A summary of the principles and ideas behind \sibyll and a review of its long history can be found in Ref.~\cite{Engel:2017wpa}.

From the beginning \sibyll{} aimed to describe a broad range of $pp(\bar{p})$ measurements at the Intersecting Storage Rings (ISR), the $Sp(\bar{p})S$ at CERN and the Tevatron at Fermilab, providing the highest interaction energies available at that time; for example, the growth of the average transverse momentum with center-of-mass (c.m.\ ) energy is adjusted according to the results of the CDF experiment at the Tevatron, UA1 at the S$p\bar{p}$S and the ISR at CERN~\cite{Abe:1988yu,Arnison:1982ed,Capiluppi:1973fz}. The hard interaction cross section is calculated in the minijet model. The Glauber scattering theory~\cite{Glauber:1970jm} is applied in hadron-nucleus collisions and extended with a semisuperposition approach~\cite{Engel:1992vf} to nucleus-nucleus collisions. 

Since the previous version 2.1~\cite{Ahn:2009wx} soft interactions and diffraction dissociation are implemented in a more sophisticated way by including multiple soft interactions and a two-channel eikonal model for diffraction, respectively. The current extension of the model is motivated by recent developments in cosmic-ray (CR) and astroparticle physics and new measurements at accelerators. At the high-energy frontier, the LHC provides for the first time constraints on extrapolation of the model to energies corresponding to cosmic rays beyond the knee. In addition, dedicated forward physics experiments (for example LHCf and CASTOR) and recent fixed target experiments (NA61) studied a larger part of the phase space that is particularly important for EAS. 

There are several challenges for the present cosmic-ray interaction models. One example arises in the interpretation of EAS data in terms of CR mass composition where  simulations predict a lower muon content than required to interpret the observations~\cite{AbuZayyad:1999xa,Aab:2014pza,Aab:2016hkv}. This challenge is specifically addressed by careful evaluation of $\rho^0$ and $p/\bar{p}$ production, both of which increase muon content in EAS.

Another example is the need to include production of charmed hadrons in event generators for EAS. The observation of high-energy astrophysical neutrinos above $100\,$TeV by IceCube~\cite{Aartsen:2013jdh,Aartsen:2014gkd} extends to the energy range where prompt muons and neutrinos from decays of charmed hadrons become larger than the conventional (light meson) channels. Eventually prompt muons and neutrinos become the main leptonic backgrounds for the astrophysical neutrino flux. Production of charm was first introduced as a modification of \sibyll~2.1~\cite{Ahn:2011wt}. Its implementation in \sibyll~2.3d is based on comparison with recent accelerator data on production of charmed hadrons and fully supports the production of charm~\cite{Fedynitch:2015zma,Fedynitch:2017fds,Fedynitch:2018cbl}. The model of the production of charm and the application of \sibyll~2.3d to the calculation of inclusive lepton fluxes is the subject of a separate paper~\cite{Fedynitch:2018cbl}.

The objective of this paper is twofold. The first is a description of the post-LHC version \sibyll~2.3d.\footnote{Preliminary versions of this model were released as \sibyll~2.3~\cite{Riehn:2015oba} and \sibyll~2.3c~\cite{Riehn:2017mfm,Fedynitch:2018cbl}.
Explanations of the changes between versions can be found in Appendix \ref{app:version_history}. } The changes to the microscopic interaction model with respect to the predecessor are detailed in \S~II. The second objective is the evaluation of the impact on EAS observables. \S~III contains the benchmark calculations and comparisons against other contemporary post-LHC models~\cite{Ostapchenko:2010vb,Pierog:2013ria} including the previous \sibyll~2.1. We conclude with a discussion in \S~IV.

\section{\label{sec:had-int}Model updates}
\subsection{\label{sec:basic}Basic model}
The aim of the event generator \sibyll is to account for the main features of strong interactions and hadronic particle production as needed for understanding air-shower cascades and inclusive secondary particle fluxes due to the interaction of cosmic rays in the Earth's atmosphere. Therefore, the focus is on the description of particle production at small angles and on the flow of energy in the projectile direction. Rare processes, such as the production of particles or jets at large $p_{\rm T}$ or electroweak processes, are either included approximately or neglected.

The model supports interactions between hadrons (mostly nucleons, pions or kaons) and light nuclei (h--A), since the targets in EAS mainly are nitrogen and oxygen. The CR flux at the top of the atmosphere contains elements up to iron, requiring a model for interactions of nuclei (A--A). Nuclear binding energies have negligible impact for high-energy interactions, allowing for the approximate construction of interactions of cosmic-ray nuclei from individual hadron-nucleon (h--N) collisions. On the target side, nucleons are combined to light nuclei on amplitude level using the Glauber model~\cite{Glauber:1955qq,Glauber:1970jm} together with the semisuperposition~\cite{Engel:1992vf} approach. This means that the interaction of an iron nucleus ($A=56$), for example, with a nitrogen nucleus in air is treated as $56$ separate nucleon--nitrogen interactions. With the exception of inelastic screening (Sect.~\ref{sec:nuc_diff}), the model extensions discussed in the following are introduced at the level of hadron-nucleon interactions.

\subsubsection{Parton level\label{sec:parton_level}}
The total scattering amplitude that determines the interaction cross sections is defined in impact parameter space by using the eikonal approximation, see Refs.~\cite{Block:1984ru,Durand:1987yv,Block:2006hy} and, for a pedagogical introduction, also Ref.~\cite{Gaisser:2016uoy},
\begin{equation}
  \label{eq:impactamp}
  a(s,\vec{b} \, ) \, = \, \frac{i}{2} \,[ 1-\exp (-\chi(s,\vec{b} \,))] \ ,
\end{equation}
where $i$ is the unit imaginary number, $\vec{b}$ is the impact parameter of the collision and $s$ is the Mandelstam variable, which for the interaction between hadrons $k$ and $l$ is defined as $s=(p_k+p_l)^2$. The eikonal function $\chi$ is given by the sum of two terms representing 
soft and hard interactions
$\chi(s,\vec{b} \,)=\chi_{\rm soft}(s,\vec{b} \,) + \chi_{\rm
  hard}(s,\vec{b} \,)$, and then unitarized as in Eq.~\eqref{eq:impactamp} ($|a| \leq 0.5$). The soft and hard eikonal functions take the form
\begin{equation}
  \chi_{\rm int}(s,\vec{b}\,) \, = \, \sigma_{\rm int}(s) \, A_{\rm int}(s,\vec{b}\,) ,
  \label{eq:eikonal}
\end{equation}
with $\int A_{\rm int}(s,\vec{b}\,)\ {\rm d}^2\vec{b} = 1$ and ${\rm int} = {\rm soft}, {\rm hard}.$

Within the parton model, there is a straightforward interpretation of Eq.~(\ref{eq:eikonal}) for hard interactions of asymptotically free partons. Then $\sigma_{\rm hard}$ is the inclusive hard scattering cross section of partons in the interaction of hadron $k$ with hadron $l$. 
The spatial distribution of partons available for hard interaction is encoded in the overlap function $A_{\rm hard}(s,\vec{b}\,)$. This overlap function between hadrons $k$ and $l$ is given by the individual transverse profile functions of partons in the scattering hadrons, $A_{k/l}(s,\vec{b}_l)$, and the transverse profile of the individual parton--parton interaction, $A_{\rm par}(s,\vec{b}_{\rm par})$,
\begin{align}
  A_{\rm hard}(s,\vec{b} \, ) \, =
  & \int \mathrm{d}^2\vec{b}_k \, \mathrm{d}^2\vec{b}_l \, \mathrm{d}^2\vec{b}_{\rm par} \\
  & \times A_{k}(s,\vec{b}_k) \, A_{l}(s,\vec{b}_l) \, A_{\rm par}(s,\vec{b}_{\rm par})  \nonumber \\
  & \times \delta^{(2)}( \vec{b}_k - \vec{b}_l + \vec{b}_{\rm par} - \vec{b}) \ , \label{eq:overlap}
\end{align}
where $\vec{b}_{k/l}$ are the positions of the interacting partons in the hadrons $k$ and $l$ and $\vec{b}_{\rm par}$ is the impact parameter between the partons, see Ref.~\cite{Ahn:2009wx}. For pointlike parton-parton interactions, $A_{\rm par}$ would be a Dirac $\delta$-function.

A geometrical (gluon) saturation condition~\cite{Ahn:2009wx,Levin98a,Levin90a} is approximated by an energy-dependent transverse momentum $p_\perp^{\rm min}(s)$ cutoff that separates soft and hard parton interactions
\begin{equation}
  p_\perp^{\rm min}(s) = p_{\rm T}^0 + \Lambda \, \exp{ [\,c \sqrt{\ln (s/\,\mathrm{GeV}^2)} \, ]} \ .
  \label{eq:pt-cut}
\end{equation}

Values of the parameters can be found in Appendix~\ref{app:model_parameters}. 
Hard interactions are calculated in leading-order quantum chromodynamics (QCD) at the minimal scale $p_\perp^{\rm min}(s)$ with a $K$-factor to account for higher-order corrections. The hard interaction is assumed to be pointlike and the partons are spatially distributed inside the hadron according to the electric form factor of the proton~\cite{Durand:1988cr}. The distribution of partons in momentum space is given by the parton distribution functions (PDFs) parameterized by Gl\"uck, Reya and Vogt~\cite{Gluck:1998xa,Gluck:1994uf}. 

The parametrization of the soft cross section is inspired by the Donnachie-Landshoff model~\cite{Donnachie:1992ny}.
The soft cross section has two components, one declining and one increasing with energy, corresponding to Reggeon and Pomeron exchange. In contrast to the hard parton interactions, the soft interactions are thought of as spatially extended, i.e.\ $A_{\rm soft}(s,\vec{b}_z\,)$ in Eq.~\eqref{eq:overlap} is given by a Gaussian profile instead of Dirac's delta function. The width of the profile is energy dependent $B_{\rm s}(s) = B_0 + \alpha^{\prime}_{i}(0)\ln{(s/s_0)}$, with $\alpha^\prime(0)$ being a parameter known from Regge phenomenology, see,
for example,~\cite{Collins:1977jy,Goulianos:1982vk}. To obtain an analytic solution for the overlap integral (Eq.~\eqref{eq:overlap}), the distribution of soft partons ($A_{x,y}(s,\vec{b}_{x,y})$) is defined as Gaussian, {\it i.e.}\ for a $pp$ collision
\begin{equation}
  A_{\rm soft}(s,\vec{b}\,) = \frac{1}{4 \pi (2B_{\rm p} + B_{\rm s}(s))} \exp{\left ( - \frac{\vec{b}^{\,2}}{4( 2 B_{\rm p} + B_{\rm s}(s))} \right )} \ . \label{eq:soft-prof}
\end{equation}
The effective width parameter $2B_{\rm p}+B_0$ is determined from a fit to cross section data and the slope of the energy dependence $\alpha^{\prime}_{i}(0)$ is given by the slope of the Pomeron (Reggeon) trajectory known from soft interactions~\cite{Donnachie:1999yb}. The interaction cross sections are calculated by integration of the above amplitude in impact parameter space, e.g.\ for the inelastic cross section
\begin{equation}
  \sigma_{\rm inel} ~=~ \int \mathrm{d}\vec{b} \, \left [ 1-e^{-2 \chi_{\rm soft}(s,\vec{b}\,) \,- 2\,  \chi_{\rm hard}(s,\vec{b}\,)} \right ] \ .
\end{equation}
The obtained values are given in Appendix~\ref{app:model_parameters}. 
A two-channel Good-Walker formalism is used for low-mass diffractive interactions, where the two channels correspond to the hadron's ground state and a generic excited state~\cite{Good:1960ba}. For simplicity, high-mass diffraction is assumed to account for $10\,$\% of the nondiffractive interactions and contributes with only a single cut. A more in depth discussion of the basic principles of the model can be found in Ref.~\cite{Ahn:2009wx}.

The partial cross sections for multiple Pomeron scattering are calculated from the elastic amplitude using unitarity cuts (Abramovsky-Gribov-Kancheli cutting rules)~\cite{Abramovsky:1973fm}. The multiple cuts (or parton interactions) are assumed to be uncorrelated and Poisson-distributed at tree level, but at later steps of the event generation correlations can arise from e.g.\ energy and momentum conservation. The cross sections for multiple cuts are calculated (neglecting diffractive channels) from
\begin{align}
  \sigma_{N_{\rm soft},\, N_{\rm hard}} ~=~ \int & \mathrm{d} \vec{b} \ \frac{ n_{\rm soft}(s,\vec{b}\,)^{N_{\rm soft}}}{N_{\rm soft} !} \, \frac{ n_{\rm hard}(s,\vec{b}\,)^{N_{\rm hard}}}{N_{\rm hard} !} \nonumber \\
  & \times \, \exp{\left(-n_{\rm soft}(s,\vec{b}\,)-n_{\rm hard}(s,\vec{b}\,)\right)} \ , \label{eq:cuts}
\end{align}
where $N_{\rm soft, hard}$ is the number of soft or hard parton scatterings in the interaction. $ n_{\rm int}(s,\vec{b}\,) ~=~ 2 \, \chi_{\rm int}(s,\vec{b}\,)$ is the average number of soft or hard interactions.

For runtime optimization the momenta of the partons in an event are sampled from approximate parameterizations instead of the full amplitude. The hard component ($\sigma_{\rm QCD}$) is calculated at leading order assuming collinear factorization, in which the full PDFs that resolve individual quark flavors and gluons are replaced by an effective PDF for all partons of the form $ f(x) ~=~ g(x) \,+\, \frac{4}{9} \left[ q(x) \,+\, \bar{q} (x) \right] \ ,$ where $q(x)$ represents the combined distribution of all quark flavors~\cite{Combridge:1983jn}. Neglecting initial transverse momentum, the transverse momentum of the partons is determined by the scattering process given by $\hat{t}^{-2}$, where $\hat{t}$ is the four momentum transfer after Mandelstam.

For the soft interaction, which are assumed to include the valence quarks, the momentum fractions are taken from the distribution 
\begin{equation}
  f_{\rm q}(x) = (1-x)^d \, (x^2+m_{\rm q}^2/s)^{-1/4} \ . \label{eq:soft-x}
\end{equation}
In case of the valence quarks, $d$ which leads to the suppression of large momentum fractions, is set to $3$ ($2$) for baryons (mesons). The pole at small momentum fractions is controlled by the choice of an effective quark mass $m_{\rm q}^2 = \unit[0.3]{GeV^2}$. For soft sea quarks and gluons, $d= \unit[1.5]{}$ and $m_{\rm   q}^2=\unit[0.01]{GeV^2}$. The conservation of energy is enforced by assigning one (the last) parton the remaining fraction. 
Since these distributions favor small momentum fractions, the remainder usually constitutes the largest fraction and thus emerges as leading particle. For baryons this fraction is always assigned to pairs of valence quarks, the so-called diquarks. For mesons one of the valence quarks is randomly selected as leading.

The excitation mass, $M_{\rm D}$, for diffractive interactions is sampled from a $M_{\rm D}^{-2}$ distribution without distinguishing between the contributions from low- and high-mass diffraction.
The minimal mass of the diffractively excited system is chosen such that the difference between the mass of the excited system and the original projectile hadron is larger than $1.5$, $0.2$ and $0.6\,$GeV for protons, pions and kaons, respectively.
The upper limit for the diffractive mass universally is set to $M^2_{\rm D}/s=0.2$.
The transverse momentum in the diffractive interaction is assumed to be exponential in $p_{\rm T}^2$ with a slope
\begin{equation}
  B(M^2_{\rm D}) = \max{(B_0, \, a + b \, \ln{ (M^2_{\rm D} c^4 / \mathrm{GeV}^2 })) } \ ,
  \label{eq:diff-slope}
\end{equation}
with $B_0=\unit[6.5]{GeV^2/c^4}$, $a=\unit[31.1]{GeV^2/c^4}$ and $b=\unit[-15.3]{GeV^2/c^4}$~\cite{Albrow76,Breakstone84a}.

\subsubsection{Hadron level}
\begin{figure}
  \includegraphics[width=0.8\columnwidth]{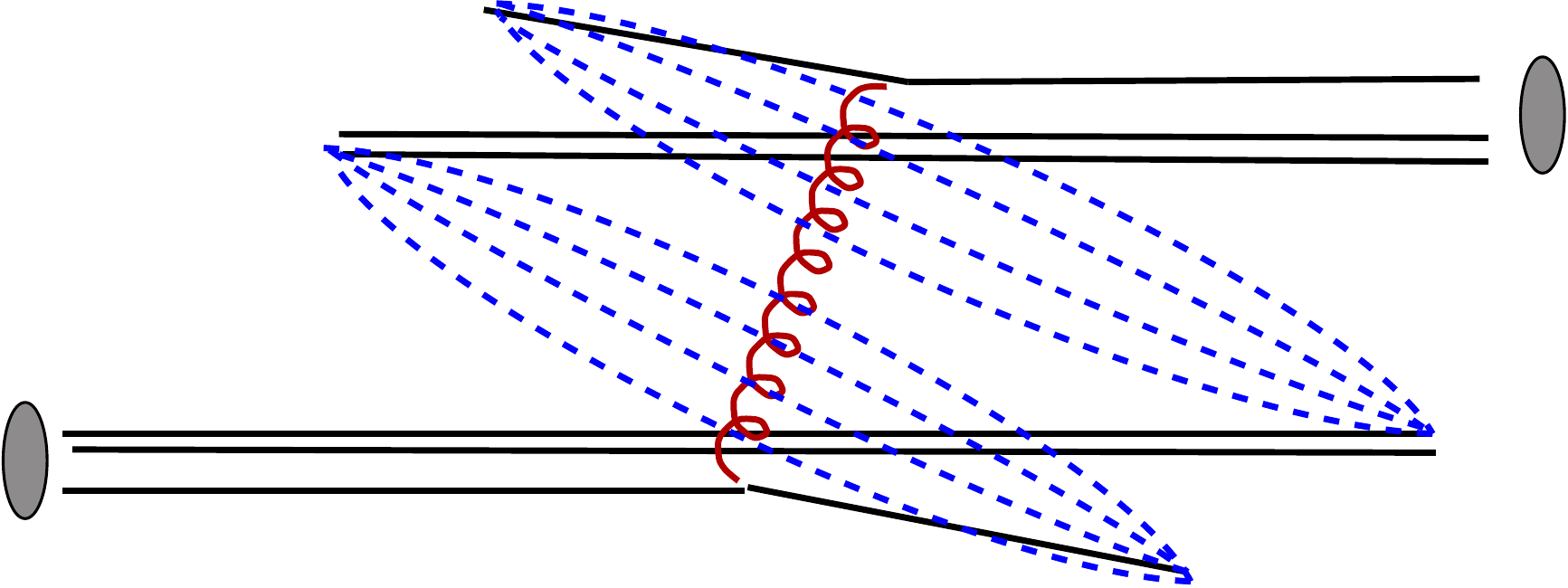}
  \caption{\label{fig:valence-string} Schematic view of the string
    configuration for the soft interaction of the valence quarks in \sibyll. Double
    lines represent diquarks. The probability of the occurrence for this event topology is determined by $\sigma_{1,0}$ from Eq.~\eqref{eq:cuts}.}
\end{figure}

\begin{figure}
  \includegraphics[width=0.9\columnwidth]{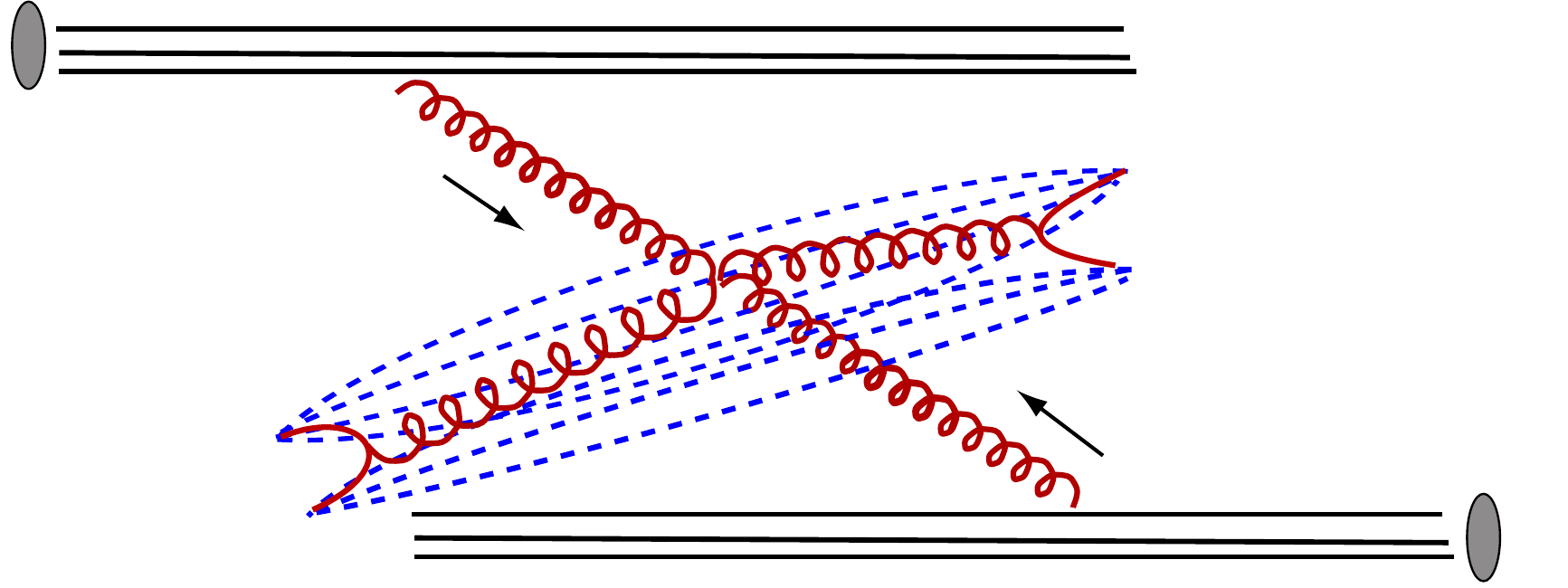}
  \caption{\label{fig:minijet-string}String configuration for a single hard interaction (minijet)
    in \sibyll. Each hadron interaction is composed of a single soft interaction between the valence quarks (Figure~\ref{fig:valence-string}) and $(n_{\rm hard}+n_{\rm soft}-1)$ additional parton interactions, resulting in $2(n_{\rm hard}+n_{\rm soft})$ strings.}
\end{figure}

The hadronization model in \sibyll is based on the Lund string fragmentation model~\cite{Bengtsson:1987kr, Sjostrand:1987xj}. Each (nondiffractive) interaction involves the exchange of color between the hadrons. For the valence quarks a single soft gluon (two colors) is exchanged forming two color fields (strings) between the two quark--diquark pairs for baryons and quark--antiquark pair for mesons, respectively (Figure~\ref{fig:valence-string}). Since gluon scattering is the dominant process at high energy, all the additional hard or soft interactions are modeled as gluon--gluon scattering. Furthermore, the color flow of the gluon scattering is approximated by a closed color loop between two gluons resulting in two strings (see Figure~\ref{fig:minijet-string}). In general, a single hadron-hadron interaction will be a complex combination of such two string configurations, where the probability density for the multiple cut (or string) topology is determined by $\sigma_{N_{\rm soft},N_{\rm hard}}$ (Eq.~\eqref{eq:cuts}).

The fraction of the string energy $z$ assigned to the quarks in each step in the fragmentation is taken from the symmetric Lund function~\cite{Andersson:1983jt} 
\begin{equation}
  f(z) ~=~ (1-z)^a \, z^{-1} \, \exp{ (-\kappa_{\rm string} \, m_{\rm T}^2 \, z^{-1}) }  \ , \label{eq:lundFunc}
\end{equation}
where $a=0.5$ and $\kappa_{\rm string}=\unit[0.8]{c^2/GeV^2}$ and $m_{\rm T}^2$ is the transverse mass $p_{\rm T}^2+m^2$. The transverse momentum of a quark-antiquark pair of flavor $i$ is sampled from a Gaussian distribution with the mean \begin{equation}
  \langle p^i_{\rm T}(s) \rangle ~=~ p^i_0 \, + \, A \log_{10} \left(
    \frac{\sqrt{s}}{30 \, \rm{GeV}} \right) \ .
\label{eq:mean-pt}
\end{equation}
The parameters $A=0.08\,\mathrm{GeV}/\mathrm{c}$ and $p^i_0$ are determined from comparisons with fixed target experiments. The $p^i_0$ take individual values for quarks, diquarks and the different quark flavors (u,d : s : qq $=$ $0.3\,$:$\,0.45\,$:$\,0.6$ (GeV$/$c)). 

Hadronic interactions with zero net quantum number exchange, and in particular no color exchange between the scattering partners, may leave one or both of the hadrons in an excited state and are referred to as low-mass diffraction. The deexcitation of this state is separated into the resonance region at the lowest masses ($M_{\rm D}<2\,$GeV), modeled with isotropic phase space decay (thermal fireball), and the continuum region where string fragmentation is used to produce the multiparticle final state. The hadron-Pomeron scattering in high-mass diffraction is approximated by $\pi^0$-hadron scattering in the rest frame of the diffractive system. 

\subsubsection{Basic model characteristics}
\label{sec:model-perf}
\begin{figure}
  \includegraphics[width=\columnwidth]{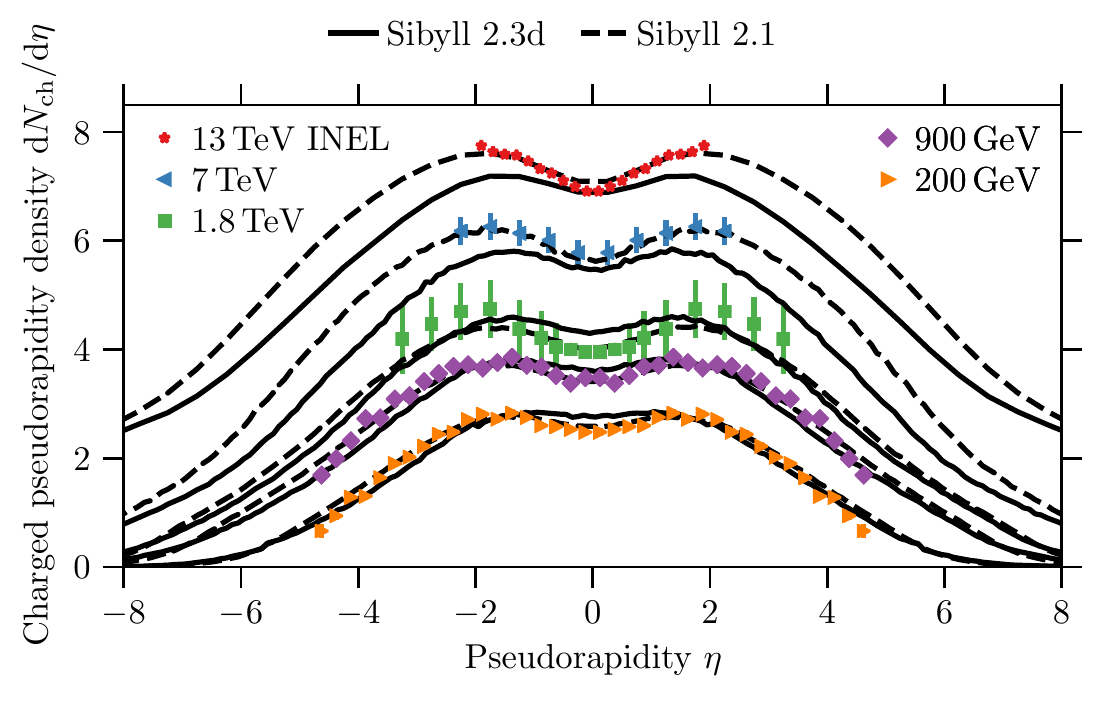}
  \caption{\label{fig:dndeta} Distribution of charged particles in
    pseudorapidity. Data are from CMS, CDF and UA5~\cite{Khachatryan:2015jna,Khachatryan:2010us,Abe:1989td,Alner:1986xu}. The $13\,$TeV data are shifted by one unit up for clarity. The $13\,$TeV measurement
    is an inelastic event selection and remaining sets are nonsingle diffractive. Note, how large parts of forward phase space fall outside of the detector acceptance as the interaction energy increases.
    The central region, that is most sensitive to the number of multiple partonic interactions, is always covered and is used to constrain the model for multiple interactions.}
\end{figure}

\begin{figure}
  \includegraphics[width=\columnwidth]{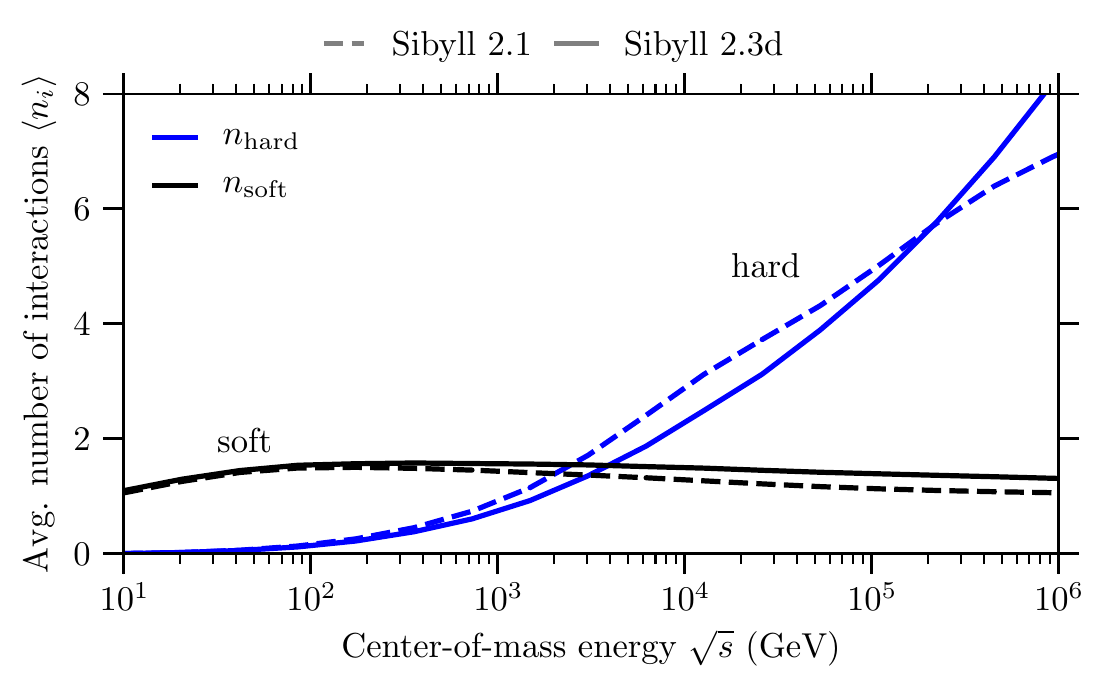}
  \caption{\label{fig:xsctn-nint} Energy dependence of soft and hard parton-parton interactions. The lower number of hard interactions at LHC energies in \sibyll~2.3d is an effect of the narrower proton profile. The change of the slope for \sibyll~2.1 at high energy is due to technical limitations that have been removed in \sibyll~2.3d.}
\end{figure}

\sibyll gives a remarkably good description of the general features of hadronic interactions. Particularly encouraging is the comparison of predictions of \sibyll~2.1 with the results from LHC run I as demonstrated, for example, in Figure~\ref{fig:dndeta} by the yield of charged particles at large scattering angles (pseudorapidity $\eta \sim \tan{\theta /2}$). The widening of the distributions is a phase space effect and arises from the available interaction energy. At central rapidities particle production increases with energy as in Figure~\ref{fig:dndeta} according to the growth of the multiple parton scattering probability. The energy dependence of the average number of soft and hard interactions in Figure~\ref{fig:xsctn-nint} shows that below $1\,$TeV mostly one soft scattering occurs. At higher energies, hard scatterings dominate due to the steep rise of the parton-parton cross section (see $\sigma_{\rm QCD}$ in Figure~\ref{fig:xsctn-inel}). In combination, these figures demonstrate the energy scaling of interaction cross sections, multiple interactions and particle production.

For the high-energy data in Figure~\ref{fig:dndeta}, the new model is underestimating the width of the pseudorapidity distribution, indicating a problem with the transition from hard (central) to soft (forward) processes. This problem is becoming more evident with the shift to PDFs in \sibyll~2.3d that include a steeper rise of the sea quark and gluon distributions toward small x values as favored by measurements at the Hadron-electron ring accelerator (HERA). The scale of the hard scatterings is integrated out for the event generation and the PDFs are evaluated at an effective scale. In nature, the separation between soft and hard scatterings is not well defined and can be thought of as a gradual transition. In principle there should be mixed processes, usually referred to as semi-hard, which are currently not included in \sibyll{} leading to a faster drop of multiplicity for rapidities around the hard-soft scale transition. The comparison to TOTEM measurements in this region ($5<\eta<6$) reveals a underestimation of the particle density of $30-40\,$\%~\cite{Chatrchyan:2014qka}. However, the more important quantity for EAS than the particle density is the energy flow. Measurements are available in the very forward region by LHCf~\cite{Makino:2266968} and at the edge of the central region by CMS and CASTOR~\cite{Chatrchyan:2011wm,Sirunyan:2017nsj}. The former is described reasonably well by the new model (see Figure~\ref{fig:lead-neutron} in \S~\ref{sec:leading} below), whereas the CASTOR measurement indicates a deficit~\cite{Chatrchyan:2011wm}. The largest part of the energy is carried by particles produced in between these regions and hence remains unobserved. Therefore it is not evident from these data that the omission of semihard processes in the model has an impact on the EAS predictions.

\subsection{\label{sec:xsctn}Interaction cross section}

\begin{figure}
  \includegraphics[width=\columnwidth]{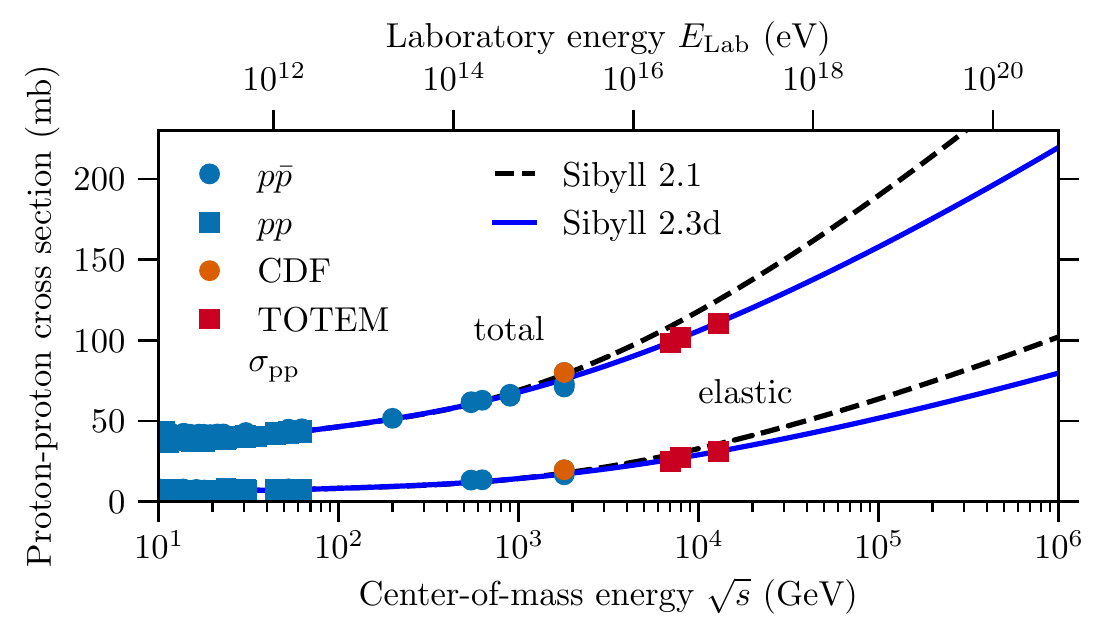}
  \caption{\label{fig:xsctn-tot} Total and elastic proton--proton cross section. \sibyll~2.1 is
    tuned to the \unit[1.8]{TeV} CDF value at the Tevatron~\cite{Abe94d,Avila99a}. The narrower hard interaction profile reduces the inelastic cross section (see Figure~\ref{fig:xsctn-inel}) in \sibyll~2.3d such that total and elastic cross sections coincide with the TOTEM measurements at the LHC~\cite{Antchev:2011zz,Antchev:2013paa}.}
\end{figure}

The parameters of the amplitude are determined by fitting the interaction cross section to measurements.
When the cross section fit was performed for \sibyll~2.1, the highest energy data points that were available were the ones obtained at the Tevatron~\cite{Abe:1993xy,Amos90a,Avila99a} (see Table~\ref{tab:xsctn}). These data suffered from an unresolved ambiguity between the measurement by CDF and the other measurements (Figure~\ref{fig:xsctn-tot}). The higher data point was supported by some cosmic-ray measurements at the time. Recent measurements at the LHC~\cite{Antchev:2011vs} agree well with each other and suggest a lower cross section. These higher-energy data impose stronger constraints on the extrapolation to UHECR energies constitute an important input in \sibyll~2.3d.

\begin{table}[t]
  \caption{Total cross section measurements at the Tevatron and LHC compared to predictions by \sibyll.\label{tab:xsctn}}
  \begin{center}
    \renewcommand{\arraystretch}{1.5}
    \begin{tabular}{cccccc}
      \hline Experiment & $\sqrt{s}$ & $\sigma_{\rm tot}$ (mb) & \sibyll~2.1 & \sibyll~2.3d & Ref. \\
      \hline CDF & \unit[1.8]{TeV} & $\unit[80.03 \pm 2.24]{}$ & & & \cite{Abe:1993xy} \\
      E-710 &  &$\unit[72.8 \pm 3.1]{}$ & \unit[78.8]{} & \unit[75.9]{} & \cite{Amos90a}   \\
      E-811 &  &$\unit[71.71 \pm 2.02]{}$ & && \cite{Avila99a}\\
      \hline
      TOTEM  & \unit[7]{TeV} & $\unit[98.3 \pm 2.9]{}$ & \unit[108.6]{} & \unit[98.8]{}  &\cite{Antchev:2011vs}\\
      ALFA  &  & $\unit[95.35 \pm 1.36]{}$ &  &  &\cite{Aad:2014dca} \\
      \hline
      TOTEM  & \unit[13]{TeV} & $\unit[110.6 \pm 3.4]{}$ & \unit[125.1]{} & \unit[111.1]{}  &\cite{Antchev:2017dia} \\
        \hline
    \end{tabular}
  \end{center}
\end{table}

\begin{figure}
  \includegraphics[width=\columnwidth]{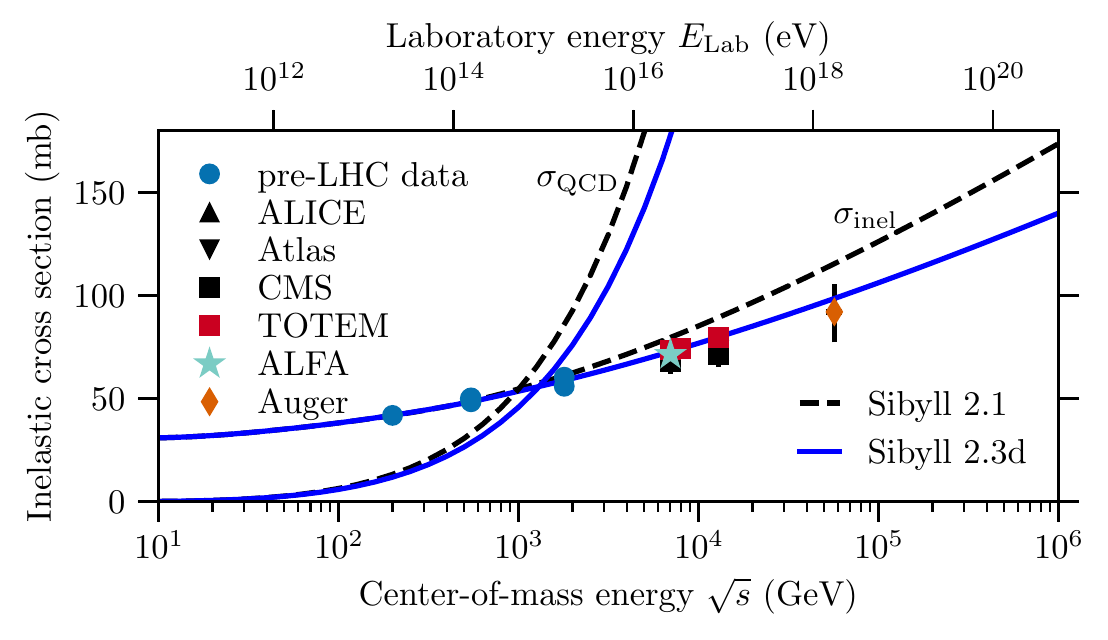}
  \caption{\label{fig:xsctn-inel}Inelastic proton--proton cross section. The data points are compiled from~\cite{Aad:2014dca,Aaboud:2016mmw,Abelev:2012sea,Chatrchyan:2012nj,Antchev:2011zz,Antchev:2013paa,Sirunyan:2018nqx}. The smaller rise of the cross section in \sibyll~2.3d agrees well with the LHC and the \unit[57]{TeV} measurement by the Pierre Auger Observatory~\cite{Auger:2012wt}. This comes mainly from the reduction of hard minijet cross section $\sigma_{\rm QCD}$. At the intersection of $\sigma_{\rm QCD}$ and $\sigma_{\rm inel}$ the probability for multiple hard interactions becomes larger than one and marks the energy range at which multiple parton-parton interactions become increasingly important.}
\end{figure}


\begin{figure}
  \includegraphics[width=\columnwidth]{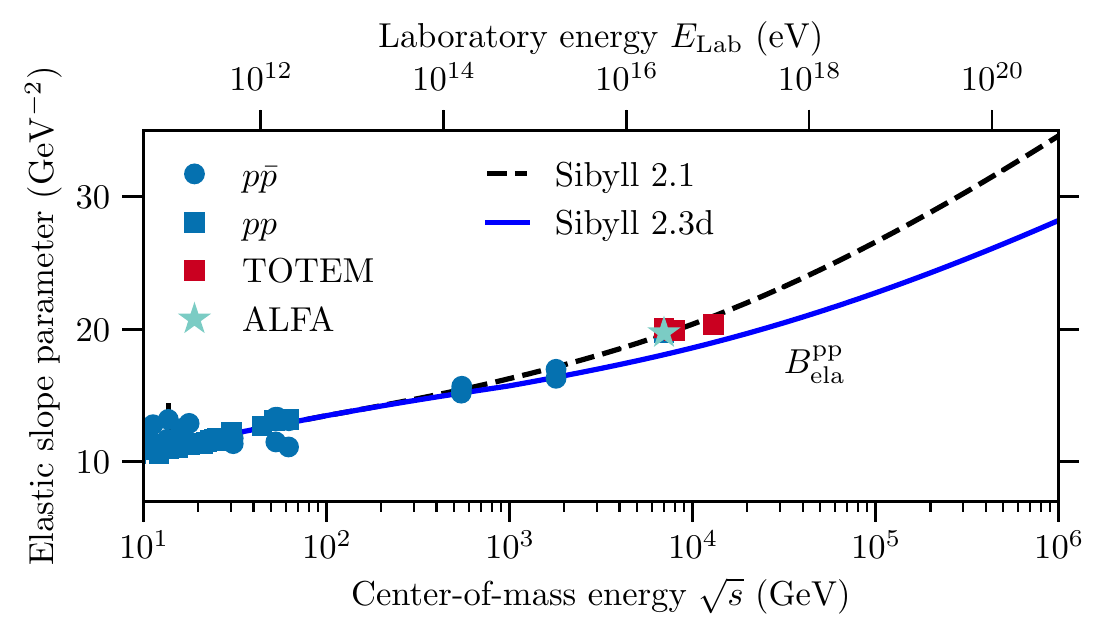}
  \caption{\label{fig:xsctn-slope}The elastic slope parameter in proton--proton interactions. The slope parameter is related to the width of the impact parameter profile. The decrease in the width of the hard profile between \sibyll~2.1 and \sibyll~2.3d, means the slope parameter decreases.}
\end{figure}

Despite an overestimation of the interaction cross section, \sibyll~2.1 gives a remarkably good description of the general features of minimum-bias data. Therefore, we aim for an evolutionary extension of the previous model, in which the hard interaction cross section is smaller. This change yields smaller total and inelastic cross sections in the TeV range and above, while at lower energies remain mostly unaffected according to Figure~\ref{fig:xsctn-tot}. Hard parton scattering is calculated in perturbative QCD, generally leaving little room for alterations. The hard cross section can be reduced by increasing the transverse momentum cutoff $p_{\rm T}^{\rm min}(s)$ that defines the transition between soft and hard interactions. However, in \sibyll{} the energy dependence is derived from a geometrical saturation condition (see Eq.~(\ref{eq:pt-cut})) and is, therefore, fixed. 

A different possibility is the modification of the opacity profile $A_{\mathrm{hard}}(\vec{b}\,)$. The overlap integral for two protons, the formal definition is given in Eq.~\eqref{eq:overlap}, in the model takes the explicit form given by
\begin{equation}
  A(\nu_{\rm h},\vec{b}\,) = { \nu_{\rm h}^2 \over 12 \pi} {1 \over 8} \, (\nu_{\rm h} b)^3 \, K_3(\nu_{\rm h} b) \ , \label{eq:hard-prof}
\end{equation}
where $K_3(x)$ is a modified Bessel function of the second kind. The parameter $\nu_{\rm h}$ determines the width of the profile that controls the share between more peripheral and central collisions, i.e.\ narrow profiles lead to a reduction of peripheral collisions. Since most collisions are peripheral, a narrower profile reduces the interaction cross section. Figure~\ref{fig:xsctn-tot} shows the new and old fits of the total and the elastic cross section after narrowing the profile function and adjusting the soft interaction parameters. The result gives a good description of the measurements at high energy~\cite{Abelev:2012sea,Chatrchyan:2012nj,Aad:2014dca,Antchev:2011zz,Antchev:2013paa}.
As shown in Figure~\ref{fig:xsctn-inel}, the inelastic cross section in the new model is compatible with that derived from an UHECR measurement~\cite{Auger:2012wt}, whereas the cross section in \sibyll~2.1 was too high. At the time of the fit the LHC run I data reached only up to $7\,$TeV c.m.\ , but nonetheless the previous parameters are compatible with LHC run II data at $13\,$TeV~\cite{Aaboud:2016mmw,Sirunyan:2018nqx,Antchev:2017dia} (see also Table~\ref{tab:xsctn}).

In the scattering of waves a refraction pattern is determined by the form of the scattering object. For hadrons, the shape of the refraction pattern in first approximation is described by the \textit{elastic slope parameter}, $B_{\rm ela}$, the slope of the forward peak of the differential elastic cross section,
\begin{equation}
  \frac{\mathrm{d}\sigma_{\rm ela}}{\mathrm{d}t} ~\sim~ e^{-B_{\rm ela}t} \ .
\end{equation}
The $-t$ is the transferred momentum squared. Decreasing the width of the proton profile, results in a broadening of the refraction pattern and hence a decrease of the slope. While the interaction cross sections are better described by the narrower profile, the measurements of the elastic slope~\cite{Antchev:2011zz} do not reflect this preference (see Figure~\ref{fig:xsctn-slope}).

More recent, LHC-constrained parameterizations of the PDFs (e.g.\ CT14~\cite{Dulat:2015mca}) instead of the older GRV98-LO~\cite{Gluck:1994uf,Gluck:1998xa} typically show a less steep rise of the gluon distribution toward small $x$ and hence result in a smaller hard scattering cross section. This would lead to a smaller rise of $\sigma_{\rm QCD}$ and hence a wider profile can be chosen to reduce the tension with data in $B_{\rm ela}$. As the integration of the new PDFs in the complete event generator requires the readjustment of almost all model parameters this endeavor is left to a future update.

These modifications to the proton--proton cross sections also affect the cross sections for hadron-nucleus and nucleus--nucleus collisions. The extension to meson-nucleus interactions is discussed in Sec.~\ref{sec:meson-nucleus}, $\sigma_{\rm p-air}$ is presented in Figure~\ref{fig:predict-air-xsctn} and the interaction lengths of iron nuclei, protons, pions and kaons in air are given in Appendix~\ref{app:eas_obs} 
and discussed in Sec.~\ref{sec:predict}.


\subsection{\label{sec:leading}Leading particles}

\begin{figure}
  \includegraphics[width=\columnwidth]{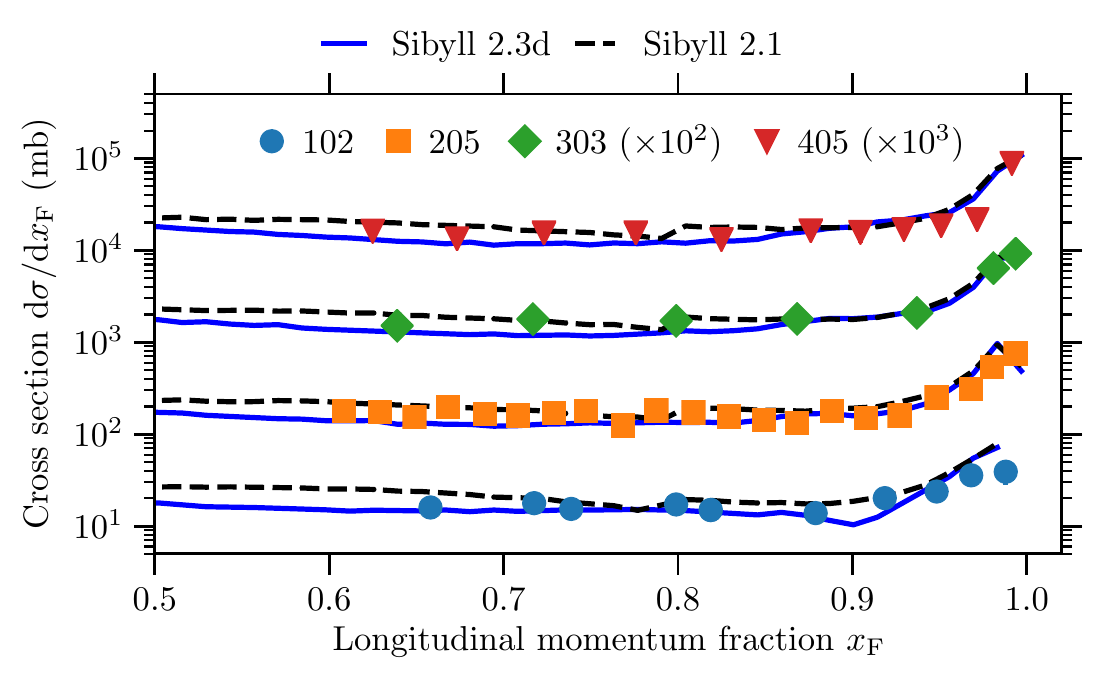}
  \caption{\label{fig:lead-fnal}Distribution of leading protons in proton--proton interactions. Data are from bubble chamber experiments at the Fermi National Accelerator Laboratory obtained with beam
    momenta $p_{\rm Lab}=102,\, 205,\,303$ and $405$\,GeV/c~\cite{Whitmore:1973ri}, and offset by $1,\,10,\,100$ and $1000$, respectively, for clarity. Longitudinal momentum fraction is expressed relative to the maximal momentum in the center-of-mass frame (Feynman $x_\text{F}$).}
\end{figure}

\begin{figure}
  \includegraphics[width=\columnwidth]{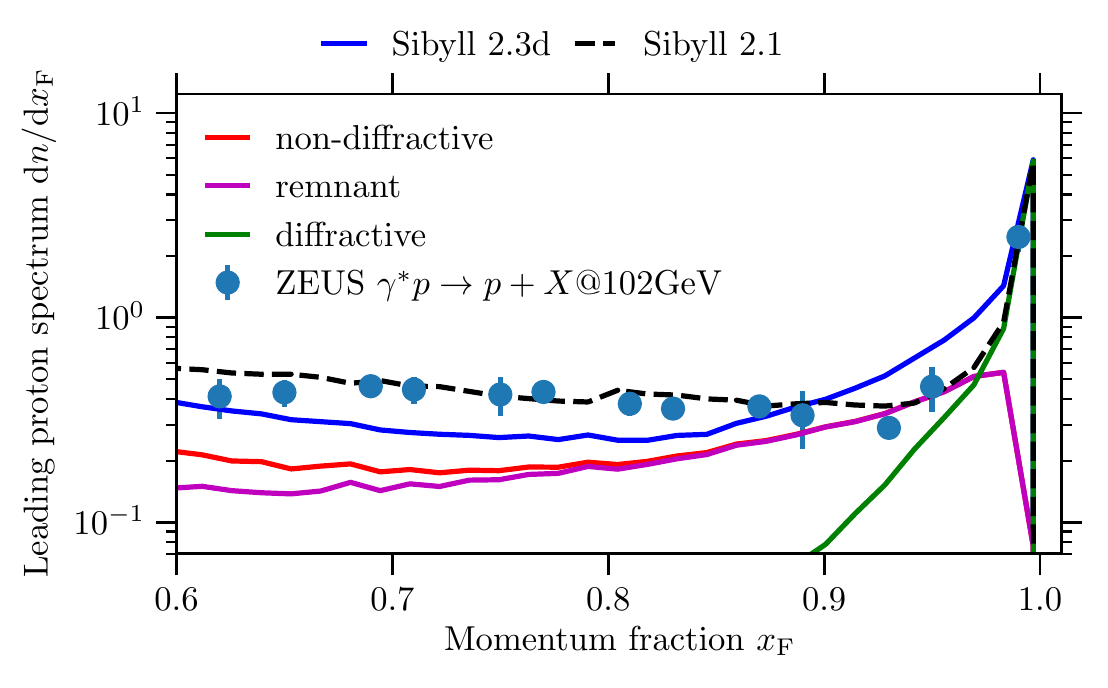}
  \caption{\label{fig:lead-zeus}Distribution of leading protons in photon-proton interactions at
    $\sqrt{s}=\unit[102]{GeV}$~\cite{Chekanov:2002yh} performed at the HERA collider in the proton fragmentation region. The equivalent interaction energy in proton--proton collisions is $\sqrt{s}=\unit[210]{GeV}$. The spectrum is a combination of the individual contributions in \sibyll~2.3d: diffractive (green), nondiffractive (red) and remnant (purple). The nondiffractive component includes the contribution from the remnant. At large Feynman $x$ above 0.75, the non-diffractive component is dominated by the fragmentation of the remnant.}
\end{figure}

\begin{figure}
  \includegraphics[width=\columnwidth]{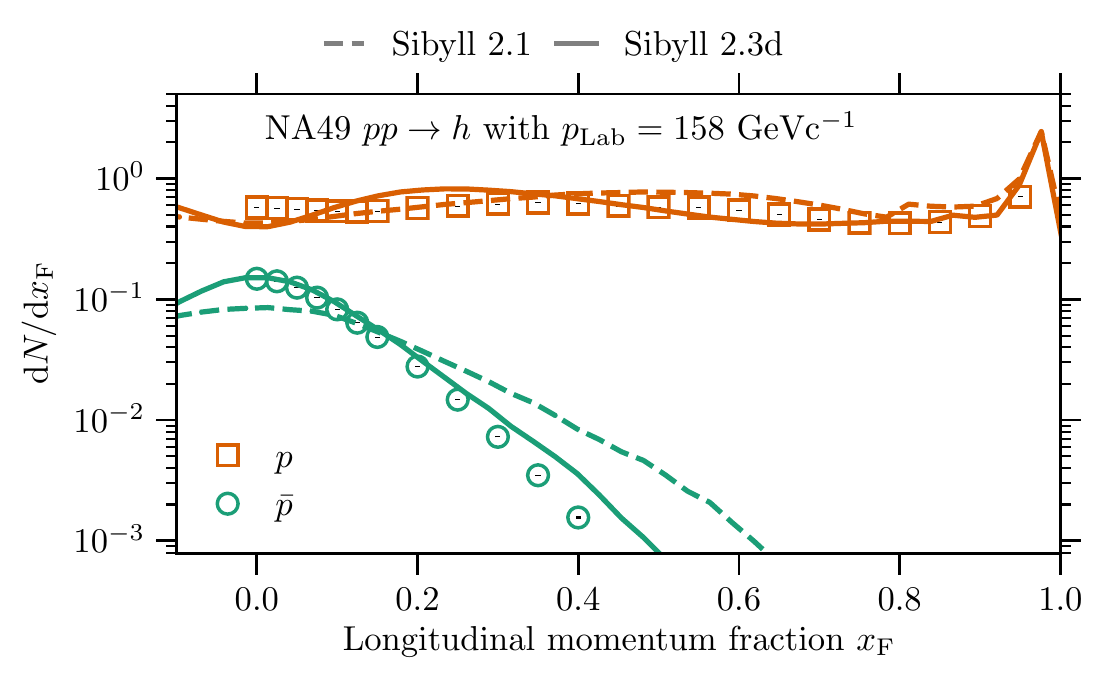}
  \caption{\label{fig:lead-p-pbar}Longitudinal momentum spectrum of protons and antiprotons in proton--proton
    collisions~\cite{Anticic:2009wd}. The flat distribution for protons is achieved using an \textit{ad hoc} mechanism in \sibyll~2.1. However, the central and the leading particles are produced by the same process, resulting in a hard spectrum for antiprotons. In \sibyll~2.3d the central and the fragmentation region are related to separate processes, leading to more accurate descriptions for longitudinal baryon spectra.}
\end{figure}


Secondary particles that carry a very large momentum fraction of the initial projectile are called leading particles. They are of utmost importance for the longitudinal development of EAS since they transport energy more efficiently into the deeper atmosphere requiring at the same time fewer interactions. The origin of leading particles is not clearly related to one hadronic or partonic process and can be thought of as a superposition of all processes contributing to the forward phase space, often involving valence quark interactions.

\subsubsection{Leading protons and hadron remnants}
\label{sec:leading}

\begin{figure}
  \includegraphics[width=0.7\columnwidth]{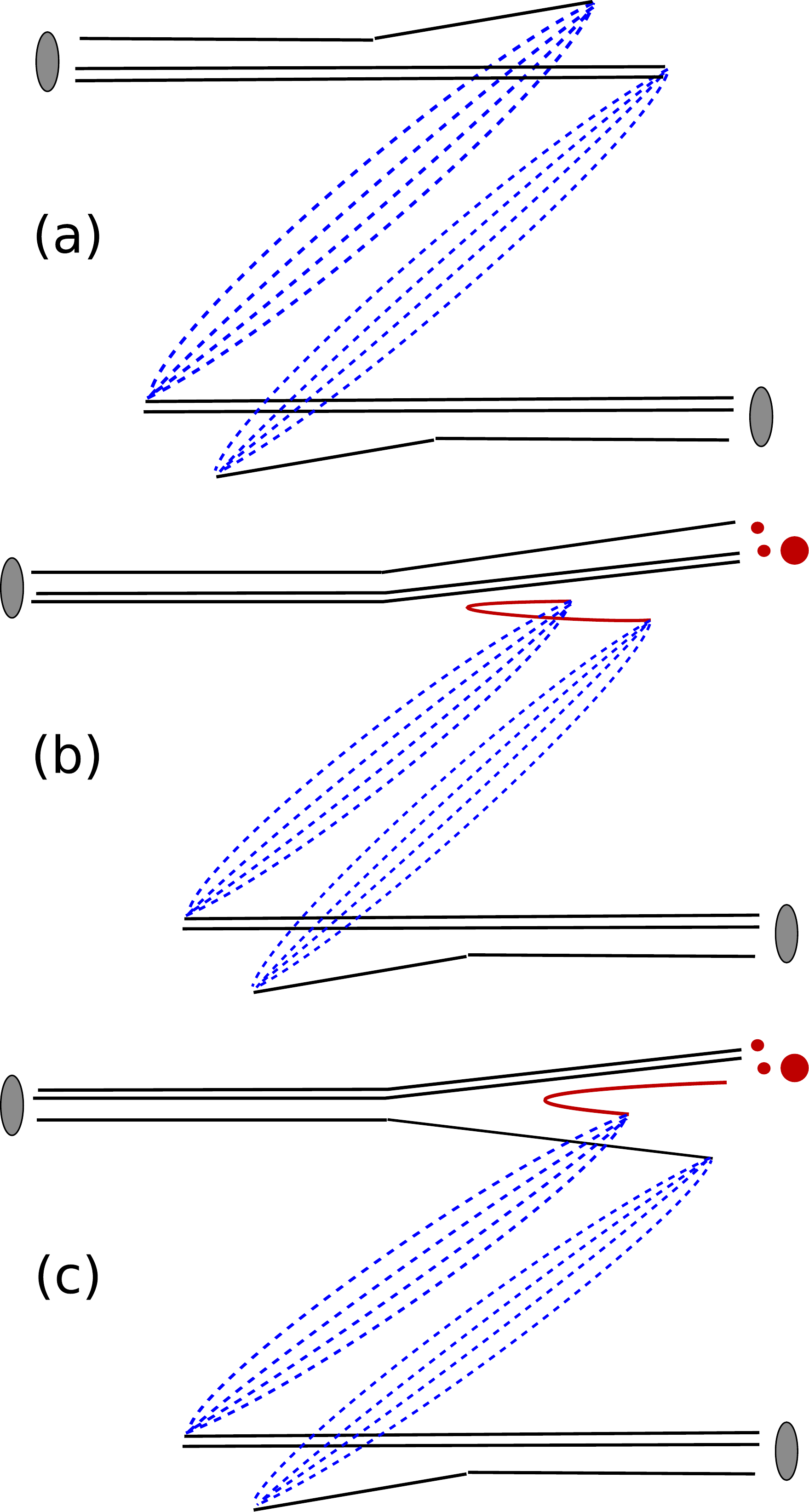}
  \caption{\label{fig:lead-hadron}Schematic view of different string
    configurations involving valence and sea quarks in \sibyll: a) default DPM scenario, configuration in \sibyll~2.1. b, c) remnant configurations without (b) and with (c) color exchange.}
\end{figure}

In the parton model the leading hadron is related to the partons with the largest momentum fractions, which in most cases are the valence quarks. Figure~\ref{fig:lead-fnal} and Figure~\ref{fig:lead-zeus} demonstrate the characteristic ``flatness'' and a diffractive peak of the longitudinal momentum distribution (in $x_{\rm F}~=~ p^{\rm CM}_{\rm z}/p^{\rm CM}_{\rm z, max}$). The latter naturally fits into the leading particle definition since in diffraction no quantum numbers are exchanged. The flat region below $0.9$ corresponds to the leading particle in nondiffractive events. The presence of this plateau in the proton spectrum and its absence for secondary particles that do not share quantum numbers with the projectile (see antiproton spectrum in Figure~\ref{fig:lead-p-pbar}) identifies the valence quarks as high momentum constituents of the projectile.

In \sibyll~2.1 the leading particles are implemented by assigning one of the valence (di)quarks a large momentum fraction. In a proton--proton interaction each proton is split into a quark-diquark pair forming a pair of strings between a quark and a diquark of the other proton, as illustrated in Figure~\ref{fig:lead-hadron}a). The momentum fraction of the quark is sampled from a soft distribution as in Eq.~\eqref{eq:soft-x}, leaving a larger fraction to the diquark. In addition, the subsequent fragmentation of the quark-diquark string is biased toward the diquark by sampling the energy fraction in the first string break next to the diquark from $f_{\rm lead}(z) \sim z$ instead of the standard Lund function (Eq.\eqref{eq:lundFunc}). This mechanism reproduces the observed flat proton spectra in Figs.~\ref{fig:lead-fnal},~\ref{fig:lead-zeus} or~\ref{fig:lead-p-pbar}.

Interactions of hadrons at low energies (e.g.\ $\sqrt{s}=\mathcal{O}(20 \, \mathrm{GeV})$) are dominated by soft parton scattering. In \sibyll, most of these interactions happen between the valence quarks (see Figure~\ref{fig:xsctn-nint}). The conservation of energy and baryon number for such systems introduces a strong correlation between the production of leading protons and central ($x_{\rm F}\sim 0$) antiprotons, as both come from the hadronization of the same valence quark system. In the leading proton scenario, where a large momentum fraction is assigned to the leading string break, an antiproton produced in a later break is necessarily slow. Often its production will be energetically forbidden because the antiproton has to be produced alongside a second baryon. The opposite case, in which the leading proton is slow ($f_{\rm Lund}(z) \sim \exp{(-1/z)} ~\ll~ f_{\rm   lead}(z)\sim z$ as $z\to 0$), is more problematic since the antiproton can carry a large momentum fraction. Measurements of $x_{\rm   F}$ spectra of protons and antiprotons in Figure~\ref{fig:lead-p-pbar} do not confirm the presence of antiprotons with large momenta (an additional discussion of baryon-pair production can be found in Sec.~\ref{sec:fragH}). By changing the momentum fraction of the leading protons the production of antiprotons with large momentum fraction cannot be avoided since the protons demonstrate a flat spectrum down to the central region.

In \sibyll~2.3d the issues with leading baryon production are addressed with the so-called \textit{remnant formation}. In this mechanism, the leading protons are produced from the remnant, while antiprotons and central protons are produced from strings that are attached to soft sea quarks (Figs.~\ref{fig:lead-hadron}b and~\ref{fig:lead-hadron}c). The momentum fraction of the sea quarks is sampled from $f_{\rm soft~q}(x) = (1-x)^{1.5} \, (x^2-m_{\rm q}^2/s)^{-1/4}$ with $m_{\rm q}=\unit[0.6]{GeV}$. The momentum fraction for the remnant (system of valence quarks) is distributed like $x^{1.5}$.

The energy and the momentum transferred in the remnant interaction are modeled similarly to diffractive interactions as discussed at the end of Sec.~\ref{sec:parton_level}. The squared mass spectrum approximately follows $\mathrm{d}N/\mathrm{d}M_{\rm r}^2 \sim 1/M_r^2$ and the slope of the $p_{\rm T}$ spectrum is
\begin{equation}
\nonumber  B_{\rm r}(M^2_{\rm r})~=~\max{ ( B_{0,\mathrm{r}}, \, a_{\rm r} + b_{\rm r} \, \ln{ ( M^2_{\rm r} c^4 / \, \mathrm{GeV}^2)})},
\end{equation}
with the parameters $B_{0,\mathrm{r}}=0.2\,$GeV$^2/$c$^4$, $a_{\rm r}=7.0\,$GeV$^2/$c$^4$ and $b_{\rm r}=-2.5\,$GeV$^2/$c$^4$. In addition to the continuous spectrum, discrete excitations of resonances are included. Due to their isospin structure, the decay channels may be weighted differently than for isotropic phase space decay. For each projectile two resonances are included (e.g.\ see Table~\ref{tab:resonances}).

When parton densities become large at high energies and the number of parton interactions increases, it is less likely that partons remain to form a remnant. In this case the situation is more similar to the two-string approach in \sibyll~2.1. This transition effect is taken into account by imposing a dependence on the sum of soft and hard parton interactions $(n_{\rm s}+n_{\rm h})$ to the remnant survival probability
\begin{equation}
  P_{\rm r}~=~ P_{\rm r,0} \, \exp{( - \left [ N_{\rm w} + \epsilon \,(n_{\rm h}+n_{\rm s}) \right ] ) } \ .  \label{eq:rmnt-prob}
\end{equation}
In nuclear interactions (even at low energies) parton densities can be large. Correspondingly, the remnant probability depends on the number of nucleon interactions $N_{\rm w}$. The relative importance of nucleon and parton multiplicity is determined by $\epsilon$ and is set to 0.2. The remnant survival probability at low energies $P_{\rm r,0}$ is 60\%.

\begin{table}
  \caption{Table of the resonances used for remnant excitations of the most common projectiles in \sibyll~2.3d (also visible in Figure~\ref{fig:lead-mass-spec}).
    \label{tab:resonances}}
  \begin{center}
    \renewcommand{\arraystretch}{1.5}
    \begin{tabular}{ccc}
      \hline
      Projectile  & Resonance  & Mass (GeV) \\
      \hline
      $p,n$ & $N(1440)^{+,0}$   &  1.44 \\
            & $N(1770)^{+,0}$   &  1.77 \\
      $\pi^{0,\pm}$  & $\rho^{0,\pm}$  & 0.76     \\
                & $\pi_{1}^{0,\pm}$  &   1.30   \\
      $K^{0,\pm}$  & $K^{*\pm}, \, K^{*0}$ & 0.89 \\
                & $K_{0}^{*\pm}, \, K_{0}^{*0}$  & 1.43 \\

      \hline
    \end{tabular}
  \end{center}
\end{table}

\begin{figure}
  \includegraphics[width=\columnwidth]{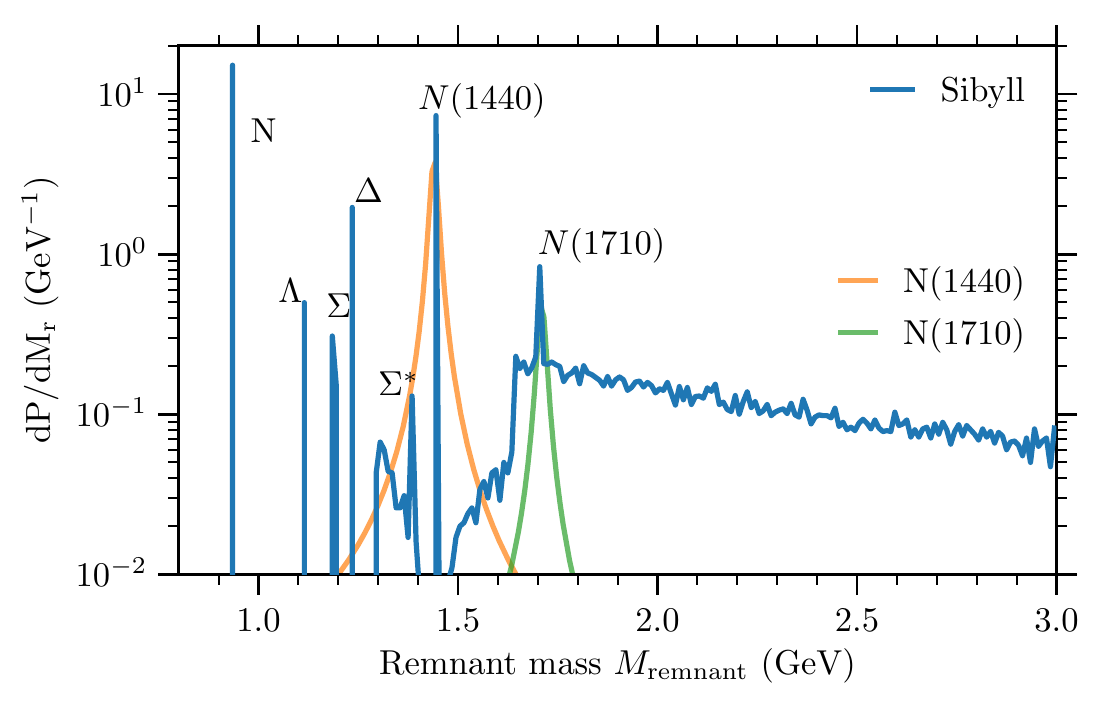}
  \caption{\label{fig:lead-mass-spec}Mass distribution of the proton remnant in the model. The continuum resembles approximately an $M_{\rm remnant}^{-2}$ distribution. The resonances at low excitation masses are taken into account according to Table~\ref{tab:resonances}.}
\end{figure}

The spectrum of the remnant excitation masses for proton interactions in Figure~\ref{fig:lead-mass-spec} demonstrates how different hadronization mechanisms apply for different regions of the mass spectrum. For large masses ($\Delta M = M_{\rm remnant}-m_{\rm projectile}>\unit[1]{GeV}$, where $m_{\rm projectile}$ is the mass of the projectile), indicating the presence of a fast valence quark, the deexcitation is very anisotropic and particles are emitted mostly in the direction of the leading quark. In this case, the hadronization of high-mass remnants is implemented as the fragmentation of a single string. At intermediate masses ($\unit[0.4]{GeV}<\Delta M<\unit[1]{GeV}$), a continuum of isotropic particles is produced by phase space decay. The number of particles produced is selected from a truncated Gaussian distribution with the mean
$n_{\rm thermal}=2\sqrt{\Delta M/\mathrm{GeV}} , \, n_{\rm thermal} > 2$. Below the threshold for the production of particles and resonances ($\Delta M < \unit[0.2]{GeV}$), the remnant is recombined to the initial beam particle. This recombination region determines the proton distribution at intermediate and large Feynman $x$. Hence, the shape of the final particle spectra depends on the combination of the separate hadronization mechanisms. The adjustment of the remnant model parameters has been mainly achieved from comparisons with the leading low-energy proton data shown in Figure~\ref{fig:lead-fnal} together with the antiprotons distribution shown in Figure~\ref{fig:lead-p-pbar}. In particular, the latter is much better described by the updated model. At the higher energies probed in the ZEUS experiment~\cite{Chekanov:2002yh} (see Figure~\ref{fig:lead-zeus}), the contribution from the remnant in the region $x_{\rm F}>0.9$ overlaps with the diffractive peak, resulting in an overestimation of the spectrum in \sibyll~2.3d, while in the region of $0.6<x_{\rm F}<0.8$ the spectrum is underestimated. This can be addressed in the future by adjusting the remnant and the diffractive mass distribution.

\begin{figure}
  \includegraphics[width=\columnwidth]{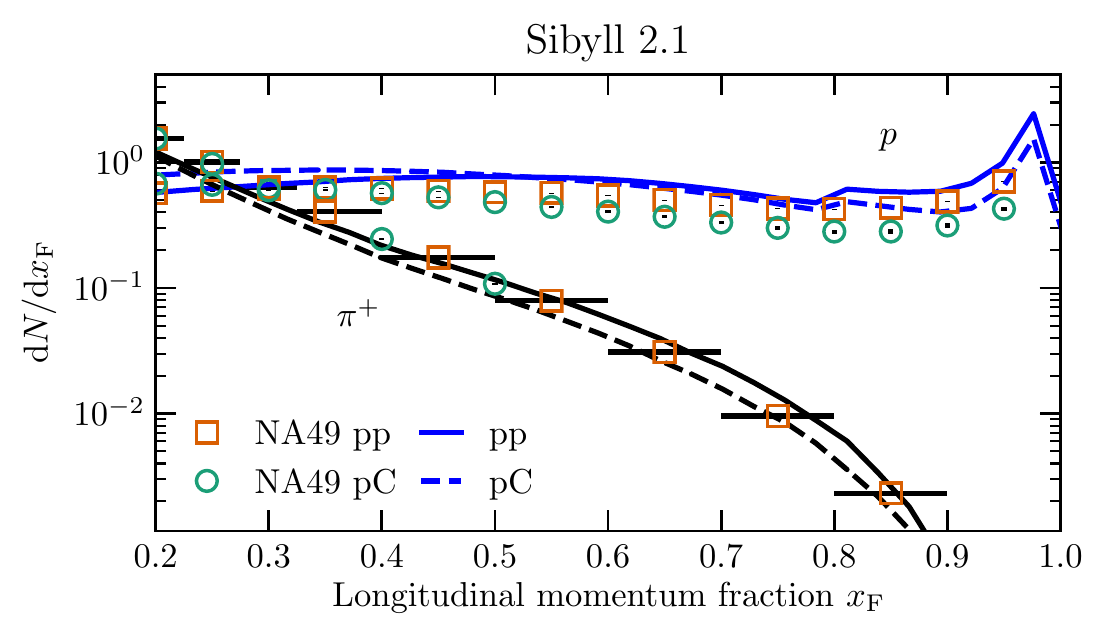}
  \includegraphics[width=\columnwidth]{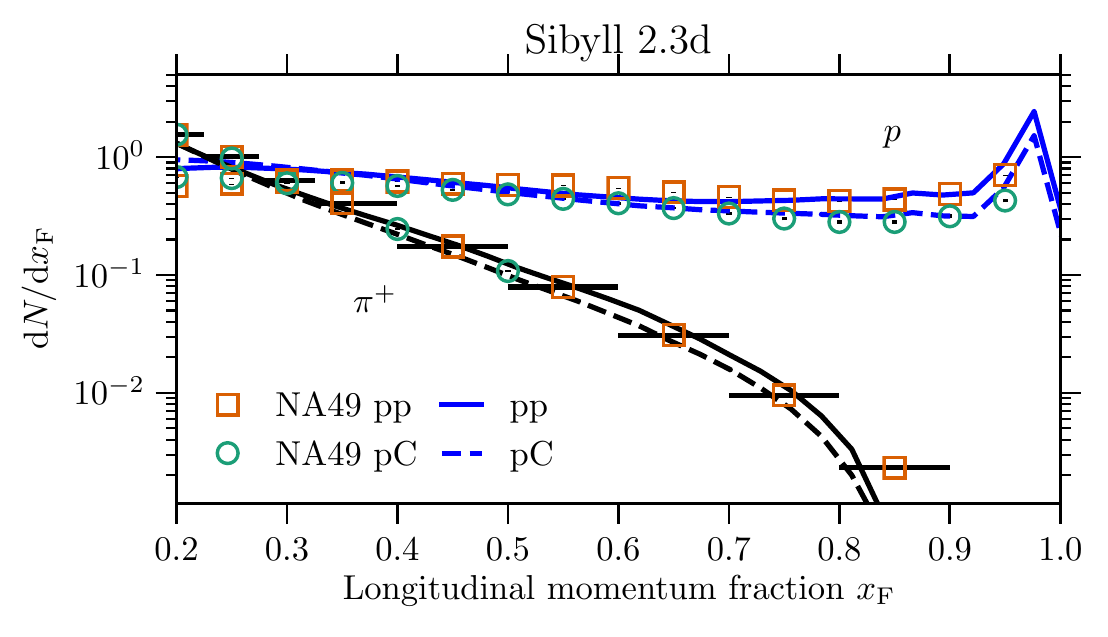}
  \caption{\label{fig:lead-p-pip}Comparison of the Feynman $x$ spectra of protons and positive pions in proton--proton and proton-carbon interactions at $p_\text{lab}=158\,$GeV/c~\cite{Anticic:2009wd,Alt:2006fr,Alt:2005zq,Baatar:2012fua}. The leading particle model in \sibyll~2.1 (upper panel), based on a fine-tuned leading fragmentation function, does not reproduce the attenuation of leading protons due to the nuclear target. In the remnant model (lower panel) the attenuation of leading protons is described correctly.}
\end{figure}

Another drawback of the model for leading particle production in \sibyll~2.1 is the insufficient attenuation of the leading particles in the transition from proton to nuclear targets (see secondary proton spectrum in Figure~\ref{fig:lead-p-pip}). While the proton spectrum is clearly affected by the number of target nucleons, this effect is much smaller for mesons (pions). The model for the reduced remnant formation probability in the presence of multiple target nucleons (Eq.~\eqref{eq:rmnt-prob}) in \sibyll~2.3d reproduces this effect correctly.

\begin{figure}
  \includegraphics[width=\columnwidth]{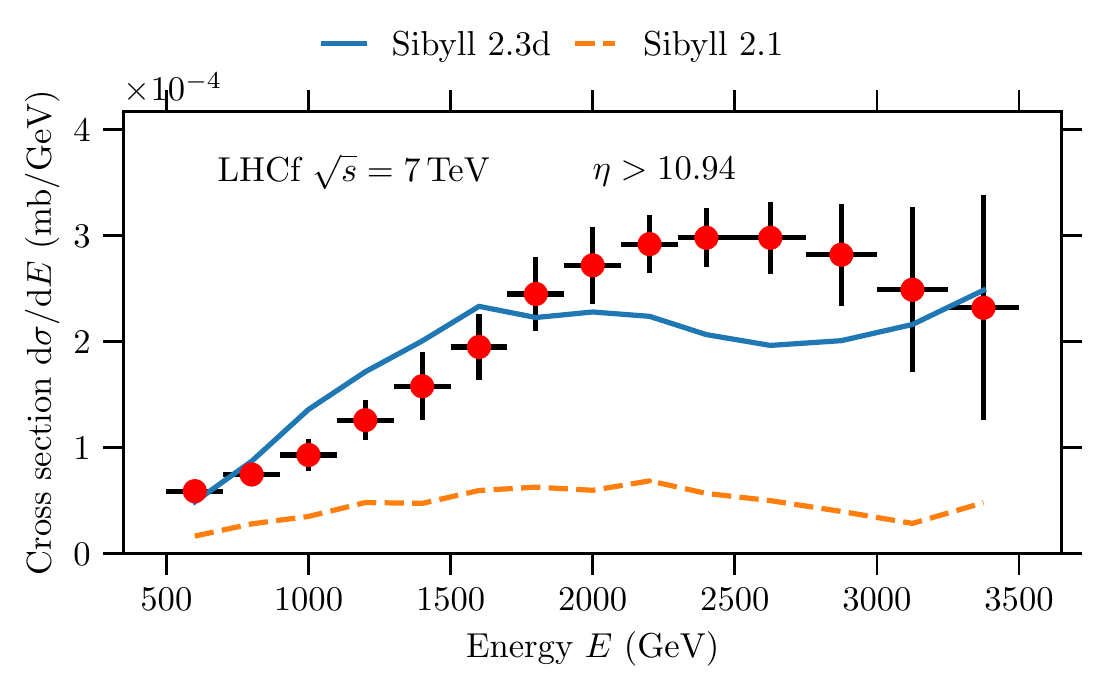}
  \caption{\label{fig:lead-neutron}Energy spectrum of leading neutrons
    in the LHCf forward calorimeter at $\sqrt{s}=7\,$TeV~\cite{Adriani:2015nwa}. The remnant model clearly improves the description by \sibyll~2.3d.}
\end{figure}

The model parameters are adjusted according to low-energy data from the NA49 experiment that provides a large $x_{\rm F}$ coverage. However, the remnant model affects high energies as well, resulting in a significant improvement of leading neutrons at LHCf~\cite{Adriani:2015nwa} ($7\,$TeV), as shown in Figure~\ref{fig:lead-neutron}.

\subsubsection{Leading mesons and $\rho^0$ production \label{sec:leading_mesons}}
A second important role of leading particles in EAS is their impact on the redistribution of energy between the hadronic and the electromagnetic (EM) shower component. Any charged pion of the hadronic cascade can transform into a neutral pion in a charge exchange interaction. Through the prompt decay of the neutral pion into two photons, all the energy is then transferred to the EM component
\begin{align}
  \pi^\pm + p \, \to \, &\pi^0 + X \nonumber \\
                        & \pi^0 \to \, \gamma \, \gamma \ .  \label{eq:pi0}
\end{align}
The influence of this reaction is largest for the leading particles and usually results in a decrease of the muon production that occurs at late stages of the EAS development~\cite{Drescher:2007hc}. A suppression of the pion charge exchange process has the opposite effect.

An example for such a competing reaction is the production of neutral vector mesons ($\rho^0 :\, I(J^{CP})=1\,(1^{--})$) from a pion beam
\begin{align}
  \pi^\pm + p \, \to \, &\rho^0 + X \nonumber \\
  & \rho^0 \to \, \pi^+ \, \pi^- \ . \label{eq:rho0}
\end{align}
Whereas a neutral pion decays into two photons, the conservation of spin requires a $\rho^0$ to decay into two charged pions.

In the Heitler-Matthews model~\cite{Matthews:2005sd} the average number of muons in an EAS initiated by a primary cosmic-ray with energy $E_0$ is given by
\begin{equation}
  N_\mu~=~\left ( \frac{E_0}{E_c}
  \right )^{\alpha} \ \mathrm{with} \ \alpha=\frac{\ln (n_{\rm ch})}{\ln(n_{\rm tot})} \ , \label{eq:nmu}
\end{equation}
and critical energy $E_{\rm c}$. The change of the number of muons per decade of energy ($\alpha$) thus depends on the total and charged multiplicities. It is evident that the ratio between $\rho^0$ and $\pi^0$ production directly affects the exponent $\alpha$.

\begin{figure}
  \includegraphics[width=\columnwidth]{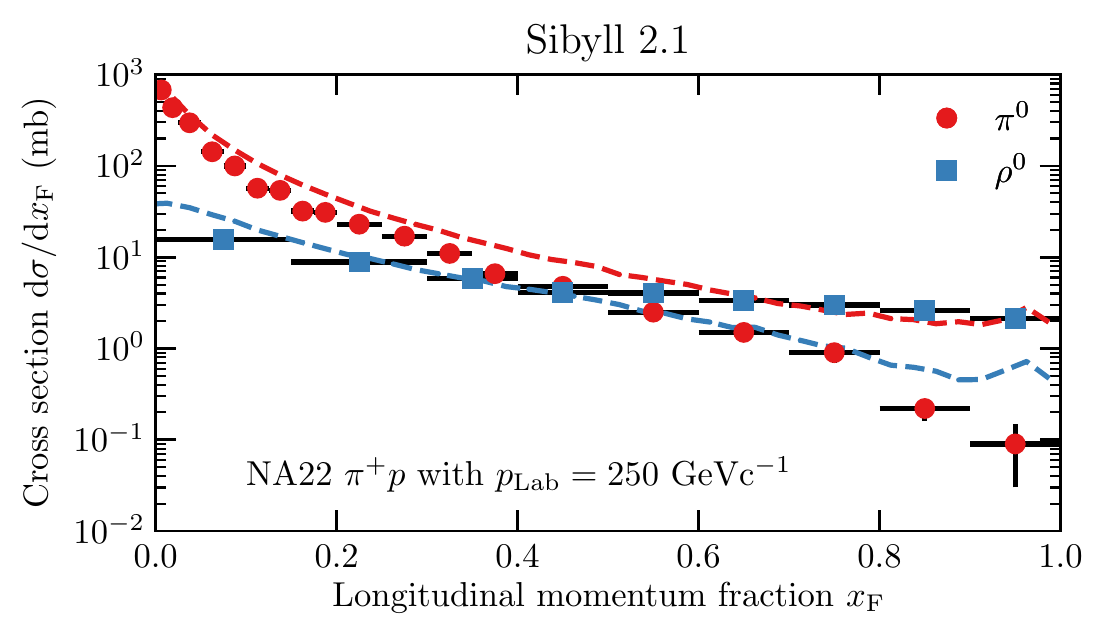}
  \includegraphics[width=\columnwidth]{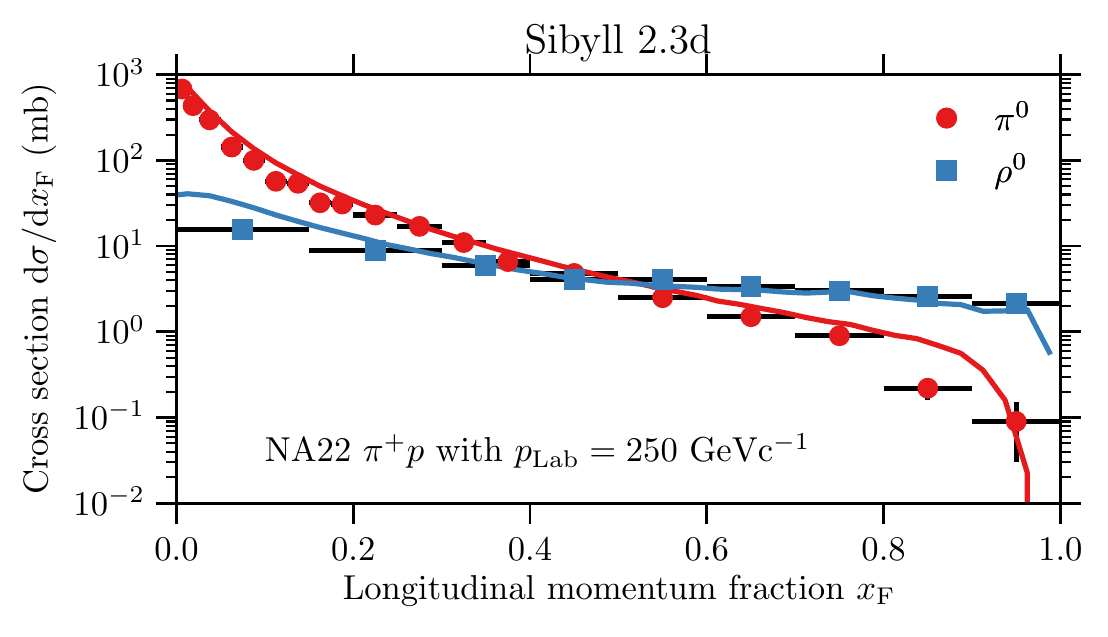}
  \caption{\label{fig:lead-na22}Feynman $x$ spectrum of neutral pions
    and their spin-$1$ resonance state $\rho^0$ in $\pi^+$--proton collisions at $p_\text{lab}=250\,$GeV/c ~\cite{Agababyan:1990df,Adamus:1986ta}. The expectation from standard quark splitting ($\pi^+:u\bar{d}$) and fragmentation is that a fixed fraction of the leading $\pi^+$ transforms into neutral
    pions ($\pi^0:(u\bar{u}-d\bar{d})/\sqrt{2}$) and a smaller fraction into the resonance state $\rho^0$ (upper figure). Data, on the other hand, show an enhancement of the production of the resonant state and a suppression of the ground state in the region of the leading particle. The effect is reproduced in
    \sibyll~2.3d (lower figure) by increasing the rate at which resonances occur in the fragmentation of diffractive processes and by including the $\rho^0$ as a resonance state in the remnant formation of the pion.}
\end{figure}

In charged pion--proton interactions the NA22 fixed target experiment found that at large momentum fractions vector mesons are more abundantly produced than neutral pions (Figure~\ref{fig:lead-na22})~\cite{Agababyan:1990df,Adamus:1986ta}. In the dual parton approach with standard string fragmentation, as it is used in \sibyll and several other models, this result is unexpected and probably cannot be reproduced without invoking an additional exchange reaction. Recent measurements by the NA61 Collaboration have confirmed the leading $\rho^0$ enhancement in case of pion nuclear interactions~\cite{Aduszkiewicz:2017anm}.

\begin{figure}
  \includegraphics[width=\columnwidth]{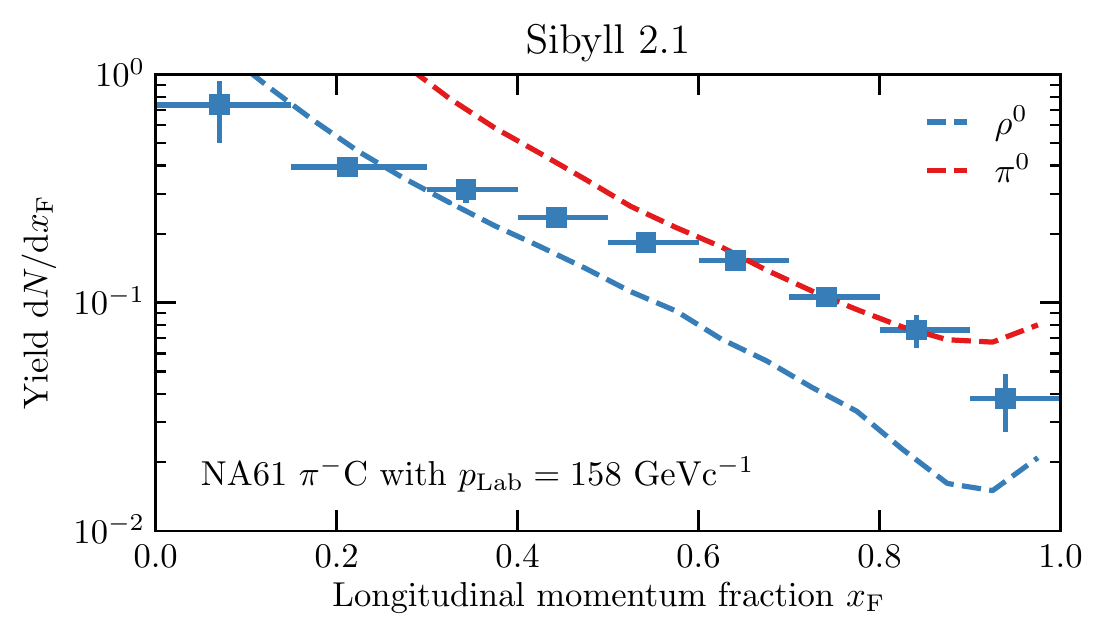}
  \includegraphics[width=\columnwidth]{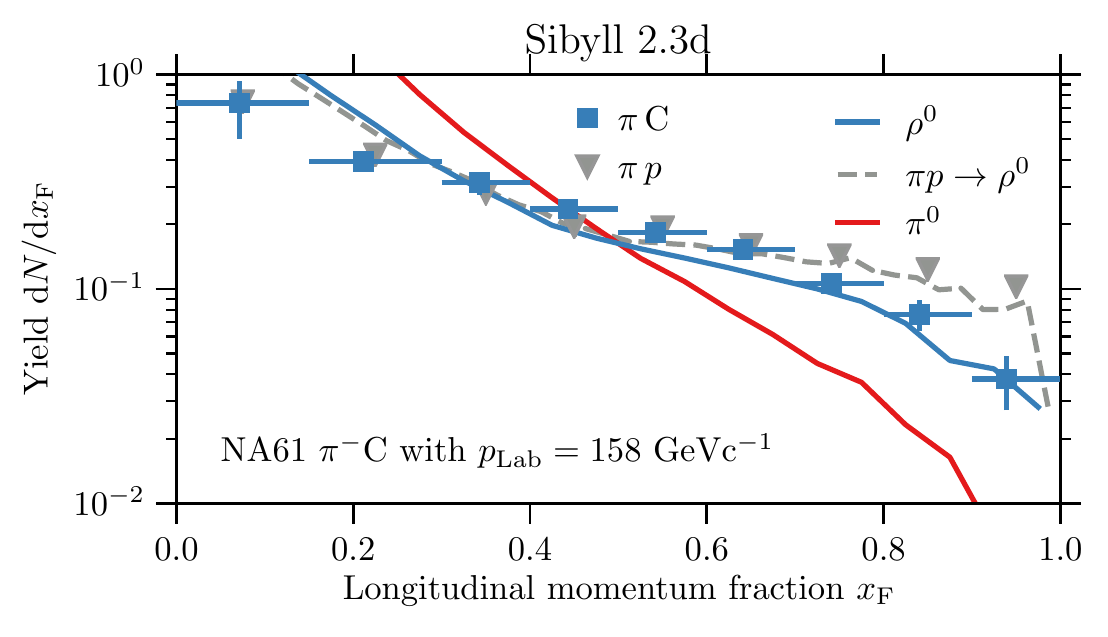}
  \caption{\label{fig:lead-na61}Feynman $x$ spectrum of neutral $\rho$ mesons in pion--carbon interactions as measured in the NA61 experiment~\cite{Aduszkiewicz:2017anm}. This measurement confirms the enhancement of leading $\rho^0$ for nuclear targets. Compared to the data obtained with a proton target (gray triangles), the carbon data (blue squares) reveal a softening of the spectrum, indicating the relevance of interactions with multiple target nucleons. The new remnant model (bottom) correctly reproduces the softening of the leading $\rho^0$ and predicts a suppression of the production of leading neutral pions (red curve).}
\end{figure}

The leading $\rho$ enhancement and $\pi^0$ suppression can be reproduced in \sibyll by adjusting the hadronization for the remnant and for diffraction dissociation. The result is shown in Figure~\ref{fig:lead-na22}. The transition from proton to nuclear targets is entirely described by the dependence of the remnant survival probability on $N_{\rm w}$ in Eq.~\eqref{eq:rmnt-prob}. As demonstrated in Figure~\ref{fig:lead-na61}, the softening of the leading $\rho^0$ spectrum in pion--carbon interactions is well reproduced by the current model. The intersection between the $\rho^0$ and $\pi^0$ spectra is predicted to occur at the same $x_{\rm F}$ in pion--proton and pion--carbon collisions ($x_{\rm F}\approx 0.5$). The position of this intersection is important for EAS since it determines the fraction of the energy that goes either into the EM or hadronic shower component. Until the spectrum of $\pi^0$ is measured for meson-nucleus interactions, this intersection is experimentally not fully determined. Thus the total effect of the leading $\rho^0$ on the number of muons in EAS remains unconstrained (this topic is further discussed in Sec.~\ref{sec:predict_nmu}).

\subsection{\label{sec:fragH}Hadronization}

\subsubsection{Baryon-pair production\label{sec:baryon}}

\begin{figure}
  \includegraphics[width=\columnwidth]{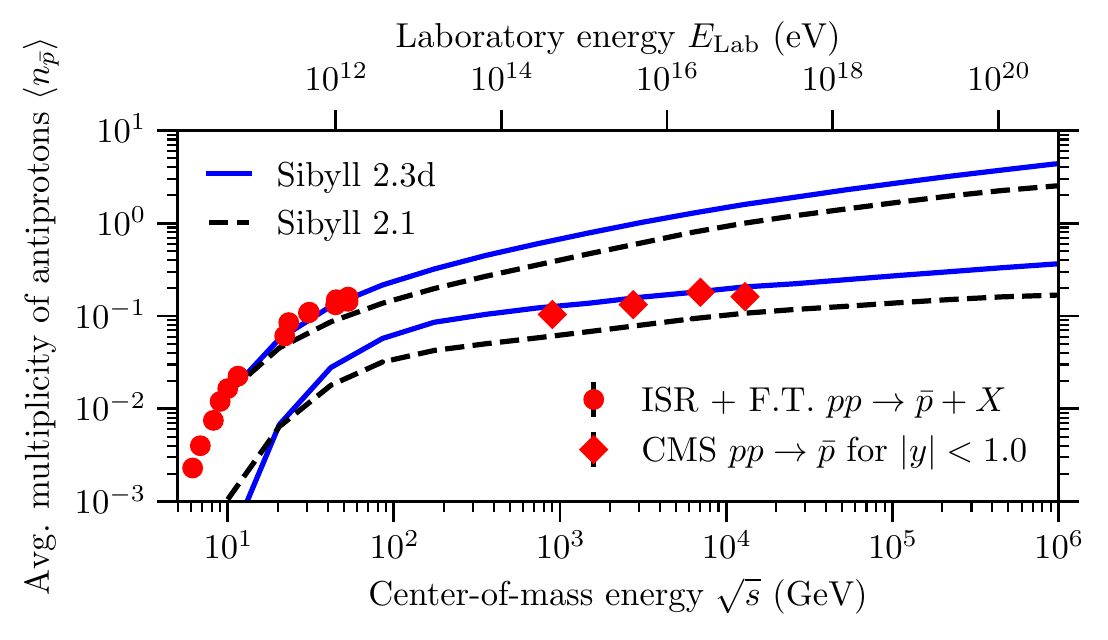}
  \caption{\label{fig:frag-pbar}Average multiplicity of antiprotons as
    a function of center-of-mass energy in proton--proton collisions. The full phase space measurements (filled circles) are obtained at fixed target or early collider experiments (ISR)~\cite{Antinucci73} and measurements at central rapidities (diamonds) from CMS~\cite{Chatrchyan:2012qb,Sirunyan:2017zmn}. The enhanced antibaryon production as implemented in \sibyll~2.3d agrees well with the data at all energies.}
\end{figure}

\begin{figure}
  \includegraphics[width=\columnwidth]{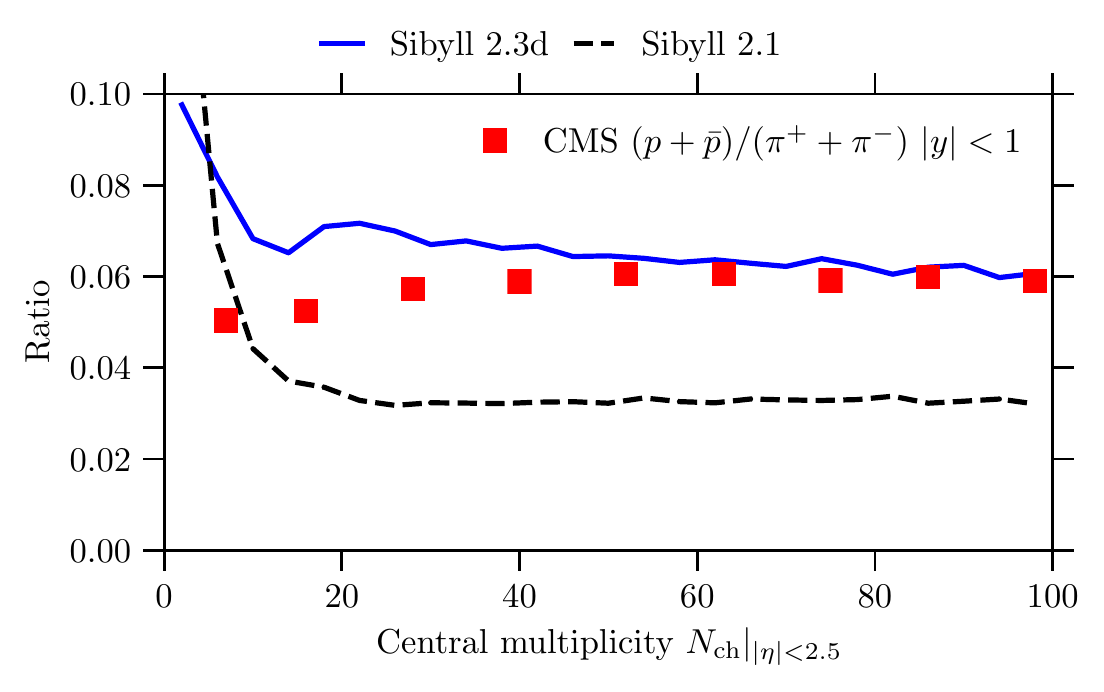}
  \caption{\label{fig:frag-bm-ratio}Ratio of baryons to mesons as a function of central multiplicity at $\sqrt{s} = 7\,$TeV in proton--proton collisions measured by CMS~\cite{Chatrchyan:2012qb}. The central multiplicity is sensitive to the number of parton interactions. High-multiplicity data suggest a constant rate of baryon production per minijet, whereas the region at low central multiplicities is populated by diffractive events and events with a single parton interaction. The substructure for \sibyll~2.3d is due to the remnant model.}
\end{figure}

\begin{figure}
  \includegraphics[width=\columnwidth]{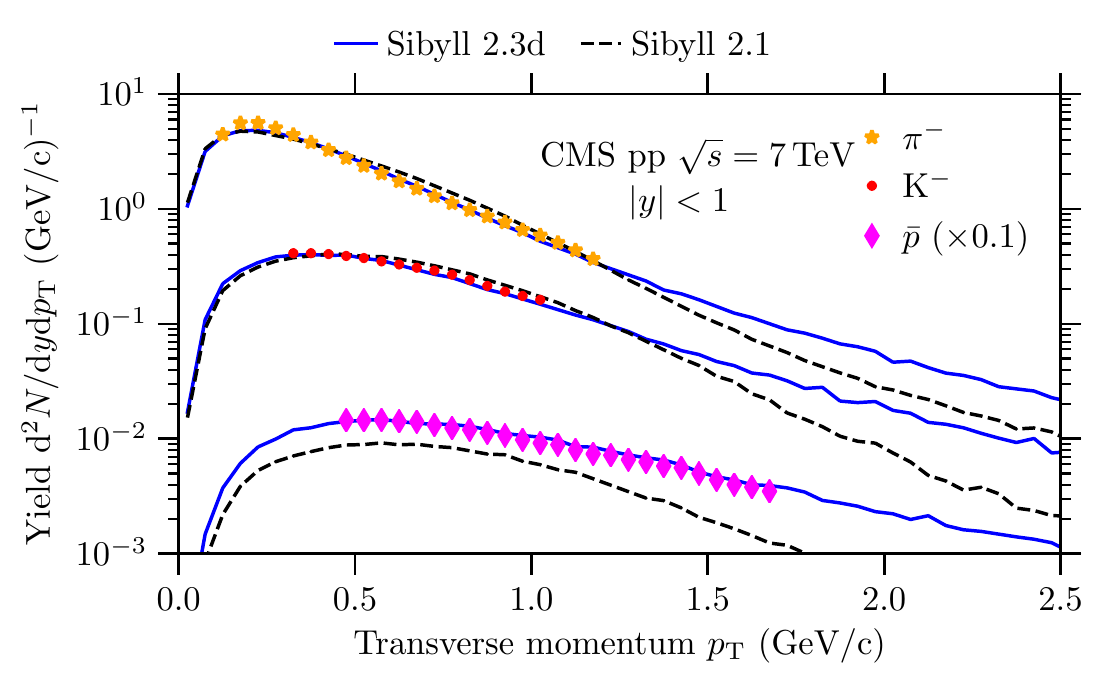}
  \caption{\label{fig:frag-pt}Spectrum of transverse momentum of
    pions, kaons and antiprotons at
    $\sqrt{s}=\unit[7]{TeV}$~\cite{Chatrchyan:2012qb}. The measurement is done for proton--proton collisions in central phase space ($|y|<\unit[1]{}$). The exponential distribution for the string $p_{\rm T}$ in \sibyll~2.3d (solid blue line) gives a much improved description of the spectrum compared to the Gaussian distribution used in \sibyll~2.1 (dashed black line). The improvement in the normalization for antiprotons is due to the enhanced production of baryon pairs in minijets.}
\end{figure}

While the importance of leading particles for the development of EAS is clear, it is not directly evident how a relatively rare process as baryon-pair production affects muon production~\cite{Grieder:1973x1,Pierog:2006qv,Drescher:2007hc}. The role the baryons play is similar to a catalyst in a chemical reaction. Any baryon produced in an air-shower will undergo interactions and produce new particles; in particular, it will regenerate at least itself due to the conservation of baryon number. The interactions continue until the kinetic energy falls below the particle production threshold. Through this mechanism any additional baryon yields more pions and kaons and hence ultimately more muons. In terms of the Heitler-Matthews model, where the number of muons is given by Eq.~\eqref{eq:nmu}, additional baryons represent an increase of the exponent $\alpha$.

In \sibyll's string model, baryonic pairs are generated through the occurrence of diquark pairs in the string splitting with a certain probability, which in \sibyll~2.1 is the global diquark rate $P_{\rm diq}/P_{\rm q}=\unit[0.04]{}$. This model works well at low energies where mostly a single gluon exchange occurs. It fails, however, in the multiminijet regime at higher energies~\cite{Ahn:2013dhz,Engel:2015dxa} (see Figure~\ref{fig:frag-pbar}). Both regimes can be jointly described by choosing a different value for the diquark pair rate in events with multiple parton interactions.
The constant ratio of baryons to mesons in the measurement that is shown in Figure~\ref{fig:frag-bm-ratio} suggests that baryon-pair production cannot depend on the number of minijets or the centrality of the interaction. In the model, the diquark probability is then
\[ P_{\rm diq}/P_{\rm q} = \begin{cases}
  P_{\rm single} & n_{\rm s}+n_{\rm h}=1 \\
  P_{\rm multi} & n_{\rm s}+n_{\rm h}>1 \\
  P_{\rm diff.} & \mathrm{diffractive} \ ,
\end{cases}
\]
where $P_{\rm single}~=~0.06$, $P_{\rm multi}~=~0.13$, $P_{\rm diff.}~=~0.04$ and $n_{\rm s}+n_{\rm h}$ is the sum of the number of soft and hard interactions.

This purely phenomenological model is inspired by the observation in $e^+ e^-$ collision experiments where it is found that baryon-pair production in the fragmentation of quarks or gluons can be different~\cite{Briere:2007ch}.

\subsubsection{Transverse momentum}

\begin{figure}
  \includegraphics[width=\columnwidth]{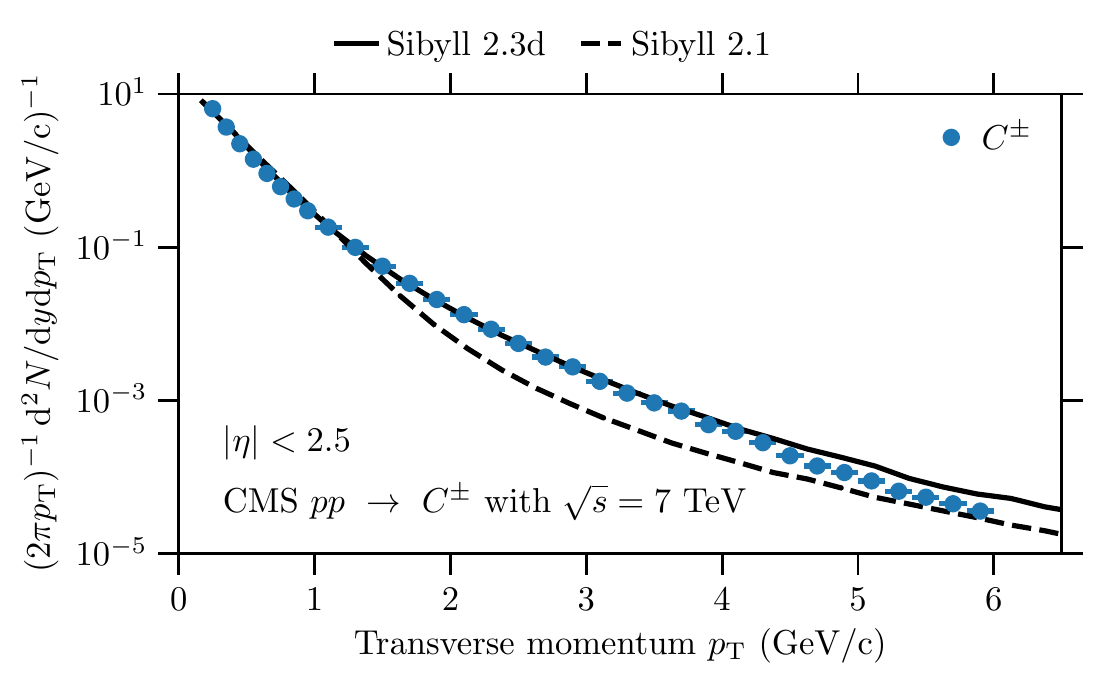}
  \caption{\label{fig:frag-chd-pt}Spectrum of transverse momentum of charged hadrons in proton--proton interactions at $\sqrt{s}=\unit[7]{TeV}$~\cite{Khachatryan:2010us}. The low-$p_{\rm T}$ region is determined by string-$p_{\rm T}$ and the region beyond \unit[2]{GeV} is also influenced by the new PDF.}
\end{figure}

\begin{figure}
  \includegraphics[width=\columnwidth]{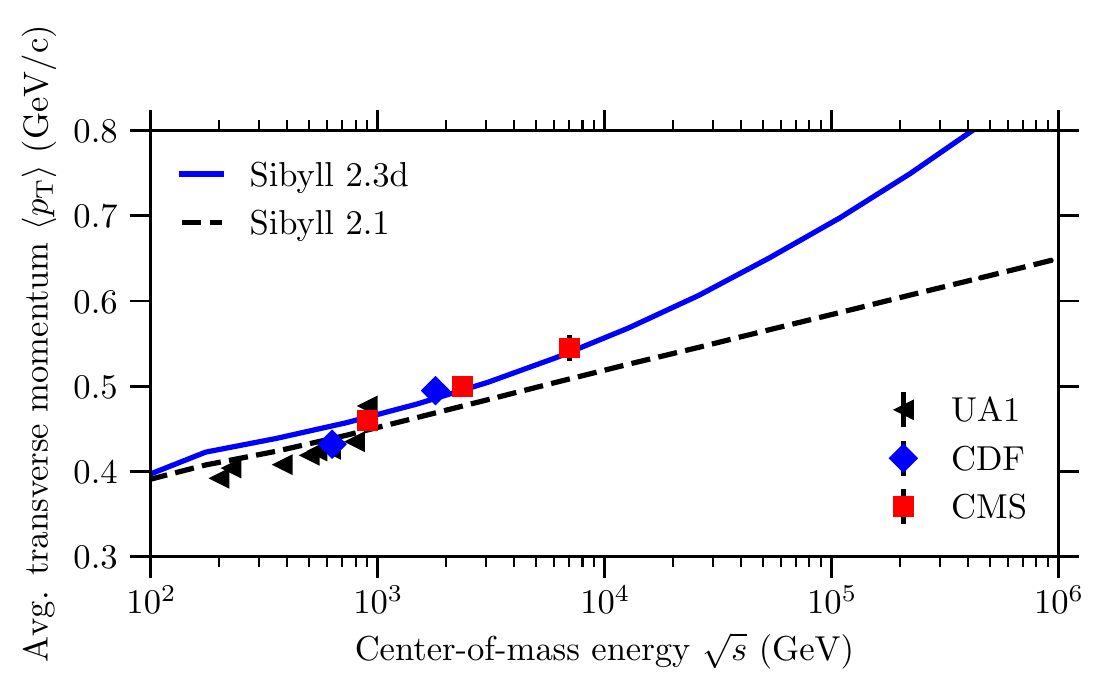}
  \caption{\label{fig:frag-sqs-pt}Average transverse momentum of charged hadrons as a function of center-of-mass energy~\cite{Albajar:1989an,Abe:1988yu,Khachatryan:2010us}. The low energy limit is given by the confinement of the partons to the hadron. The increase with energy is in part due to the increase in
    the hard scattering (jets) threshold ($p_{\rm T}^{\rm min}$) and in part due to the hardening of the string-$p_{\rm T}$ spectrum according to Eq.~\ref{eq:mean-mt}. While the rise in the $p_{\rm T}$-cut is given by QCD and saturation, the rise of string-$p_{\rm T}$ is entirely phenomenological.}
\end{figure}

The transverse momentum in the string fragmentation model (string $p_{\rm T}$) is usually derived from the tunneling of the quark pairs in the string splitting, which results in a Gaussian distribution~\cite{Bialas:1999zg}. However, the observed distribution of transverse momenta in hadron collisions~\cite{Adamus:1988xc,Adare:2011vy} more closely resembles an exponential distribution as predicted by models of ``thermal'' particle production~\cite{Becattini:1995if}, motivating us to distribute the string $p_{\rm T}$ in \sibyll~2.3d according to
\begin{equation}
  f(m_{\mathrm{T},i}) \sim \exp{ [ - (m_{\mathrm{T},i}-m_i)/ \langle m_{\mathrm{T},i} \rangle ] } \ , \nonumber
\end{equation}
where $i$ denotes different flavors of quarks and diquarks. The energy dependence of the average transverse mass $\langle m_{\mathrm{T},i} \rangle$ is parameterized as
\begin{equation}
  \langle m_{\mathrm{T},i}(s) \rangle ~=~ m_{0,i} \, + \, A_{\mathrm{T},i} \log_{10} \left( \frac{\sqrt{s}}{30 \, \rm{GeV}} \right)^2 \ ,
  \label{eq:mean-mt}
\end{equation}
with the parameters $A_{\mathrm{T},i}$ and $m_{0,i}$. The values are given in Table~\ref{tab:mt-pars}.
\begin{table}
  \centering
  \caption{Parameters of the average transverse mass for the different quark flavors in string fragmentation. The diquark masses are computed from the sum of the quark masses. \label{tab:mt-pars}}
    \renewcommand{\arraystretch}{1.5}
    \begin{tabular}{cccc}
      \hline
      Parton & $m_i$ (GeV) & $m_{0,i}$ (GeV) &  $A_{\mathrm{T},i}$ (GeV) \\
      \hline
      u,d & 0.325 & 0.18 & 0.006 \\
      s & 0.5 & 0.28 & 0.007 \\
      c & 1.5 & 0.308 & 0.165 \\
      diq & \ldots & 0.3 & 0.05 \\
      c-diq & \ldots & 0.5 & 0.165 \\
      \hline
    \end{tabular}
\end{table}

These values are derived from the measured $p_{\rm T}$ spectra of pions, kaons and protons at low (NA49) and high energies (CMS, see Figure~\ref{fig:frag-pt}). In addition to the string $p_{\rm T}$, the hadrons acquire their transverse momentum from the initial partonic interaction. As previously mentioned, the parton kinematics in \sibyll~2.3d are determined from post-HERA PDFs (GRV98-LO~\cite{Gluck:1994uf,Gluck:1998xa}), which predict a steeper rise of the gluon density at small $x$, when compared to the old parameterization in \sibyll~2.1. With the new parameterizations the transition between the regions dominated by soft scattering ($p_{\rm T}<3\,$GeV) and hard scattering is described better (see Figure~\ref{fig:frag-chd-pt}).

While the new PDFs help in describing the transition region, the rise of the average transverse momentum with energy is not described well (not shown). To account for the rapid rise with energy seen in the data (see Figure~\ref{fig:frag-sqs-pt}), the energy dependence of the average transverse mass in Eq.~\eqref{eq:mean-mt} is set to be quadratic in $\log{(\sqrt{s})}$. The integration of post-LHC PDFs, in which the small $x$ gluon densities tend to be smaller than in the GRV98 parameterizations, is not expected to help with this.

\subsection{\label{sec:nuc_diff} Nuclear diffraction and inelastic screening}

\begin{figure}
  \includegraphics[width=\columnwidth]{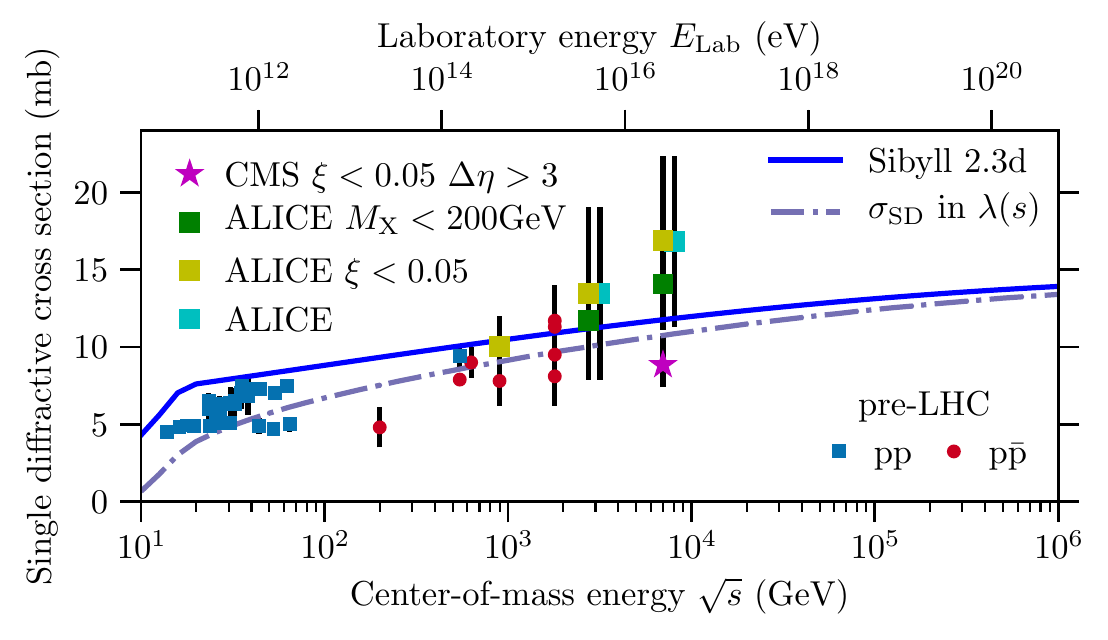}
  \caption{\label{fig:screen-diff-xsctn-pp}Single diffractive cross section
    in proton--proton and proton--antiproton interactions~\cite{Amos:1992jw,Abelev:2012sea,Khachatryan:2015gka}. The cross section in \sibyll~2.3d and the one used in the coupling $\lambda(s)$ are based on the same parameterization with different upper mass limits~\cite{Goulianos:1982vk}.}
\end{figure}

\begin{figure}
  \includegraphics[width=\columnwidth]{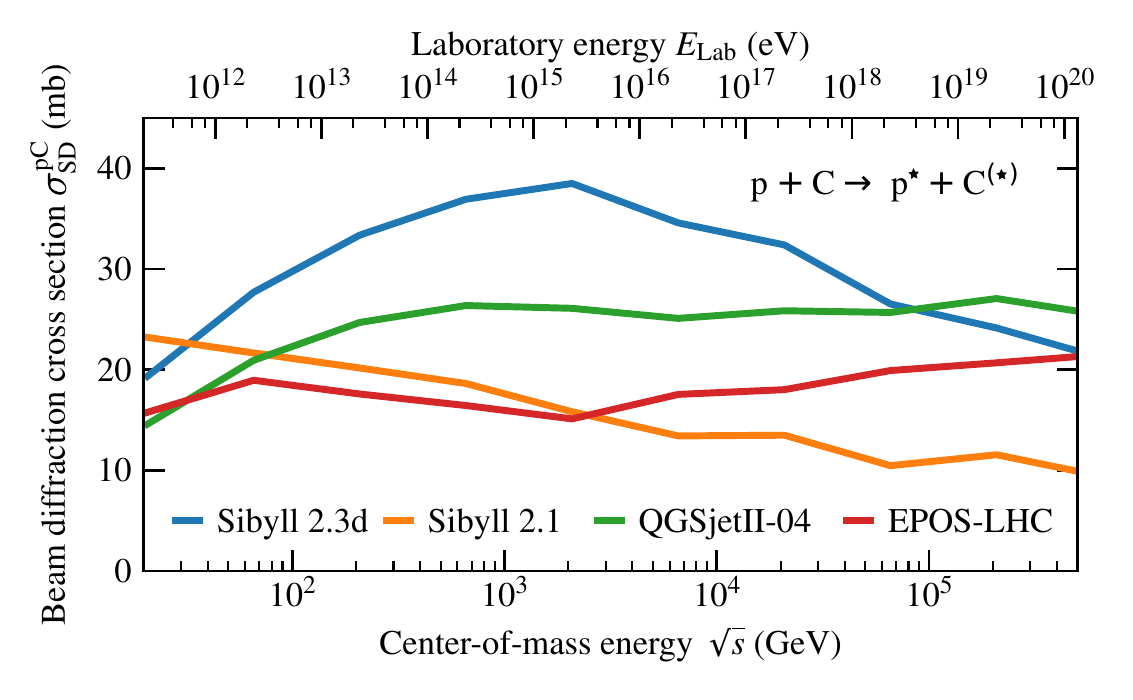}
  \caption{\label{fig:screen-diff-xsctn}Cross section for diffraction
    dissociation in proton--carbon interactions for different CR
    models. Coherent diffraction in \sibyll~2.3d results in the
    increased cross section at low energies.}
\end{figure}

Nuclear cross sections in \sibyll~2.1 are calculated with the Glauber model~\cite{Glauber:1955qq,Glauber:1970jm} neglecting screening effects due to inelastic intermediate states~\cite{Kalmykov:1993qe} in which an excited nucleon may reinteract and return to its ground state. Also, diffraction dissociation in hadron--nucleus interactions is restricted to the incoherent component. \sibyll~2.3d also makes use of the Glauber model but includes screening and the diffractive excitation of the beam hadron in a coherent interaction~\cite{Engel2012,Kaidalov:1979jz}.

In analogy to diffraction dissociation in hadron--nucleon interactions~\cite{Good:1960ba,Ahn:2009wx}, the coherent diffractive excitation of a hadron by a nucleus is implemented using a two-channel formalism with a single effective diffractive intermediate state, where the shape of the transition amplitude to the excited state is equal to the elastic amplitude. The remaining free parameter of the model is the coupling between the states $\lambda$. In the following, we will limit the discussion to proton--nucleus interactions and substitute the nucleon with a proton. With
\begin{equation}
  | p \rangle ~=~ \begin{pmatrix} 1 \\ 0 \end{pmatrix} ~ \textrm{and} ~
  | p^{\star} \rangle ~=~ \begin{pmatrix} 0 \\ 1 \end{pmatrix} \ ,
\end{equation}
where $|p \rangle$ represents the proton and $|p^{\star} \rangle$ is the effective intermediate state or diffractive final state, the generalized amplitude for the described model of proton--proton interactions is
\begin{equation}
  \hat{\Gamma}_{\rm pp} ~=~
  \begin{pmatrix}
    1 & \lambda \\ \lambda & 1 \\
  \end{pmatrix}
  \Gamma^{\, \text{ela}}_{\rm pp} \ . \label{eq:amp-matrix}
\end{equation}
The proton--nucleus cross sections $\sigma_{\text{p}A}$ are calculated with the standard Glauber expressions using the proton--proton amplitude $\hat{\Gamma}_{\rm pp}$, projected onto the desired transition $\langle p | \, \cdots \, | p \rangle$. The diffractive cross sections are calculated in the same way but for the projection $\langle p^\star | \, \cdots \, | p \rangle$~\cite{Engel2012,Riehn:2015vxz}.

The assumed equivalence of the elastic and diffractive amplitude ($ \Gamma_{{\rm pp}\to   \mathrm{p}^{\star}\mathrm{p}}=\lambda \, \Gamma_{{\rm pp}\to {\rm pp}}$) implies for the energy dependence of the coupling $\lambda$
\begin{equation}
   \lambda^2(s) = \frac{ \sigma^{\rm SD}_{\rm pp}(s,M^2_{\rm D, max}) }{\sigma^{\rm ela}_{\rm pp}(s)} \ ,
  \label{eq:nuc-diff-lam}
\end{equation}
where $M^2_{\rm D, max}$ is the upper limit for the excitation mass in diffraction dissociation motivated by the coherence limit~\cite{Goulianos:1982vk} and $s$ is the square of the center-of-mass energy. We assume the coupling $\lambda(s)$ to be universal for different hadrons. The cross sections in Eq.~\eqref{eq:nuc-diff-lam} are taken from parameterizations~\cite{Goulianos95a,PDG96}. The single diffractive cross section used in proton-proton collisions and the parametrization of the coupling $\lambda(s)$ are shown in Figure~\ref{fig:screen-diff-xsctn-pp}. The difference is due to the larger value for the upper mass limit of $M^2_{\rm D,max}/s~=~0.1$ for hadron targets, whereas a lower value of $M^2_{\rm D,max}/s = \unit[0.02]{}$ was found to give the best description of the production cross sections in proton--carbon and neutron--carbon interactions~\cite{Bellettini:1966x1,Dersch:1999zg,Murthy:1975qb}. Although, the description of data in Figure~\ref{fig:screen-diff-xsctn-pp} does not look ideal, one shall consider that several shown data points are extrapolations of rapidity gap data from limited detector acceptance and must not represent accurately $\sigma^{\rm SD}$. A more accurate description of rapidity gaps~\cite{Aad:2012pw,CMS:2013mda} and particle production with diffractive cuts~\cite{Zhou:2019pkz} has to be addressed in a different revision of the model.

The cross section for the diffractive dissociation of the projectile proton in proton--carbon interactions is shown together with the predictions from commonly used interaction models in Figure~\ref{fig:screen-diff-xsctn}. The diffractive cross in \sibyll~2.1 section drops toward high energies, whereas the contribution from coherent diffraction in \sibyll~2.3d compensates this trend. \qgsjet{}II-04~\cite{Ostapchenko:2010vb} and \eposlhc~\cite{Pierog:2013ria} predict almost constant cross sections. Since the diffractive cross section is small relative to the production cross section of $\mathcal{O}(\unit[400]{mb})$, the differences among the models are not expected to be important in EAS.

\subsection{Meson-nucleus interactions}
\label{sec:meson-nucleus}

\begin{figure}
  \centering
  \includegraphics[width=\columnwidth]{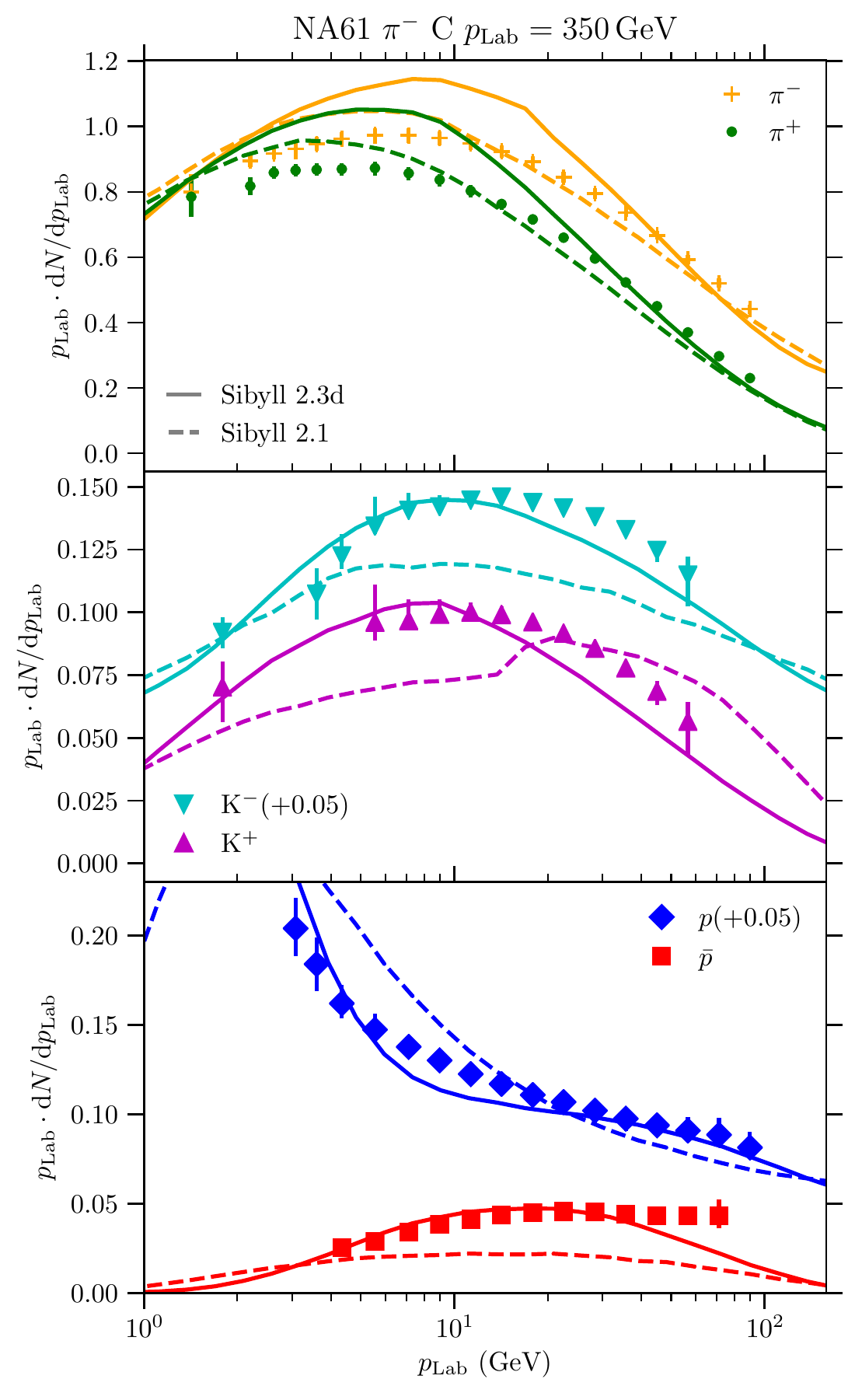}
  \caption{Secondary particle spectra in pion--carbon interactions with $p_{\rm Lab}=350\,$GeV measured by NA61~\cite{Prado:2017hub,Unger:2019nus} and shown together with predictions from \sibyll{}~2.3d (full line and \sibyll{}~2.1 (dashed line). Note that these newer data were not yet available during the development of the models. Some aspects of the distributions are better described by the newer model, in particular the antiprotons, \sibyll~2.3d is far from perfect. Although \sibyll{}~2.3d lacks forward kaons, the description of the charge ratio is improved, resulting in a positive impact on the atmospheric muon charge ratio.\label{fig:predict-na61}}
\end{figure}

The extension of the model from proton--nucleon collisions (as discussed Sec.~\ref{sec:xsctn}) to pion-- and kaon--nucleon collisions is straightforward, since at the microscopic level the interactions are treated universally as scatterings of quarks and gluons. Differences, in particular at low energies, arise from the different profile functions~\cite{Durand:1990qa}, momentum distributions (PDFs)~\cite{Gluck92d} and Regge couplings in the soft interaction cross section (see Appendix~\ref{app:model_parameters}). 

Since the measurements~\cite{Prado:2017hub,Unger:2019nus} from Figure~\ref{fig:predict-na61} were not yet available during the development of the model, the distributions obtained with \sibyll{}~2.3d and \sibyll{}~2.1 are predictions.
Some improvement is observed in the distributions of baryons and kaons. However, the production of central pions, forward kaons and antiprotons clearly demonstrates that the model requires more work.


\section{\label{sec:predict}Air-shower predictions}

\begin{figure}
  \includegraphics[width=\columnwidth]{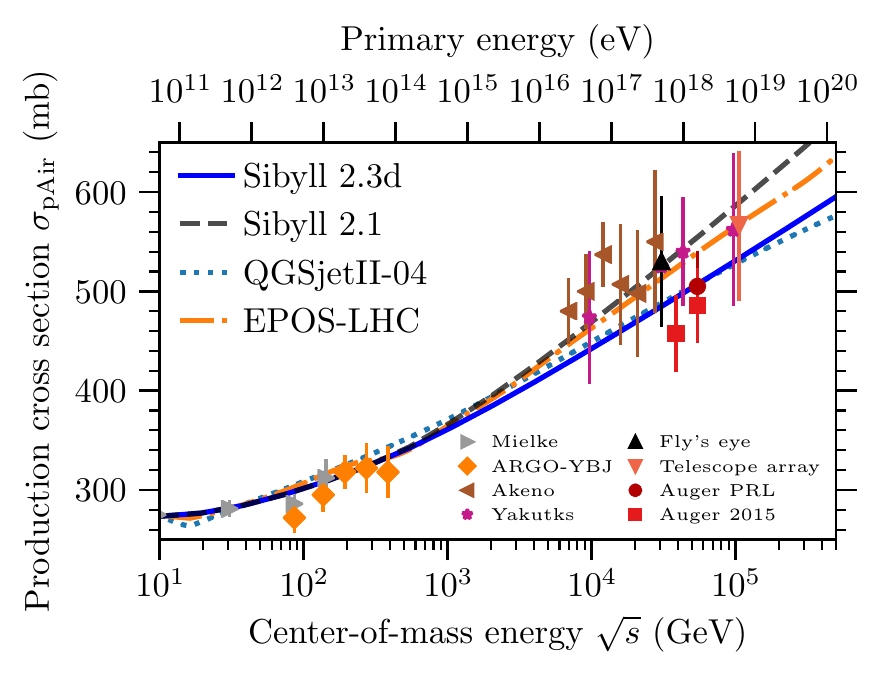}
  \caption{\label{fig:predict-air-xsctn}Energy dependence of the proton--air production cross
    section. The measurements are based on cosmic-ray detections~\cite{Nam:1975aa,Siohan:1978zk,Baltrusaitis:1984ka,Honda:1993kv,Aielli:2009ca,Mielke94,Abbasi:2015fdr,Auger:2012wt,Ulrich:2015yoo}. The
    reduction between the versions of \sibyll comes mainly from the updated proton--proton cross section, whereas the correction due to inelastic screening is small. The most precise measurement at the highest energies by the Pierre Auger Observatory also favors a lower cross section~\cite{Auger:2012wt,Ulrich:2015yoo} in agreement with the extrapolations of the LHC measurements.}
\end{figure}

Some relations between air-shower observables and specific properties of hadronic interactions have been studied in the past~\cite{Ulrich:2010rg}. Here we focus on the depth of shower maximum \xmax and the number of muons \nmu. The calculations are obtained with CONEX~\cite{Bergmann:2006yz}, using FLUKA~\cite{Ferrari:2005zk,fluka2014} to simulate interactions at $E_{\rm kin}<80\,$GeV. The employed scheme is hybrid, meaning that all subshowers with less than $1\,$\% of the primary energy are treated semianalytically using numerical solutions of the average subshower. We compare the predictions from \sibyll~2.3d with the previous \sibyll~2.1 and two other post-LHC models, \eposlhc~\cite{Pierog:2013ria} and \qgsjet{}II-04~\cite{Ostapchenko:2010vb}. In addition, we calculate some of the observables with modified versions of \sibyll~2.3d to show the impact of individual extensions introduced in Sec.~\ref{sec:had-int}. The extensions are labeled in Table~\ref{tab:eas-models} and will be used throughout the next sections. Tables with the predictions for \xmax, \nmu and $\lambda_{\rm int}$ can be found in Appendix~\ref{app:eas_obs}.

\begin{table*}
  \centering
  \caption{Summary of the modified versions of \sibyll{}~2.3d. The modifications correspond to switching off one of the extensions discussed in Sec.~\ref{sec:had-int}.
    \label{tab:eas-models}}
    \renewcommand{\arraystretch}{1.5}
    \begin{tabular}{cc}
      \hline
      Label & Description: \sibyll~2.3d with ...\ \\      
      \hline
      no coherent diffraction & no coherent diffraction in $h$--nucleus collisions (Sec.~\ref{sec:nuc_diff}).\\
      $\lambda_{\rm int,p}$ & proton interaction length as in \sibyll{}~2.1 (Sec.~\ref{sec:xsctn}). \\
      no $\rho^0$ enhancement & no enhanced leading $\rho^0$ in $\pi$--nucleus interactions (Sec.~\ref{sec:leading_mesons}). \\
      no $\bar{p}$ enhancement & no enhanced production of baryons (Sec.~\ref{sec:baryon}). \\
      \hline
    \end{tabular}
\end{table*}

\subsection{\label{sec:predict_sigair}Interaction length and $\sigma_{\rm air}$}

The simplest and most direct connection between the development of an air-shower and hadronic interactions is governed by the interaction length $\lambda_{\rm int}(E)= \langle m_{\rm air} \rangle / \sigma_{\rm prod}(E)$.
It determines the position of the first interaction in the atmosphere and thus directly influences the position of the shower maximum ($X_{\rm max}$).
In the Glauber model~\cite{Glauber:1955qq}, the inelastic cross section in proton--air interactions, $\sigma_{\rm prod}$ is derived from the proton--proton cross section $\sigma_{\rm pp}$. A smaller $\sigma_{\rm pp}$, as in \sibyll~2.3d (Sec.~\ref{sec:xsctn}), translates into a smaller proton--air cross section. The effect on $\sigma_{\rm prod}$ is less than proportional since $\sigma_{\rm pp}$ is only a small contribution to the overall value that is mostly defined by the nuclear geometry.
An additional small reduction of the cross section originates from inelastic screening (Sec.~\ref{sec:nuc_diff}).
The updated proton--air cross section results in a better compatibility with observations as can be seen in Figure~\ref{fig:predict-air-xsctn}.
The impact of the updated interaction length on \xmax is demonstrated in Figure~\ref{fig:predict-dxmax-proton}.
The reduction of the cross section at high energy leads to a shift of \unit[5]{}-$\unit[10]{g/cm^2}$.
Interaction lengths for different primary nuclei and secondary mesons in air are listed in the appendix.

\subsection{\label{sec:predict_xmax} \xmax and $\sigma(X_{\rm max})$}
\begin{figure}
\includegraphics[width=\columnwidth]{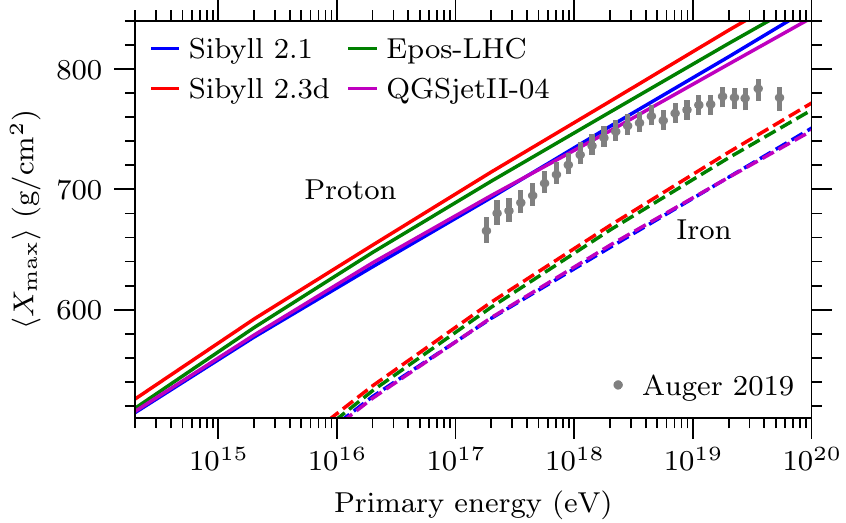}
\caption{\label{fig:predict-xmax-sib}Average depth of air-shower maxima \xmax for different models compared to recent data from the Pierre Auger Observatory~\cite{Aab:2014kda,Yushkov:2019aaa} obtained with the fluorescence detectors. The model lines represent the expectations for a pure proton and iron composition, respectively. The deviation of the data from the pure composition indicates a change toward a mixed composition, i.e.\ cosmic-ray consist of a combination of light and heavier nuclei. The modifications in \sibyll~2.3d drive the interpretation toward heavier nuclei since the \xmax becomes deeper.}
\end{figure}
\begin{figure}
\includegraphics[width=\columnwidth]{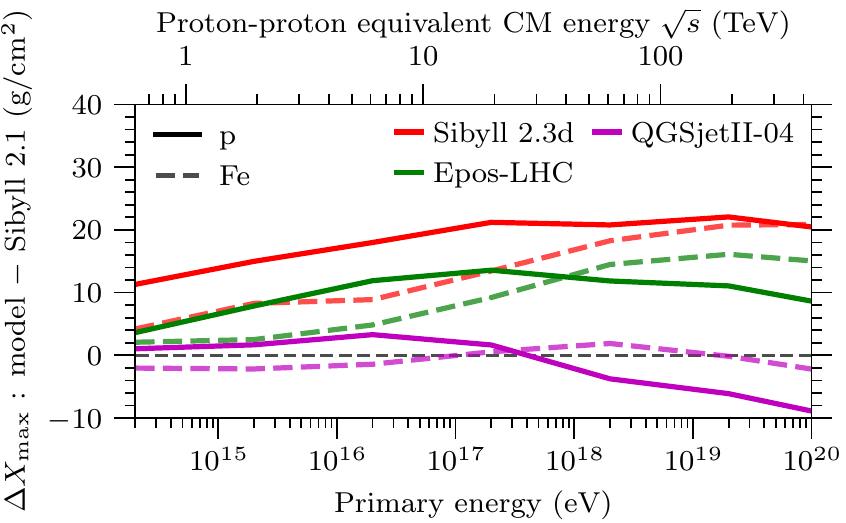}
\caption{\label{fig:predict-dxmax-sib}Difference in the prediction of the average depth of shower maximum between the latest hadronic interaction models (\eposlhc (green), \qgsjet{}II-04 (purple), \sibyll~2.3d (red)) and \sibyll~2.1.}
\end{figure}

\begin{figure}
\includegraphics[width=\columnwidth]{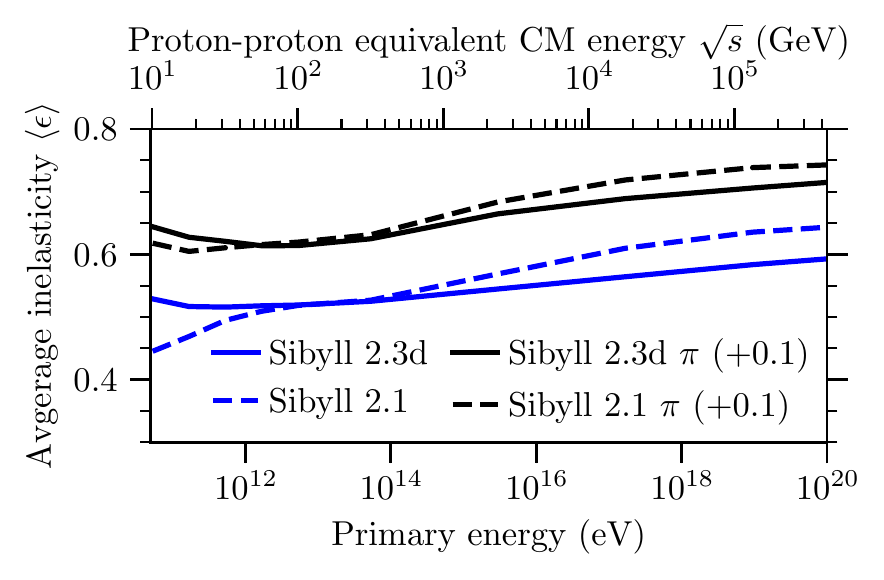}
\caption{\label{fig:predict-inelasticity} Inelasticity in interactions of protons and pions with air. The curves for pions are offset by $+0.1$ for clarity. The interactions of protons and pions are more elastic in \sibyll~2.3d leading to an increased \xmax.}
\end{figure}

The depth at which an individual shower reaches the maximum number of particles is determined by the depth of the first interaction and the subsequent development of the particle cascade.
In very general terms, the development of the cascade is influenced by how the energy of the interacting particle is distributed among the secondaries, in particular by how energy is shared among electromagnetic and hadronic particles. The average shower maximum for proton initiated showers in \sibyll~2.3d is almost $\unit[20]{g/cm^2}$ deeper than that in \sibyll~2.1 (see Figure~\ref{fig:predict-xmax-sib} and Figure~\ref{fig:predict-dxmax-sib}) and on average $10$ to $\unit[20]{g/cm^2}$ deeper compared to other contemporary models.
A large part of this difference comes from the shift in the depth of the first interaction due to the larger interaction length of protons in air. Another contribution to the difference in \xmax is the decreased inelasticity of the interactions (see Figure~\ref{fig:predict-inelasticity}).

\begin{figure}
  \includegraphics[width=\columnwidth]{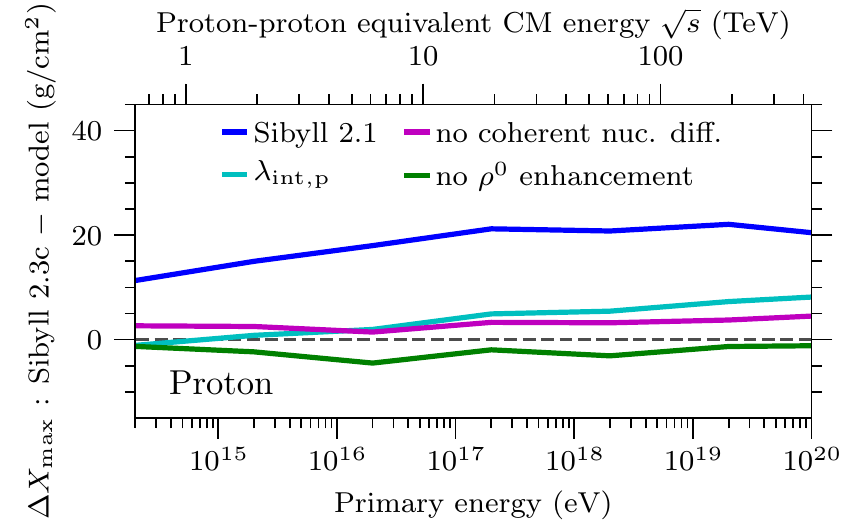}
  \caption{\label{fig:predict-dxmax-proton} Effect of model modifications in \sibyll~2.3d on \xmax. The labels for the modifications are explained in Table~\ref{tab:eas-models}. The change of the cross section for coherent diffraction as described in Sec.\ref{sec:nuc_diff} increases the \xmax by $5\,$g$/$cm$^2$. The change in $\lambda_{\rm int, p}$ due to the smaller proton--proton cross section amounts to another $10\,$g$/$cm$^2$. $\rho^0$ production has a negligible effect on \xmax.}
\end{figure}

Figure~\ref{fig:predict-dxmax-proton} illustrates the effect of the individual modifications on the shift in \xmax. This comparison is produced by individually switching off the model extensions introduced in Sec.~\ref{sec:had-int} and summarized in Table~\ref{tab:eas-models}.
The change in the interaction length (cyan line) is responsible for $10\,$g$/$cm$^2$ out of the $20\,$g$/$cm$^2$ difference between \sibyll~2.1 and \sibyll~2.3d at high energy. Coherent diffraction on the nuclei in the air (purple line), contributes another $5\,$g$/$cm$^2$. The remaining $7\,$g$/$cm$^2$ cannot be attributed to a single feature and emerge from the combination of the model modifications.

The enhanced $\rho^0$ production (green line) and the improved baryon-pair production (not shown) have a small effect on \xmax. These processes mostly affect the later stages of EAS that are more important for muon production (see the next section for more details).

The overall effect of the changes in the multiparticle production between the 2.1 and 2.3d versions result in a decreased inelasticity in Figure~\ref{fig:predict-inelasticity} for proton and pion interactions. Compared to \sibyll~2.1, the inelasticity increases less steeply with energy and should have impacted the elongation rate for protons. This effect seems to have been compensated by the change in the energy dependence of the interaction lenght or cross section (cyan line in Figure~\ref{fig:predict-dxmax-proton}).


\begin{figure}
\includegraphics[width=\columnwidth]{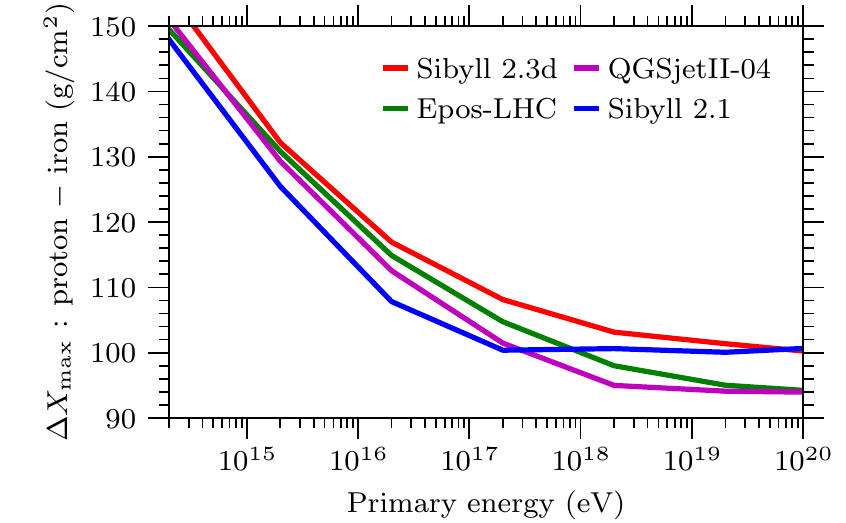}
\caption{\label{fig:predict-dxmax-all-iron} Difference in the \xmax between proton- and iron-induced showers. This observable is relevant for measurements of the cosmic-ray mass composition that are based on observations of \xmax.}
\end{figure}
The separation between proton and iron showers in \xmax at lower energies is larger in \sibyll~2.3d (see Figure~\ref{fig:predict-dxmax-all-iron}), since coherent diffraction only deepens the proton showers and has no effect for nuclear projectiles. This effect is expected to have a higher impact on the measurements of the cosmic-ray composition that were previously interpreted using predictions from \sibyll~2.1.

\begin{figure}
\includegraphics[width=\columnwidth]{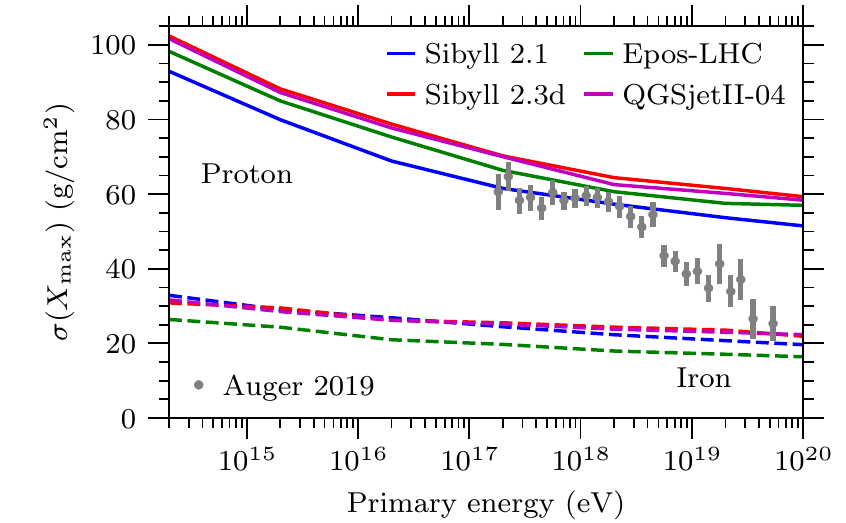}
\caption{\label{fig:predict-rms-all} The width of the $X_{\rm max}$ distribution expected from models using a pure composition compared to data from the Pierre Auger Observatory~\cite{Aab:2014kda,Yushkov:2019aaa}. The $\sigma(X_{\rm max})$ plays an important role in the determination of the mixture of different mass groups at a particular energy.}
\end{figure}

\begin{figure}
\includegraphics[width=\columnwidth]{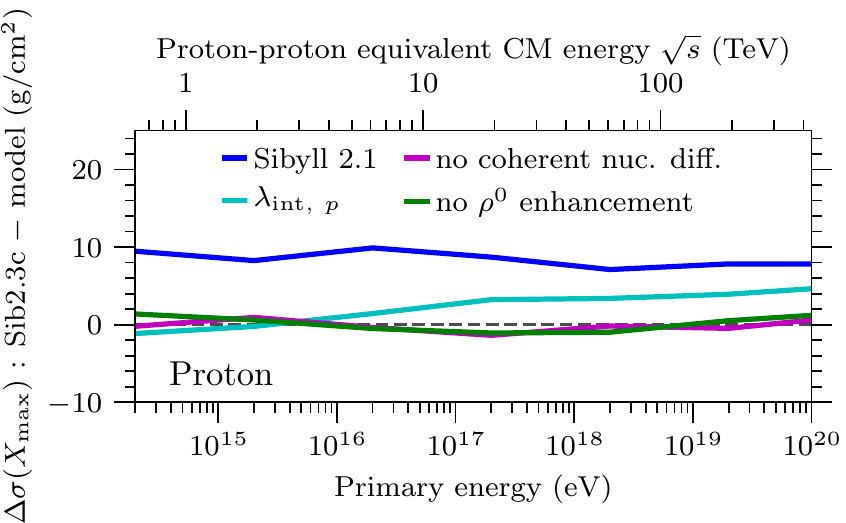}
\caption{\label{fig:predict-drms-sib-p}Effect of model modifications in \sibyll~2.3d on the fluctuations of $X_{\rm max}$. The labels for the modifications are explained in Table~\ref{tab:eas-models}. The fluctuations are most strongly affected by the change in the interaction length. Since the nuclear cross sections are not very sensitive to changes of $\sigma_{\rm pp}$, the impact is highest for proton primaries. This is clearly seen for the iron predictions in Figure~\ref{fig:predict-rms-all}.}
\end{figure}

The width of the distribution of shower maxima $\sigma(X_{\rm max})$ in Figure~\ref{fig:predict-rms-all} increased by $\unit[10]{g/cm^2}$ between the versions, becoming the largest of all CR models.
This change is dominated by the increased interaction length, as is shown Figure~\ref{fig:predict-drms-sib-p}.
Note, that the $\sigma(X_{\rm max})$ increases only for protons, widening the distance between the pure protons and other masses.
This behavior has an important impact on the theoretical interpretation of the measurements in terms of cosmic-ray sources and it has been shown that \sibyll~2.3d produces distinctly different results compared to other contemporary interaction models~\cite{Heinze:2019jou}.


\subsection{\label{sec:predict_nmu}Muons in EAS}
\subsubsection{Number of muons}

\begin{figure}
\includegraphics[width=\columnwidth]{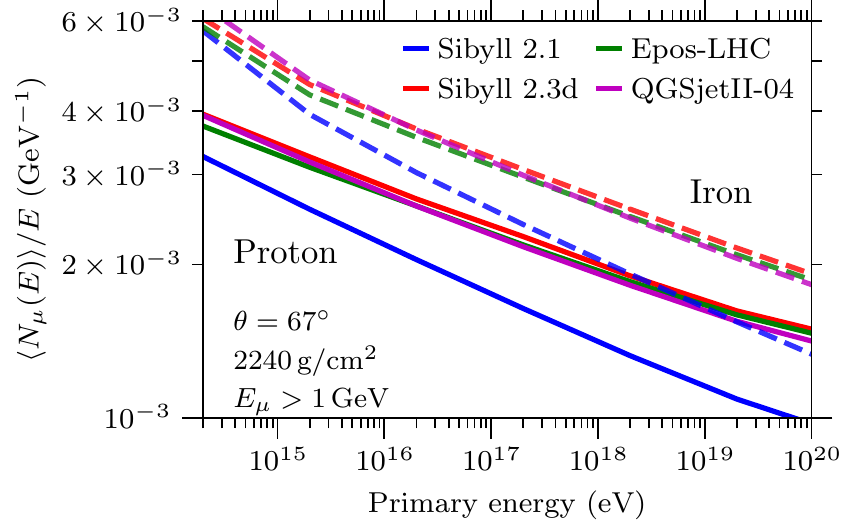}
\caption{\label{fig:predict-nmu-all}Average number of muons at ground in proton and iron showers in air for $E_\mu > 1\,$GeV. It is remarkable that at $10^{17}$ eV, the expectation from \sibyll~2.3d for protons overtakes iron in \sibyll~2.1.}
\end{figure}

\begin{figure}
\includegraphics[width=\columnwidth]{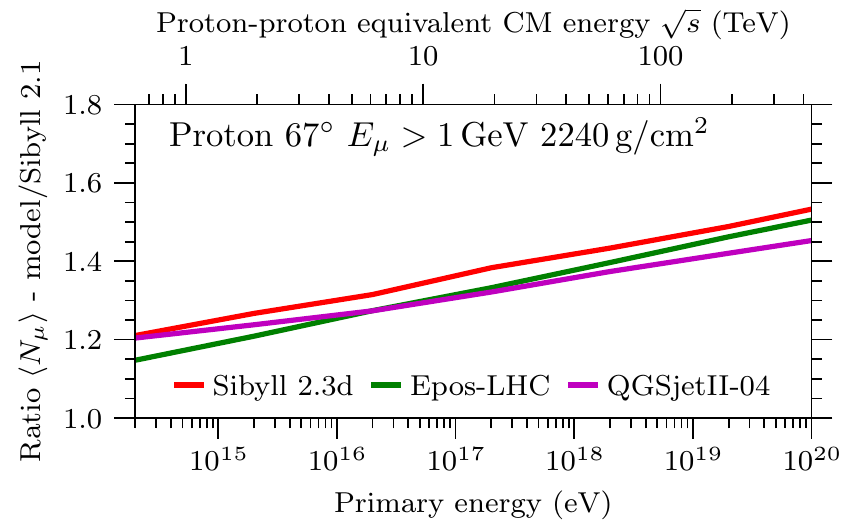}
\caption{\label{fig:predict-dnmu-all}Ratio of the average number of muons between post-LHC models and \sibyll~2.1. The energy dependence of the muon number is similar between the post-LHC models.}
\end{figure}

\begin{figure}
  \includegraphics[width=\columnwidth]{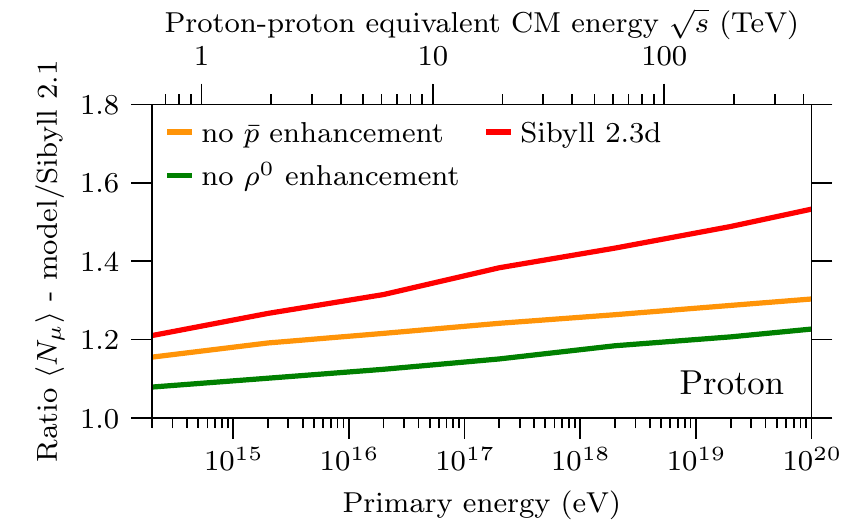}
  \caption{\label{fig:predict-dnmu-sib} Ratio of the average number of muons at ground between \sibyll~2.3d and \sibyll~2.1. The modified versions refer to \sibyll~2.3d where the enhanced $\rho^0$ and baryon production have been switched off (see Table~\ref{tab:eas-models}).}
\end{figure}


In recent years it became evident that the muon content observed in air showers differs from the predictions of the interaction models~\cite{Gaisser:2016ddr}. Recently the Pierre Auger Observatory quantified this ``muon excess'' at ground to be at the order of $30$-$60\,$\%~\cite{Aab:2016hkv}. This result is in agreement with the numbers obtained by the Telescope Array \cite{Abbasi:2018fkz}. In contrast to the \xmax, the production of muons is very sensitive to hadronic particle production at all stages of the shower. It is therefore legitimate to attribute the muon excess to a combination of flaws in the modeling of hadronic interactions. Alternatively, the excess could also be seen as the signature of a new physical phenomena beyond the scales probed by current colliders~\cite{AlvarezMuniz:2012dd,Farrar:2013sfa}.

Most muons in EAS originate from decays of hadrons, most abundantly of pions and kaons. Due to their relatively long lifetime, especially at high energy, these mesons reinteract with air molecules and initiate additional cascades, copiously creating more mesons. The large dependence of the number of muons \nmu on hadronic interactions can be understood by considering that any flaw in the production spectrum of secondaries that persists across multiple generations of reinteractions has a multiplicative effect at the final stages of the shower. In fact, most muons are produced at the end of the cascade where the energies of mesons are low enough to allow a significant fraction to decay before the next interaction. This cascade process leads to a power law relation between the number of muons and the primary energy as shown in Figure~\ref{fig:predict-nmu-all} and by Eq.~\eqref{eq:nmu}. The slope corresponds to the exponent $\alpha$ that depends on the fraction of hadrons that effectively participate in the production of muons. The enhanced baryon-pair and leading $\rho^0$ production in \sibyll~2.3d result in a higher number of charged pions and hence a higher value of $\alpha$. Relative to \sibyll~2.1 (see Figure~\ref{fig:predict-dnmu-all}) the new version has at least 30\% more muons at PeV energies, which increases to $\sim 60\%$ at the highest energies due to a steeper slope. The other post-LHC models include similar extensions and therefore show the same behavior in the muon number. 

The influence of baryon-pair production and $\rho$ production on the number of muons is shown in Figure~\ref{fig:predict-dnmu-sib}, from which the contribution from each enhancement can be seen individually. A reduction of the baryon-pair production to the level of \sibyll~2.1 results in only 10\% less muons at ground. As discussed in Sec.~\ref{sec:leading_mesons}, the ratio between $\rho^0$ and $\pi^0$ is more important for muon production. This is confirmed by Figure~\ref{fig:predict-dnmu-sib} where the difference is at the level of 25\%. With such large variations to the observable number of muons induced by qualitative improvements to the physics of the model, in contrast to just parameter settings, it appears likely that the muon excess in UHECR interactions originates from the shortcomings of the current hadronic interaction models.

\subsubsection{Muon energy spectrum}

\begin{figure*}
  \includegraphics[width=\columnwidth]{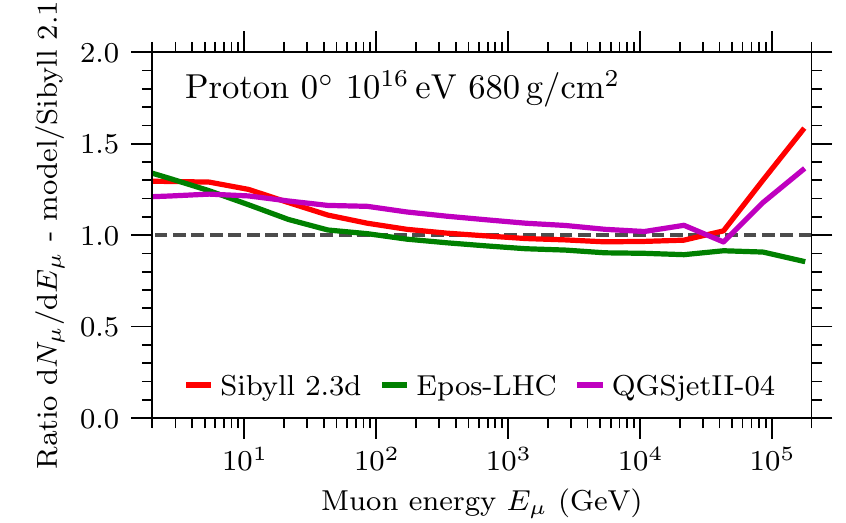}
  \includegraphics[width=\columnwidth]{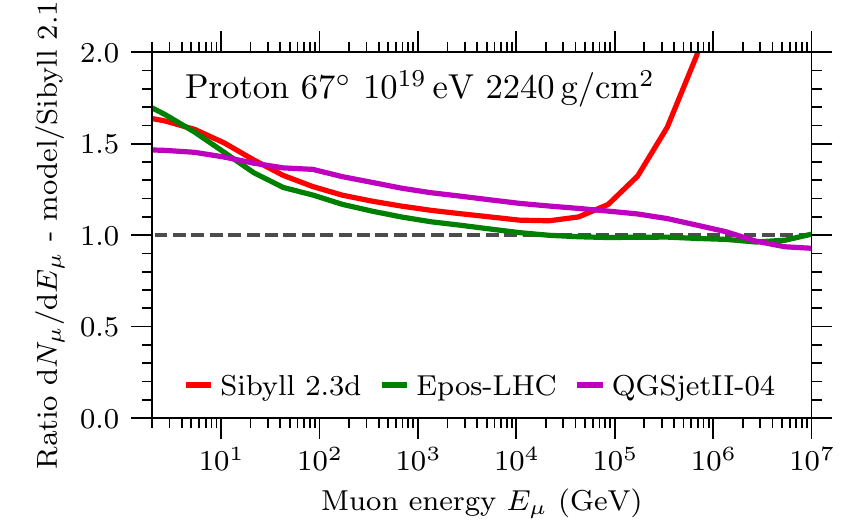}
  \caption{\label{fig:predict-dmu-spec-all}
    Ratio of the muon energy spectrum between the post-LHC interaction models and \sibyll~2.1. Primary particles are protons.
    Left: Vertical showers with primary energy $10\,$PeV, corresponding to the showers studied in IceTop and IceCube~\cite{Aartsen:2017upd}.
    Right: Showers at $10\,$EeV are simulated with a zenith angle of $67^{\circ}$ as they are observed at the Pierre Auger Observatory~\cite{Aab:2014pza}. The increased number of PeV muons in \sibyll~2.3d is due to the prompt decay of charmed hadrons not present in any of the other models~\cite{Engel:2015dxa,Fedynitch:2015zma}.}
\end{figure*}

\begin{figure}
  \includegraphics[width=0.95\columnwidth]{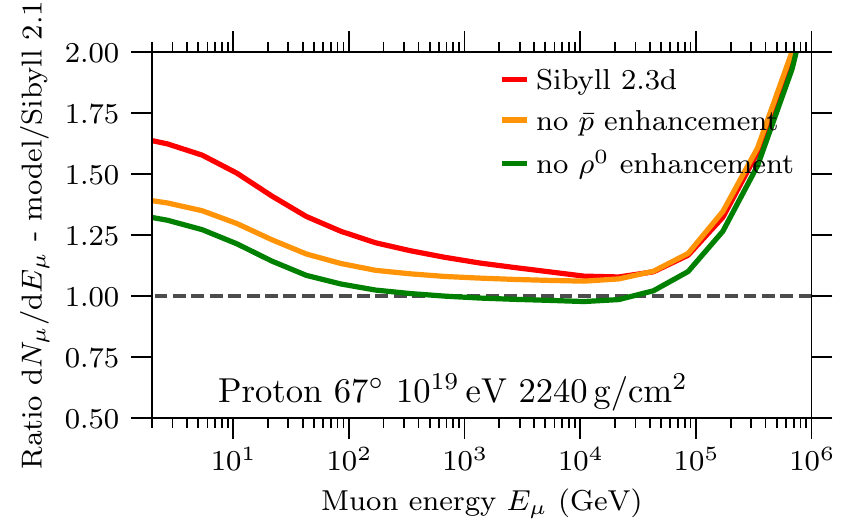}
  \caption{\label{fig:predict-dmu-spec-sib} Ratio of the muon energy spectrum between the versions of \sibyll~2.3d and \sibyll~2.1 for $10\,$EeV proton showers. The models labeled ``off'' refer to modified versions of \sibyll~2.3d where the extensions for enhanced $\rho^0$ and baryon production have been switched off (see Table~\ref{tab:eas-models}). Baryon-pair production enhances mostly the number of low-energy muons, while $\rho^0$ production also affects high-energy muons.}
\end{figure}

The energy spectra of muons for the post-LHC interaction models relative to \sibyll~2.1 are shown in Figure~\ref{fig:predict-dmu-spec-all}.
The clear rise in the number of low-energy muons predominantly originates from the increased number of cascading hadrons due to the modified baryon-pair and $\rho$ production.
The enhancement of muons at high energies originates from decays of charmed hadrons which are an exclusive feature of \sibyll~2.3d in current air-shower simulations. The number of these, so-called, prompt muons is very low and hence no impact is expected for air-shower observations since experimentally an energy threshold around a few $\mathrm{PeV}$ is required. Muons with an energy in excess of $1\,$TeV ($100\,$TeV) constitute only $0.1\,$\% ($3.1 \cdot 10^{-5}\,$\%) of all muons at ground for a $10^{19}\,$eV shower (see also Appendix~\ref{app:eas_obs}). For inclusive lepton fluxes this contribution has important implications as discussed in Ref.~\cite{Fedynitch:2018cbl}.

In the left panel of Figure~\ref{fig:predict-dmu-spec-all} the energy and incident angle of the primary CR resemble the typical experimental conditions of IceTop and IceCube~\cite{IceCube:2012nn,Aartsen:2017upd}, whereas the right panel resembles typical conditions at the Pierre Auger Observatory~\cite{Aab:2014pza}.
It is remarkable that the model-specific features of the spectrum are present across very different primary energies.

Another observation is that the current models predict different shapes of the muon spectrum. With a combination of the surface air-shower array IceTop and the main instrumented IceCube volume deep in the Antarctic ice, the IceCube Observatory has the potential to discriminate among the interaction models by measuring the muon content of a single air-shower at two different energy regimes simultaneously. IceTop is sensitive to the low-energy muons while only the muons with $E_\mu \sim$~TeV can penetrate the ice deep enough to generate the ``in-ice'' muon signal. The preliminary results clearly indicate that \sibyll~2.1 has too many high- and too few low-energy muons~\cite{DeRidder:2017aa}. The discrepancy is expected from the discussion of Figure~\ref{fig:predict-dmu-spec-all} above, since \sibyll~2.1 neither describes the baryon-pair production nor the $\rho$ production very well. The same analysis shows that \sibyll~2.3d accurately reproduces both low- and high-energy muons. The result is, however, difficult to translate into constraints on the hadronic parameters since the (unknown) mass composition has to be simultaneously taken into account. The impact of each modification on the muon spectrum is illustrated in Figure~\ref{fig:predict-dmu-spec-sib}. According to the figure baryon-pair production contributes dominantly at low energies, while the contribution from $\rho$ affects all energies.

\subsubsection{\label{sec:nmu_compo}Effect of the projectile mass on muon production}

\begin{figure}
  \includegraphics[width=\columnwidth]{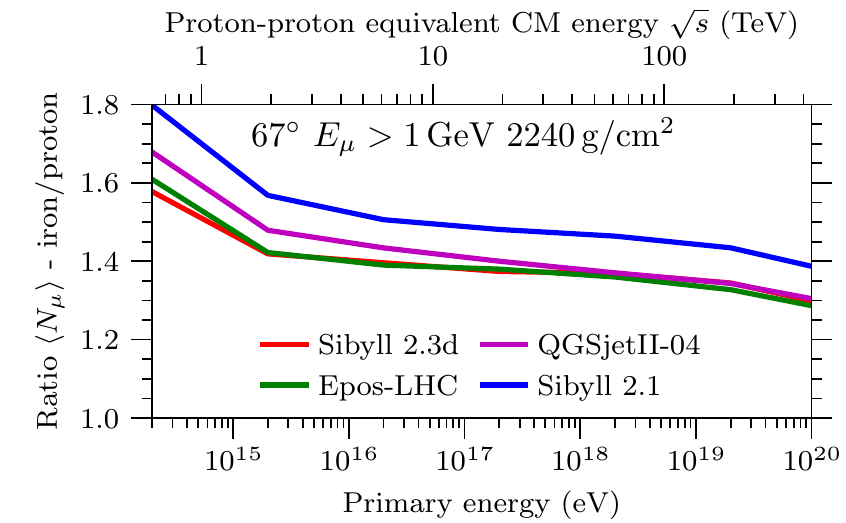}
  \caption{\label{fig:predict-dnmu-all-primary}Ratio of the average number of muons between proton- and iron-induced showers for post-LHC
    models and \sibyll~2.1. As the number of muons increases in the
    models the difference between p and Fe showers decreases.}
  \end{figure}

\begin{figure}
  \includegraphics[width=\columnwidth]{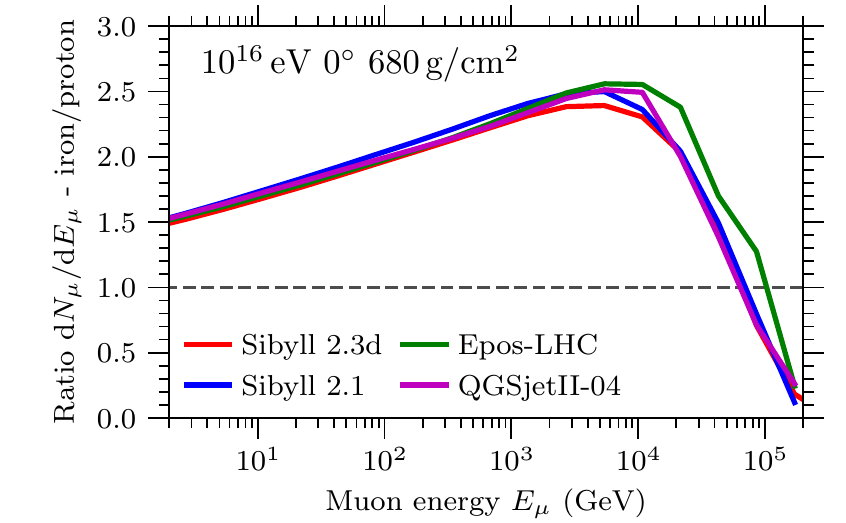}
  \hfill
  \includegraphics[width=\columnwidth]{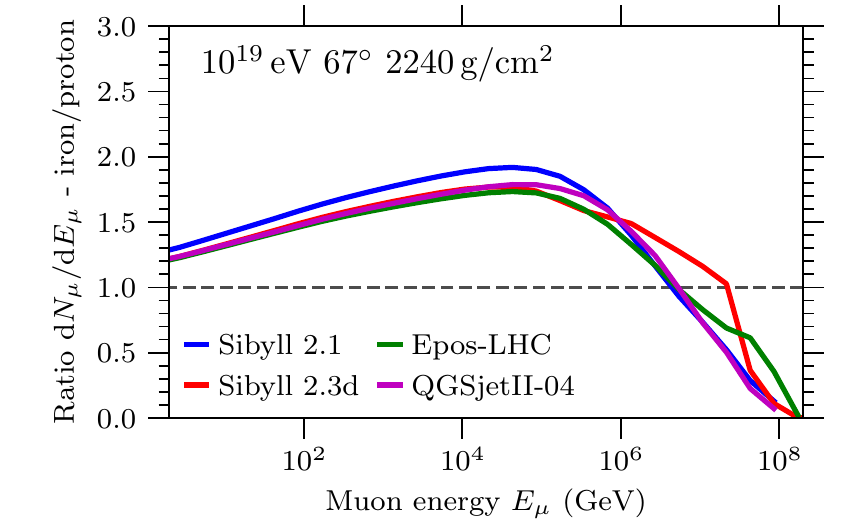}
  \caption{\label{fig:predict-dmu-spec-all-primary}Ratio of the energy spectrum of muons in iron and proton induced air showers. The upper panel shows vertical air showers at the depth of the IceTop array ($680\,$g$/$cm$^2$)~\cite{Aartsen:2017upd}. The figure on the bottom is calculated for the depth of $2230\,$g/cm$^2$, corresponding to the inclined air showers measured at the Pierre Auger Observatory~\cite{Aab:2014pza}. 
}
\end{figure}

\begin{figure}
  \includegraphics[width=\columnwidth]{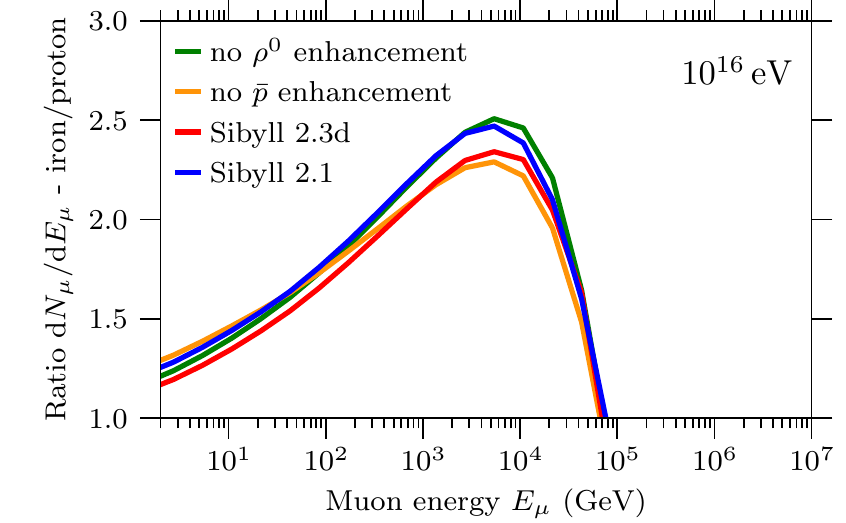}
  \includegraphics[width=\columnwidth]{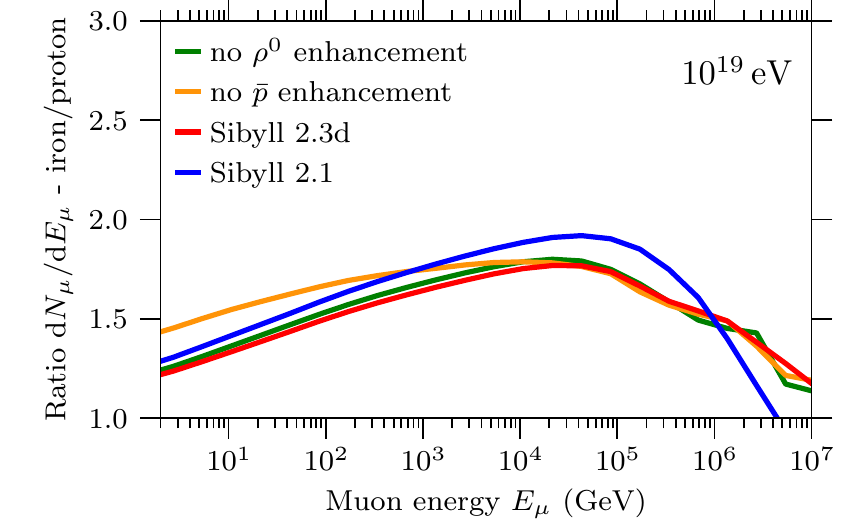}
  \caption{\label{fig:predict-dmu-spec-sib-primary} Effect of new processes in \sibyll on the separation between proton and iron, shown as the ratio of muon spectra ($\mathrm{d}N_\mu / \mathrm{d}E_\mu$) in iron-induced showers divided by that in proton showers. Due to higher hadron and muon numbers in \sibyll~2.3d the proton and iron separation decreases with respect to version 2.1 and becomes comparable to the other interaction models.}
\end{figure}
  
The spectra for the individual mass groups of cosmic-ray nuclei are not well known across the entire energy range of the indirect air-shower measurements \cite{Kampert:2012mx}. The main source of this systematic uncertainty stems from ambiguities among the interpretations of EAS observables with different hadronic interaction models. At present, at ultrahigh energies the most robust method to estimate the composition relies on the electromagnetic component only. Recent attempts to use the surface detector and exploit the muon content as a sensitive variable, often result in incompatible results \cite{Sanchez-Lucas:2017nhr}.

We study the ratio of the muon energy spectra for the two extreme composition assumptions, pure protons and pure iron. The ratios in Figure~\ref{fig:predict-dmu-spec-all-primary} demonstrate that the difference in the number of GeV muons is small between UHE protons and iron nuclei ($\sim 20\%-40\%$). As discussed in the previous section, similar variations are expected just from swapping the interaction model. At higher muon energies ($E_\mu>100\,$GeV) protons and iron are well separated. The shape comes from two effects: the earlier development of iron showers due to the shorter interaction length of the primary nucleus and the lower energy carried by the individual nucleons in the iron nucleus. If one would take the muon energy spectrum from iron primaries with $E_{\rm Fe} = 56 E_{\rm p}$ and compare with the spectrum in proton showers at the shower maximum they would have identical shapes.

The superposition ansatz ($E_0 \to E_0/A$ and $N^{\rm A}_\mu=A N^1_\mu$) in the Heitler-Matthews model of Eq.~\eqref{eq:nmu} yields for the composition dependence of the total muon number an additional multiplicative term $(1-\alpha)\,\ln (A)$. If $\alpha$ approaches unity, as is the case for the current model extensions, the difference between protons and nuclei decreases. This expectation is confirmed by full model calculations in Figure~\ref{fig:predict-dnmu-all-primary}, in which the muon number varies by only 35\% between proton and iron for post-LHC models, while for \sibyll~2.1 the difference is almost $50\%$. However, the ratio of iron to proton spectra from different interaction models agree remarkably well (see Figure~\ref{fig:predict-dmu-spec-all-primary}).

The influence of individual model processes on the separation between proton and iron are demonstrated in Figure~\ref{fig:predict-dmu-spec-sib-primary}. Both baryon-pair production and $\rho$ production enhance low-energy muons and essentially reduce this separation through a more elongated hadronic cascade (or in other terms, a larger $\alpha$ in the Heitler-Matthews model).
However there are subtle differences. At $10^{16}\,$eV only enhanced $\rho$ production is important for the difference between the primaries in TeV muons, while low-energy muons are affected by both mechanisms. At $10^{19}\,$eV, the difference between primaries is not much affected by $\rho$ production and baryon-pair production and other changes in the model seem to play more central roles.


\section{\label{sec:discuss}Discussion and conclusion}

This paper documents the latest extensions to the hadronic interaction model \sibyll and discusses their impact on extensive air showers. The model update is motivated through the availability of recent particle accelerator measurements, where measurements from experiments at the LHC and those from fixed-target experiments are equally important. The goal is to improve the consistency in the description of extensive air showers, in particular related to the muon content that impacts the interpretation of the mass composition of the primary cosmic rays. A tabulated overview of the changes between the \sibyll~2.1 and \sibyll~2.3d is available in Appendix~\ref{app:model_parameters}.

The interaction cross sections from measurements at the LHC point towards lower total and inelastic proton--proton cross sections that favor the low data points from measurements at the Tevatron. Our new fits take the measurements up to $\sqrt{s} = 13\,$TeV into account, reducing the extrapolation uncertainties up to ultrahigh cosmic-ray energies. The effect on the proton--air cross section is a reduction of the tension between \sibyll and the cross section measurement derived from UHECR observations at the Pierre Auger Observatory. The spectra of identified particles, measured in central phase space at the LHC, allow us to adjust the hadronization to account for a higher baryon-pair production compared to the previous version. Together with the updated PDFs, the high-energy data constrains the shape and energy dependence of transverse momentum distributions.

On the other hand, the fixed-target measurements in p--p, p--C, $\pi$--p and $\pi$--C beam configurations yield enough information to identify the shortcomings of the previous model version and entirely revise the leading particle production. We implement a model that makes use of the remaining hadron content in the beam remnants that can undergo further excitation and hadronization processes. This mechanism adds necessary degrees of freedom to decouple very forward particle production from central.

None of the new features requires drastic changes in the underlying principles and assumptions that were defining \sibyll during the last decades. Microscopically, the main picture is still a combination of the dual Parton and the minijet model, a fusion of perturbative QCD (hard component) and elements of the Gribov-Regge field theory (soft component).

We identified, however, a number of problems that indicate a necessity to depart from these well-explored principles in future versions. One of these problems is related to the growth of the multiplicity distribution that rises faster in the model than in data. A second problem is the narrow width of the pseudorapidity distributions that most likely is an effect of the missing contribution from semihard processes. Both aspects are related to the underlying partonic picture, and a permanent solution will require an overhaul of several old principles in the code base.

On the nuclear side, the previous Glauber-based model is extended to include screening corrections on the production cross section due to inelastic intermediate states. The updated model for diffraction dissociation now incorporates the process of coherent diffraction, in which the beam hadron transitions to an excited state without the target side nucleus loosing its coherence.

Charm hadron production is added explicitly for particle astrophyics applications. In particular this affects calculations of atmospheric neutrinos at very high energies, where the flux of atmospheric leptons competes with that of astrophysical origin. The details of this topic are discussed in a separate publication~\cite{Fedynitch:2018cbl}.


Regarding air showers, several of the changes to the hadronic interaction model impact the simulations. The showers reach their maximum deeper by $20\,$g$/$cm$^2$ with respect to \sibyll~2.1, mainly due to the modifications to nuclear diffraction and the updated interaction cross sections for protons and pions. The fluctuations of the $X_{\rm max}$ in proton showers are almost $10\,$g$/$cm$^2$ larger as an effect of the increased interaction length and elasticity. Both modifications are likely to yield a notably heavier composition in the interpretation of the flux of UHECR.

The muon number in \sibyll~2.3d drastically increases by $20\%-50\%$ relative to \sibyll~2.1, which was previously known to yield too few muons. Compared to the other interaction models the new version has the highest number of muons  but only exceeding the numbers from \eposlhc and \qgsjet{-II-04} by $\sim 1\%-5\%$. This change will certainly reduce the muon excess seen by the Pierre Auger Observatory and the Telescope Array, but will most likely not be sufficient to remove entirely the tension between simulation and data.
We demonstrated that the forward spectrum of $\pi^0$ and leading $\rho$ mesons in $\pi$--nucleus interactions effectively modulates the total muon number and that a constraining measurement of the $\pi^0$ is one of the leading uncertainties. 

We expect that the combined measurements with the IceCube and IceTop detectors at two energy regimes, and, the event-by-event composition sensitivity of the upgrade of the Pierre Auger Observatory (AugerPrime)~\cite{Aab:2016vlz}, will help to resolve the mysteries around the muon component in EAS.

\begin{acknowledgments}
We thank F.\ Penha, H.\ P.\ Dembinski, T.\ Pierog, S.\ Ostapchenko and our many colleagues from the IceCube, KASCADE-Grande, LHCf, and Pierre Auger Collaborations and the CORSIKA 8 development team for their feedback and discussions.
This work is supported in part by the KIT graduate school KSETA, in part by the German Ministry of Education and Research (BMBF), grant No.\ 05A14VK1, and the Helmholtz Alliance for Astroparticle Physics (HAP), which is funded by the Initiative and Networking Fund of the Helmholtz Association and in part by the U.S.\ National Science Foundation (PHY-1505990). The authors are grateful to the Mainz Institute for Theoretical Physics(MITP) of the DFG Cluster of Excellence PRISMA+ (Project ID39083149), for its hospitality and its support during the completion of this work. This project received funding through the contribution of A.\ F.\ from the European Research Council (ERC) under the European Unions Horizon 2020 research and innovation program (Grant No.\ 646623). The work of T.K.\ G.\ and T.\ S.\ is supported in part by Grants from the U.S.\ Department of Energy (DE-SC0013880) and the U.S.\ National Science Foundation (PHY 1505990). The work of F.\ R.\ is supported in part by OE - Portugal, FCT, I.\ P.~, under Project No.\ CERN/FIS-PAR/0023/2017 and OE - Portugal, FCT, I.\ P.~,  under Project No.\ IF/00820/2014/CP1248/CT0001. F.~R.\ also acknowledges the financial support of Ministerio de Economia, Industria y Competitividad (FPA 2017-85114-P), Xunta de Galicia (ED431C 2017/07). This work is supported by the Mar\'\i a de Maeztu Units of Excellence Program No.\ MDM-2016-0692 and the Spanish Research State Agency. This work is co-funded by the European Regional Development Fund (ERDF/FEDER program). A.F.\ completed parts of this work as JSPS International Fellow supported by JSPS KAKENHI Grant No.\ 19F19750.

\end{acknowledgments}

%

\onecolumngrid
\clearpage
\begin{appendix}
  
  \section{\sibyll version history}
  \label{app:version_history}
  Several versions of \sibyll (see Table~\ref{tab:sib-models}) have become publicly available throughout the development cycle and were available in the air-shower simulator CORSIKA (versions 7 and 8)~\cite{Heck98a,Engel:2018akg} and the cascade equation code MCEq~\cite{Fedynitch:2018cbl}. In this section, we give a brief overview of the changes and estimate the quantitative impact on $\xmax$ and the number of muons in air showers. 
  
  \sibyll~2.1 is the basic implementation of the hadron interaction model and was outlined in Sec.~\ref{sec:basic} and described in detail elsewhere~\cite{Ahn:2009wx}. The first public release of the \sibyll~2.3~\cite{Riehn:2015oba} model improved the compatibility with LHC measurements and astroparticle experiments as described in the main text. The model exhibited a stronger violation of Feynman scaling in the fragmentation region than supported by data~\cite{Riehn:2017mfm,Fedynitch:2018cbl} that has been addressed in \sibyll~2.3c.
  
\begin{figure}
    \includegraphics[width=0.48\columnwidth]{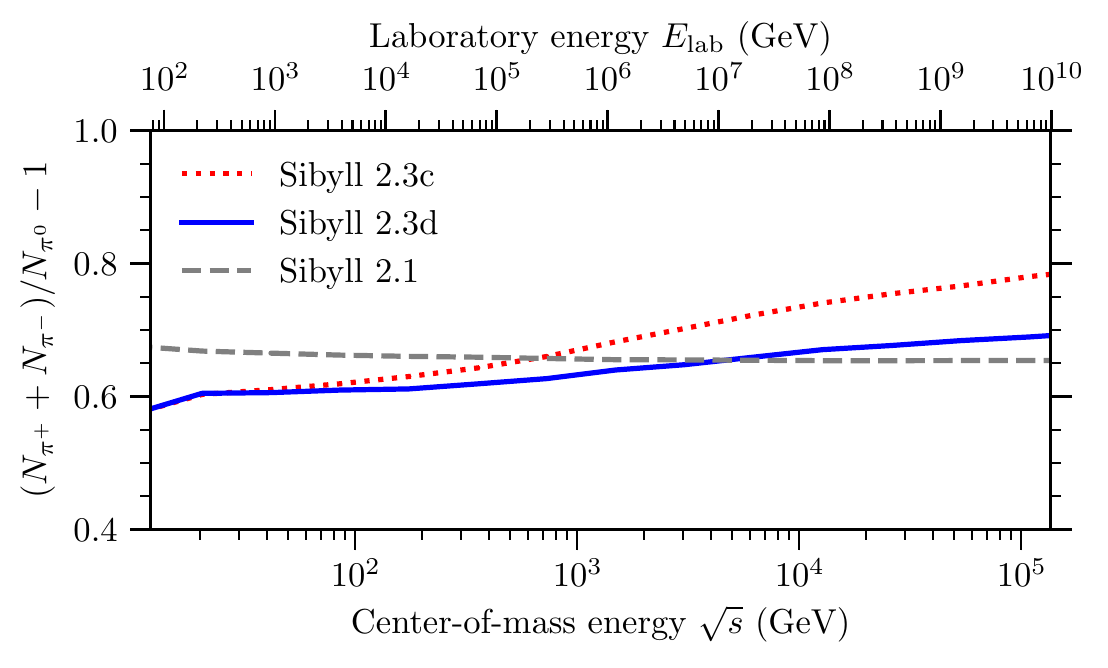}
    \includegraphics[width=0.48\columnwidth]{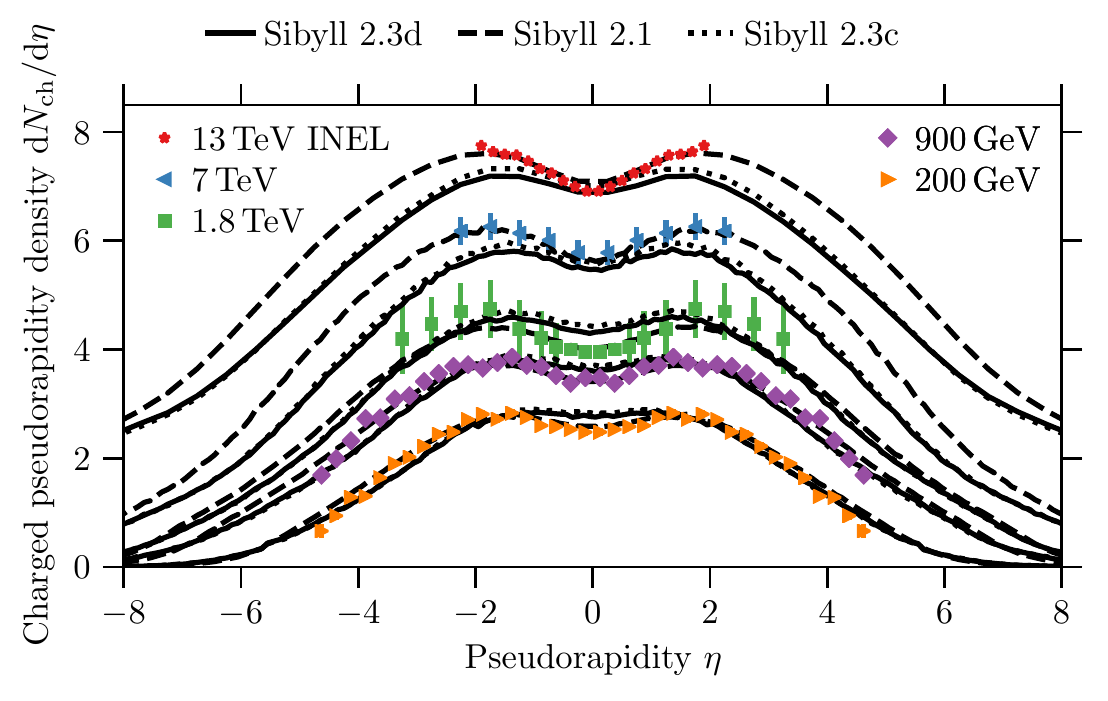}
    \caption{\label{fig:pi-ratio} Left panel: The ratio of the number of charged and neutral pions, $\pi^\pm/\pi^0-1$, in proton--proton interactions as a function of the energy. Right panel: Distribution of charged particles over pseudorapidity in \sibyll~2.1, \sibyll~2.3c and \sibyll~2.3d.}
  \end{figure}

  In a recent publication the behavior of the $\pi^\pm$ to $\pi^0$ ratio in different mechanisms of hadronization and the role in muon production in air showers were discussed~\cite{Baur:2019cpv}.
  In \sibyll~2.3c this ratio has a stronger than expected energy dependence (see left panel of Figure~\ref{fig:pi-ratio}), because a part of the model responsible for the leading $\rho^0$ (Sec.~\ref{sec:leading_mesons}) interfered with the fragmentation of minijets.

  Although this behavior increases the number of muons in air showers and reduces the tension with the observations, it was unintended and has been addressed in \sibyll~2.3d. The maximal effect occurs in the central phase space but as shown by the distribution of charged particles over pseudorapidity in the right panel in Figure~\ref{fig:pi-ratio}, the impact is small.
  
\begin{figure}
    \includegraphics[width=0.48\columnwidth]{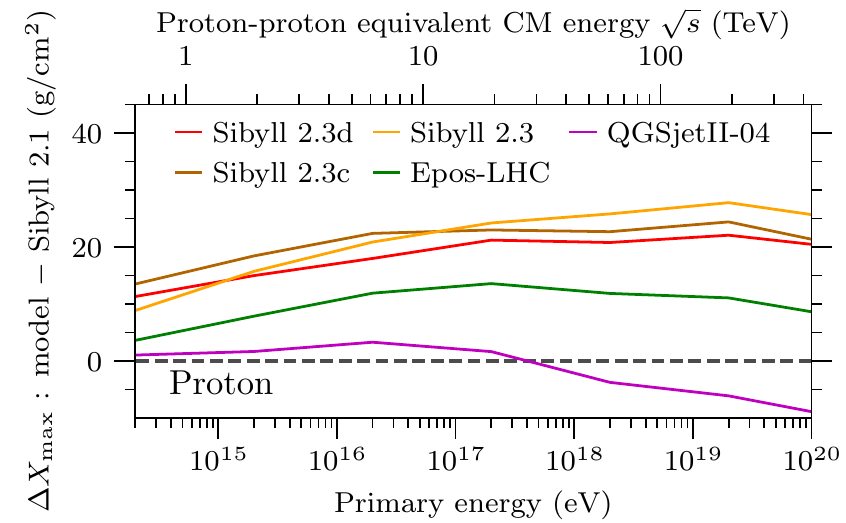}
    \includegraphics[width=0.48\columnwidth]{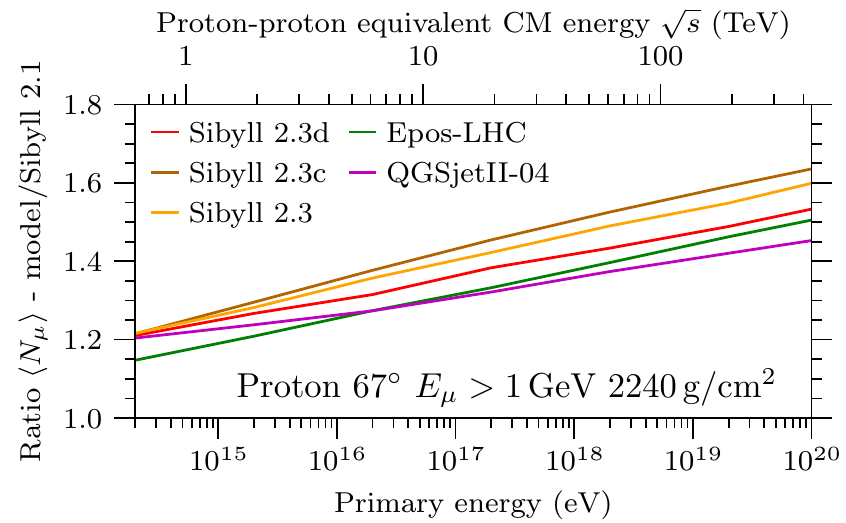}
    \caption{\label{fig:predict-sib-versions} Difference of $\xmax$ and $\langle \nmu \rangle$ predictions in the different versions of \sibyll relative to \sibyll~2.1.}
  \end{figure}

  In general the different versions of \sibyll~2.3 have rather small effects on air-shower observables. The differences in $\xmax$ for proton induced showers is shown in the left panel in Figure~\ref{fig:predict-sib-versions}, which are up to $5\,$g$/$cm$^2$ at high energies. The muon number at $10^{19}\,$eV is $\approx 7\,$\% smaller in \sibyll~2.3d than in \sibyll~2.3c (right panel in Figure~\ref{fig:predict-sib-versions}). We verified that these two releases have almost identical inclusive lepton fluxes (as in~\cite{Fedynitch:2018cbl}).

  \begin{table*}
    \centering
    \caption{Summary of the recent versions of \sibyll{}. 
      \label{tab:sib-models}}
    \renewcommand{\arraystretch}{1.5}
    \begin{tabular}{ccc}
      \hline
      Label & description & detailed description \\      
      \hline
      \sibyll~2.1 & initial implementation of the model described in Sec.~\ref{sec:basic} & ~\cite{Ahn:2009wx} \\
      \sibyll~2.3 & significant model extension (Sec.~\ref{sec:xsctn}-Sec.~\ref{sec:meson-nucleus}) & \cite{Riehn:2015oba} \& this publication \\
      \sibyll~2.3c & restored Feynman scaling in frag.\ region & \cite{Riehn:2017mfm,Fedynitch:2018cbl} \\
      \sibyll~2.3d & restored $\pi^\pm/\pi^0$ in minijets & this publication \\
      \hline
    \end{tabular}
  \end{table*}

  \section{Tables of typical air-shower observables}
  \label{app:eas_obs}
  \begin{table}[h!]
    \centering
    \caption{Predictions for the depth of shower maximum, the fluctuations thereof and the number of muons in \sibyll~2.3d for proton and iron induced showers. $X_{\rm max}$ is calculated by fitting a parabola to the profile of energy deposit in the atmosphere. The number of muons is taken at a depth of $\unit[2030]{g/cm^2}$, counting all muons with an energy exceeding \unit[1]{GeV}. Showers were simulated with an inclination of $\unit[67]{^\circ}$ using CONEX hybrid simulations~\cite{Bergmann:2006yz}. The Monte Carlo to cascade threshold was set to $E_{\rm thr}/E_0=\unit[10^{-2}]{}$.\label{tab:predict}}
    \renewcommand{\arraystretch}{1.5}
    \begin{tabular}{ccccccc}
      \hline
      $\log_{10} (E_0/\unit[]{eV})$ &  $\langle X_{\rm max}\rangle$ $\unit[]{(g/cm^2)}$ & & $\sigma(X_{\rm max})$ $\unit[]{(g/cm^2)}$ & & $\ln{ N_\mu(E_\mu>\unit[1]{GeV})}$ & \\
      & p & Fe & p & Fe & p & Fe \\
      \hline
      14.3  &  530.52  &  370.01  &  104.09  &  32.45 & 6.92 & 7.32	   \\ 
      15.3  &  596.72  &  457.36  &  89.84  &  29.53  & 9.04 & 9.35	  \\ 
      16.3  &  655.97  &  538.14  &  77.49  &  27.06  & 11.18& 11.47  \\ 
      17.3  &  715.34  &  607.59  &  72.12  &  25.18  & 13.32& 13.6	  \\ 
      18.3  &  775.18  &  671.34  &  63.41  &  23.42  & 15.45& 15.73  \\ 
      19.3  &  833.58  &  732.12  &  62.09  &  21.83  & 17.6 & 17.87	  \\ 
      20.3  &  892.46  &  791.7  &  61.26  &  20.6    & 19.79& 20.01   \\ 
      \hline
    \end{tabular}
  \end{table}

  \begin{table}[h!]
    \caption{Prediction of the interaction length of various particles
      in the atmosphere in \sibyll. The relative increase with respect
      to \sibyll~2.1 in percent is given in
      parentheses. \label{tab:int_length}}
    \begin{center}
      \renewcommand{\arraystretch}{1.5}
      \begin{tabular}{cccccc}
        \hline
      $\log_{10} (E_{\rm Lab}/\mathrm{TeV})$  &  &  & $\lambda_{\rm int}(E_{\rm Lab})$ (g$/$cm$^2$)  &  & \\
        & Fe & N & p & $\pi$ & K  \\
        \hline
0.0 & 13.02 (0.7) & 24.57 (0.7)  & 84.62 (3.6)  & 110.94 (3.6)  & 121.93 (3.4)  \\ 
1.0 & 12.67 (0.8) & 23.63 (0.5)  & 78.69 (4.8)  & 101.39 (3.7)  & 110.02 (4.6)  \\ 
2.0 & 12.10 (0.3) & 22.36 (0.2)  & 72.17 (5.5)  & 87.13 (0.2)  & 94.75 (1.9)  \\ 
3.0 & 11.56 (0.9) & 21.00 (1.3)  & 65.27 (7.0)  & 72.91 (-0.3)  & 76.54 (-0.3)  \\ 
4.0 & 11.03 (2.4) & 19.62 (3.1)  & 58.89 (8.5)  & 63.61 (1.4)  & 66.35 (1.6)  \\ 
5.0 & 10.48 (3.8) & 18.25 (4.7)  & 53.34 (9.9)  & 56.23 (2.6)  & 58.35 (2.8)  \\ 
6.0 & 9.93 (5.1) & 16.96 (6.1)  & 48.61 (11.4)  & 50.32 (3.4)  & 52.01 (3.6)  \\ 
7.0 & 9.42 (6.2) & 15.93 (7.9)  & 44.57 (12.9)  & 45.47 (4.0)  & 46.87 (4.2)  \\ 
8.0 & 8.94 (7.0) & 15.10 (9.9)  & 41.07 (14.3)  & 41.40 (4.6)  & 42.60 (4.7)  \\ 
      
\hline
      \end{tabular}
    \end{center}
  \end{table}
  
  \clearpage
  \section{Tables of interaction model parameters}
  \label{app:model_parameters}
  \begin{table}[h!]
    \caption{Summary of the differences between \sibyll~2.1 and
    \sibyll~2.3d.\label{tab:models}}
  \begin{center}
    \renewcommand{\arraystretch}{1.5}
    \begin{tabular}{ccc}
      \hline
      &  \sibyll~2.1 & \sibyll~2.3d \\
      \hline
      Valence quarks and leading particles & ``valence string'' model \&  & Remnant model \\
      & leading fragmentation & \\
      Lund string parameters & $a=0.3$, $b=0.8\,$c$/$GeV$^{-2}$ &  $a=0.8$, $b=0.8\,$c$/$GeV$^{-2}$\\
       &  ($a=0$ : leading qq, $a=a+3$ : s quarks) & (universal) \\
      String $p_{\rm T}$ & Gaussian & Exponential \\
      Flavors in hadronization & u, d, s & u, d, s, c \\
      Beam particles & p, n, $\pi$, K &  p, n, $\pi$, K + $\Sigma^{\pm}$, $\Lambda^0$, $\rho^{0}(\gamma)$, charm \\
      Interaction cross sections & p, $\pi$, K & p, $\pi$, K \\
      Target nuclei & Air & Air, $A=2$-$18$ \\
      Nuclear diffraction & Incoherent & Coherent + incoherent \\
      
      \hline
    \end{tabular}
  \end{center}
\end{table}

\begin{table}[h!]
  \caption{Summary of the amplitude parameters in \sibyll~2.1 and \sibyll~2.3d. Wherever the parameters remain unchanged only \sibyll~2.1 is reported.  \label{tab:amplitude}}
  \begin{center}
    \renewcommand{\arraystretch}{1.5}
    \begin{tabular}{ccc}
      \hline
       & \sibyll~2.1 & \sibyll~2.3d \\
      \hline
      & & \\
      Hard minijets & Leading-order QCD with energy-dependent $p_{\rm T}$-threshold & \\
      PDF: cross section & GRV-98LO~\cite{Gluck:1994uf,Gluck:1998xa} & GRV-98LO \\
      PDF: sampling & Eichten \textit{et al.}~\cite{Eichten85a} & GRV-98LO \\
      Higher-order correction ($K$-factor) & $2.0$ & \\
      $p_{\rm T}$ cut ($p^0_{\rm T}$, $\Lambda_{\rm QCD}$, $c$ in Eq.~\eqref{eq:pt-cut}) &  $1.0\,$GeV$/$c, $0.065\,$GeV$/$c, 0.9 & \\
      Profile width ($\nu_{\rm h}$ in Eq.~\eqref{eq:hard-prof}) & $0.77\,$GeV$^2/$c$^2$ & $1.0\,$GeV$^2/$c$^2$ \\
      & & \\
      Soft minijets & Gribov-Regge parameterization: $\mathcal{X}\,(s/s_0)^{\Delta}+\mathcal{Y}\,(s/s_0)^{-\epsilon}$ & \\
      Pomeron parameters ($\Delta$, $\mathcal{X}$) & 0.025, 49.9$\,$mb & 0.051, 39.2$\,$mb   \\
      Reggeon parameters ($\epsilon$, $\mathcal{Y}$) & 0.4, $8.2 \cdot 10^{-5}\,$mb & 0.4, 42.1$\,$mb \\
      Profile width, Pomeron: $ B_{\rm eff} + \alpha^{\prime}_{\Pom}(0) \ln(s) $ & 3.2$\,$GeV$^{-2}$, 0.25$\,$GeV$^{-2}$ & \\

      Profile width, Reggeon: $ B_{\rm eff} + \alpha^{\prime}_{\Reg}(0) \ln(s) $ & 0.5$\,$GeV$^{-2}$, 0.9$\,$GeV$^{-2}$ & \\
      Soft PDF ($d$, $m_{\rm q}^2$ in Eq.~\eqref{eq:soft-x}) & 0, $1.0\,$GeV$^2$ & 3, $1.0\,$GeV$^2$ \\
      \hline
    \end{tabular}
  \end{center}
\end{table}

\end{appendix}

\end{document}